\documentclass[acmsmall,screen,nonacm]{acmart}

\setcopyright{none}

\usepackage{xspace}
\usepackage{booktabs}
\usepackage{listings}
\usepackage{tabularx}
\usepackage{ragged2e}
\usepackage{tikz}
\usetikzlibrary{positioning,fit,arrows.meta,backgrounds}
\usetikzlibrary{patterns}
\usepackage{pgfplots}
\pgfplotsset{compat=1.18}
\usepackage[capitalise,nameinlink,noabbrev]{cleveref}
\crefname{section}{\S}{\S\S}
\Crefname{section}{\S}{\S\S}
\crefname{subsection}{\S}{\S\S}
\Crefname{subsection}{\S}{\S\S}
\crefname{subsubsection}{\S}{\S\S}
\Crefname{subsubsection}{\S}{\S\S}
\newcommand{\sectionrefstyle}[1]{%
  \crefformat{#1}{##2\S##1##3}%
  \Crefformat{#1}{##2\S##1##3}%
  \crefrangeformat{#1}{##3\S\S##1##4\crefrangeconjunction##5##2##6}%
  \Crefrangeformat{#1}{##3\S\S##1##4\crefrangeconjunction##5##2##6}%
  \crefmultiformat{#1}{##2\S\S##1##3}{ and~##2##1##3}{, ##2##1##3}{ and~##2##1##3}%
  \Crefmultiformat{#1}{##2\S\S##1##3}{ and~##2##1##3}{, ##2##1##3}{ and~##2##1##3}%
}
\AtBeginDocument{%
  \sectionrefstyle{section}%
  \sectionrefstyle{subsection}%
  \sectionrefstyle{subsubsection}%
}

\newif\ifdrafttodos
\drafttodostrue
\ifdefined\nodrafttodos\drafttodosfalse\fi
\ifdrafttodos
  \newcommand{\todo}[1]{{\color{red}\sffamily\small[TODO: #1]}}
  
  \newcommand{\planned}[1]{{\color{blue}\sffamily\small[PLANNED: #1]}}
\else
  \newcommand{\todo}[1]{}
  
  \newcommand{\planned}[1]{}
\fi

\newcommand{\sys}{\textsc{buildlog}\xspace}

\newcommand{\role}[1]{\emph{#1}}
\newcommand{\supp}[1]{\S\ref{#1}}
\newcommand{\unv}{\ensuremath{^\dagger}}

\definecolor{lstcomment}{gray}{0.42}
\definecolor{lstkeyword}{rgb}{0.05,0.20,0.55}
\definecolor{lststring}{rgb}{0.10,0.42,0.12}
\lstdefinelanguage{Datalog}{
  morekeywords={.decl,.type,.input,.output},
  sensitive=true,
  morecomment=[l]{//},
  morestring=[b]",
  literate={:-}{{\textcolor{lstkeyword}{\textbf{:-}}}}2,
}
\lstdefinelanguage{TOML}{
  morekeywords={true,false},
  sensitive=true,
  morecomment=[l]{\#},
  morestring=[b]",
}
\lstdefinelanguage{CMake}{
  morekeywords={add_library,add_executable,target_include_directories,target_link_libraries},
  morekeywords=[2]{PUBLIC,PRIVATE,INTERFACE},
  sensitive=true,
  morecomment=[l]{\#},
}
\begin{document}

\title{Meta Build Systems as Logic Programs}

\renewcommand{\email}[2][]{}
\author{Alex Reinking}
\orcid{0000-0003-3285-2112}
\email{alex.reinking@gmail.com}
\affiliation{\institution{Adobe Research}\country{USA}}

\authorsaddresses{}

\begin{abstract}
Meta build systems such as CMake, Meson and Bazel derive a build's command graph from
project descriptions, user options and facts about the machine, reconciling what a project
requires, what its dependencies contribute and how the user wants it built.
Because existing systems scatter this logic across imperative scripts, search paths and
special cases, routine decisions interact poorly: search priorities can resolve compatible
packages without an explicit project choice, authors hard-code flag preferences that
packagers must patch out, and cross-builds require manual annotations or nested build trees
for tools run during the build.

Our central insight is to express configuration as relational inference: objects follow
from sources, flags and libraries from dependencies, and language standards from
transitive requirements.
We introduce \sys, a meta build system whose stratified, logical core isolates the build
derivation from impure environmental probes.
We treat ambiguity in toolchain or package selection as an error, and provide a tool to
explain any derived fact from its provenance.
When cross-compiling, \sys infers the necessary build-machine configurations through
simple use of a program in the definition of a build rule.
Author constraints and user preferences are modeled as distinct input relations, and their
separation is statically checked.
Libraries can ship domain-specific compilation rules to their consumers, including Qt's
\texttt{moc}/\texttt{rcc} rules, and installed and built libraries are described by the
same facts.

We evaluate \sys on eleven case studies of ten projects, validating the ported
configurations through execution and artifact checks while reporting remaining flag
differences and platform limits.
In our largest study, Halide, generators link an LLVM-based compiler and run during the
build to emit WebAssembly or Android code; where CMake requires nested builds, \sys
derives the build-machine and target-platform configurations in a single graph.
A supporting formal account establishes termination, within-round probe-order independence
and agreement between fresh and validated cached runs under explicit finiteness and probe
obligations.
The results show that confining impurity to fact discovery and treating the rest of
configuration as logical entailment can replace a collection of special cases with a
small, uniform and explainable semantics.
\end{abstract}



\maketitle

\section{Introduction}
\label{sec:intro}

A build system runs commands, but something must first decide which commands to run.
A \emph{meta build system} such as CMake, Meson or Bazel computes this command graph from
project descriptions, user choices and facts about the machine, specifying which artifacts
to produce, with which flags and libraries, and for which platform.
An executor such as Ninja~\cite{martin2013ninja} or Bazel's runner then executes the graph;
we study the configuration step before it, deriving a graph whose decisions are explicit,
explainable and properly divided among the parties to a build.

Most of this computation is relational inference: objects follow from sources and target
configurations, a library's \emph{usage requirements} (the flags and libraries its
consumers need) propagate along public or private dependencies
(\cref{sec:background:closure}), and a target's language standard follows from its own and
its dependencies' minimum requirements.
In effect, the configured graph is a logic program's model.
Existing systems leave that program implicit, mixing derivation with imperative probes,
search and ad hoc policy, and three recurring problems follow.

\emph{Selection can hide a policy decision.}
When several packages or toolchains are compatible, CMake's \texttt{find\_package} takes
the first viable package in search order~\cite{cmake-find-package} and Bazel selects the
first compatible registered toolchain~\cite{bazel-toolchains}.
Although these first-match priorities are documented, they resolve ambiguity by search
order rather than requiring an explicit choice among candidates.

\emph{Constraints and preferences become entangled.}
Because build languages do not separate author constraints from user preferences, authors
routinely hard-code both.
LASzip's build fixes \texttt{-std=c++17}~\cite{laszip-cmake} when C++17 is only the
minimum its code needs, forcing vcpkg and Conan to patch it out instead of letting the
standard follow from the target and its dependencies (\cref{sec:design:lattice}).
More broadly, Linux and BSD packaging policies require undoing upstream rules that override
user compilers, flags or dependency choices (\cref{sec:background:patches}).

\begin{figure}[t]
\centering
\begin{tikzpicture}[
  font=\footnotesize,
  >=Stealth,
  art/.style={draw, rounded corners=2pt, align=center, text width=1.6cm, minimum height=0.95cm,
    inner sep=2pt, fill=white},
  runstep/.style={draw, rounded corners=9pt, align=center, text width=1.6cm, minimum height=0.95cm, inner sep=2pt, fill=white,
    thick},
  lab/.style={font=\scriptsize, align=center, inner sep=1pt},
]
\node[art] (src) at (0, 0) {generator\\[-1pt]\scriptsize (C++)};
\node[art] (exe) at (2.8, 0) {executable\\[-1pt]\scriptsize for the build machine};
\node[runstep] (run) at (5.6, 0) {run on the build machine};
\node[art] (obj) at (8.4, 0) {machine code\\[-1pt]\scriptsize for the device};
\node[art] (app) at (11.2, 0) {app\\[-1pt]\scriptsize for the device};
\node[art, minimum height=0.6cm] (own) at (11.2, -1.5) {app's own code};

\begin{scope}[on background layer]
  \node[fill=black!10, rounded corners=3pt, fit=(src)(exe)(run), inner xsep=6pt,
    inner ysep=13pt, yshift=6pt] (host) {};
  \node[draw, dashed, rounded corners=3pt, fit=(obj)(app)(own), inner xsep=6pt,
    inner ysep=13pt, yshift=6pt] (target) {};
\end{scope}
\node[lab, anchor=north west] at ([shift={(3pt,-3pt)}]host.north west) {\textbf{build-machine side}};
\node[lab, anchor=north west] at ([shift={(3pt,-3pt)}]target.north west) {\textbf{target-platform side}};

\draw[->] (src) -- node[lab, above] {compiler} (exe);
\draw[->] (exe) -- (run);
\draw[->, thick] (run) -- node[lab, above] {writes} (obj);
\draw[->] (obj) -- node[lab, above] {linker} (app);
\draw[->] (own) -- node[lab, left, align=right] {compiler} (app);
\end{tikzpicture}
\caption{How an app uses Halide. A C++ program (a generator) is compiled for and run on the build
machine; the machine code it writes is linked into the app by the target-platform toolchain.
Halide's CMake build configures the two sides separately; \sys derives both configurations in one graph.}
\Description{A left-to-right flow in two regions. In a shaded build-machine region, a C++ generator is
compiled into an executable for the build machine, which is then run there.
The run step writes machine code for the device into a dashed target-platform region,
where the target-platform linker combines it with the app's own code, compiled for that platform,
into the app.}
\label{fig:halide-pattern}
\end{figure}
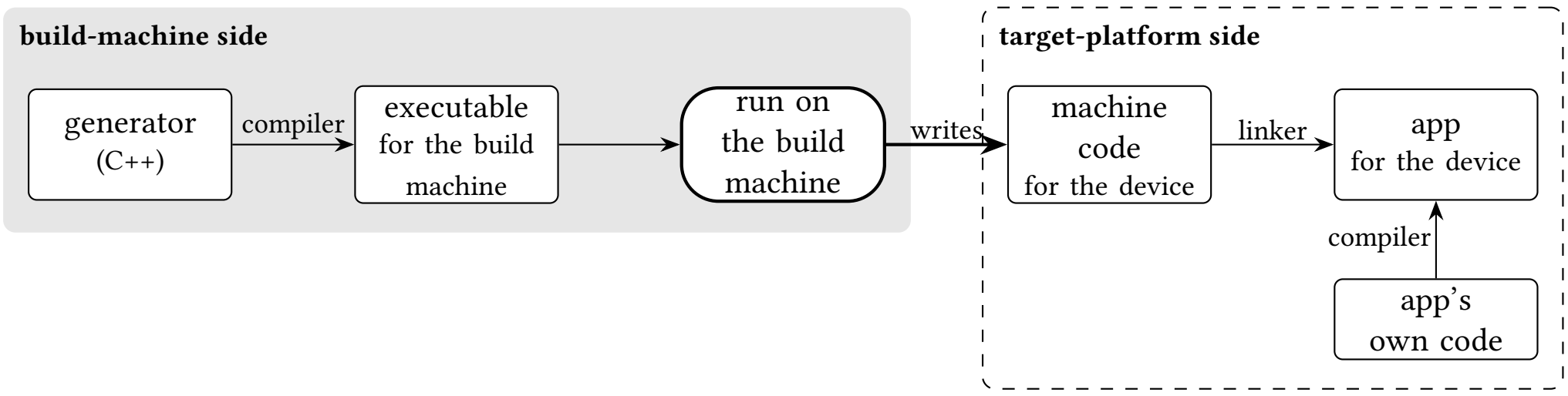

\emph{Programs run during a build need special handling.}
A Halide~\cite{ragankelley2013halide} cross-build compiles a C++ generator linked to Halide
and LLVM, runs it on the build machine, and links its output into an app for the target
platform (\cref{fig:halide-pattern}).
Although the generator's execution platform follows from the step that runs it, Meson and
Bazel require explicit tool annotations (\texttt{native:~true} and
\texttt{cfg = "exec"}~\cite{meson-cross,bazel-config}), and CMake's single-toolchain-per-tree
limit forces separate build trees and nested configuration---even to cross-build libHalide
itself from macOS to AArch64 Linux.

Although all three problems concern decisions about one command graph, existing systems
lack a shared account of how that graph follows from its inputs, and prior formal work
models graph execution or CMake scripts rather than treating the configured graph as a
logic program's model (\cref{sec:related}).
We make that program explicit in \sys, a meta build system whose stratified Datalog core
is evaluated by Souffl\'e~\cite{jordan2016souffle} from TOML or Datalog specifications and
paired with a Python driver that emits Ninja.
Two principles guide its design.

First, \emph{separate discovery from derivation}.
Plugins probe the machine only for environmental facts the program requests, and pure rules
derive the build from their answers; ambiguous toolchain or package candidates are errors
listing the alternatives, not implicit selections.

Second, \emph{each party states its own decisions}.
Authors state constraints and users state preferences through distinct, statically checked
input relations, while libraries ship domain-specific consumer rules (such as Qt's
\texttt{moc}/\texttt{rcc} rules) using the same facts for installed and built code, and
every derived fact has a provenance proof.
Running a compiled program in a build step likewise entails its build-machine
configuration, so \sys derives Halide's generator and target product in one graph;
\cref{sec:cs:first} validates a libHalide cross-build against CMake on a 32-test subset.

Our contributions are:
\begin{itemize}
  \item \emph{A shared configuration model} (\cref{sec:design}) unifying library consumer
    rules, inferred tool configurations, checked author/user inputs and provenance in one
    composition (\cref{sec:design:loop}).
  \item \emph{A supporting formal account} (\cref{sec:formal,tab:guarantees}) proving
    termination, within-round probe-order independence and validated-cache agreement under
    stated obligations, mechanized in Lean.
  \item \emph{Eleven case studies of ten projects} (\cref{sec:case-studies,sec:evaluation}),
    comparing builds, generated commands and configuration costs against CMake, Meson and
    Bazel.
\end{itemize}

\section{Background and Motivation}
\label{sec:background}

\subsection{Build Systems and Meta Build Systems}
\label{sec:background:meta}

Following \citet{mokhov2018bsalc}, a \emph{build system} brings a store of keys and values up
to date using a task description. A \emph{meta build system} computes that task graph from
build code, user choices and environmental probes. We study this latter computation, whether
its executor is separate (CMake, Meson, GN) or integrated (Bazel, Buck2, Pants).

The \role{author} writes the build code; the \role{user}, such as a packager, invokes and
configures it. Some inputs are \emph{hard constraints} only the author can know; others are
\emph{preferences} of the user (\cref{sec:background:patches}). Conflating them forces users to
become authors.

\subsection{The Closure at the Core}
\label{sec:background:closure}

Consider a library \texttt{z} exporting an \texttt{include} directory, a library \texttt{png}
that depends publicly on \texttt{z}, and an executable \texttt{app} that depends privately on
\texttt{png}. In CMake:
\begin{lstlisting}[language=CMake,xleftmargin=0.12\linewidth]
add_library(z ...)        target_include_directories(z PUBLIC include)
add_library(png ...)      target_link_libraries(png PUBLIC z)
add_executable(app ...)   target_link_libraries(app PRIVATE png)
\end{lstlisting}
\texttt{app} receives \texttt{z}'s include directory and links \texttt{png}, then \texttt{z}.
CMake defines this through the transitive closure of
\texttt{INTERFACE\_LINK\_LIBRARIES}~\cite{cmake-buildsystem}.
As Datalog: interfaces propagate through public edges; linking follows all edges.
\begin{lstlisting}[language=Datalog,xleftmargin=0.22\linewidth]
sees(t, d)        :- dep(t, d, _).
sees(t, e)        :- sees(t, d), dep(d, e, "public").
links(t, d)       :- dep(t, d, _).
links(t, e)       :- links(t, d), dep(d, e, _).
use(t, k, v)      :- own_use(t, k, v).
use(t, k, v)      :- sees(t, d), iface(d, k, v).
compile_arg(o, v) :- object(o, t, _), use(t, "include_dir", v).
\end{lstlisting}
Here \texttt{use} combines own properties (\texttt{own\_use}) and public dependency properties
(\texttt{iface}); \texttt{object} relates objects to targets.
In \sys a shared library hides its private dependencies from consumers' links (\cref{sec:design:linkage}).
GN~\cite{gn-reference}, Bazel~\cite{bazel-ccinfo} and Buck2~\cite{buck2-tsets} propagate
according to the same visibility model, but none as a logic program. \supp{sec:survey}\footnote{Lettered
sections and S-numbered figures and tables are in the supplement.} compares twelve build and
meta build systems along eleven axes, from discovery to compatibility.

\subsection{Where Meta Build Systems Depart from Logic}
\label{sec:background:departures}

Today's systems give the following policies separate mechanisms; a logic program expresses
them as rules over shared relations.

\paragraph{Discovery is entangled with derivation}
CMake's \texttt{find\_package} and \texttt{try\_compile} run in imperative control order,
with \texttt{find\_package} taking the first viable package file~\cite{cmake-find-package};
Bazel likewise selects the first compatible toolchain for each execution
platform~\cite{bazel-toolchains}, so the outcome can depend on a search order the project
never states.
In \sys, probe requests are relations over finite domains, separating the rules that ask
questions from the probes that answer them (\cref{sec:design:loop}).

\paragraph{Properties are untyped strings}
CMake computes the closure with generator expressions~\cite{cmake-genex}. Targets are a
distinguished type there, but their properties are strings, and mistakes in them are silent:
zlib~1.3.2 writes~\cite{zlib-cmake} \verb|$<$<BOOL:NOT:${HAVE_FSEEKO}>:NO_FSEEKO>|, which tests
the non-empty string \texttt{"NOT:1"} and is always true.
In a logic program a condition is a term over declared names, and \sys reports a condition it
does not know as an error.

\paragraph{Imported libraries and build-time tools are special cases}
CMake distinguishes imported from built targets and generates no rules for imported
libraries~\cite{cmake-add-library,cmake-imported-location}. Halide's generated libraries
therefore need object-as-source conventions and extra ordering targets~\cite{halide-cmake}.
Build-time tools also need explicit declarations (Bazel's \texttt{cfg = "exec"}, Meson's
\texttt{native:~true}~\cite{meson-cross}), although their execution platform follows from how they are used.
\sys models built, imported and generated artifacts uniformly and derives tool configurations
from typed uses (\cref{sec:design:generators}).

\paragraph{Configurations and compatibility are second-class}
CMake has one toolchain per language per build tree (\cref{sec:intro}), and Debug and Release are
privileged names. Compatibility has no algebra, with CMake's compatible interface
properties~\cite{cmake-compatible-interface} the clearest exception. In \sys a
configuration is a value like any other and compatibility uses declared lattices
(\cref{sec:design:lattice}).

\paragraph{Third-party library logic is adopted into the core}
\label{sec:background:qt}
CMake absorbed Qt's consumer-side logic as \texttt{AUTOMOC}~\cite{neundorf2011automoc}
and added imperative deferral for Qt~6 finalizers~\cite{qt-add-executable,cmake-issue-19575},
though code deferred later can still mutate the target.
Meson absorbs dependency-specific logic as a matter of policy~\cite{meson-faq}: its core
special-cases 38 dependencies~\cite{meson-dependencies} and ships built-in Qt, GNOME and
CUDA modules~\cite{meson-modules}. Rejecting user-defined functions to keep build files
simple merely moves reusable abstraction into Meson itself, relocating complexity rather than
bounding it. Restricted computation need not exclude abstraction: \sys's finite, stratified
rule compositions let libraries and projects supply their own reusable build logic.
In a logic program the model is a fixpoint, so a rule that negates or aggregates over a target's
sources sees all of them, and in \sys Qt's rules ship with Qt.

\paragraph{Overridden choices are defects}
\label{sec:background:patches}
Packaging policies protect user choices: FreeBSD requires respecting user compilers and
flags~\cite[Sections~13.8--13.9]{freebsd-porters-handbook}, Gentoo rejects upstream
\texttt{-Werror}~\cite{gentoo-qa-flags}, Fedora requires system rather than bundled
libraries~\cite{fedora-packaging-guidelines}, and vcpkg patches builds lacking static/shared
selection~\cite{vcpkg-maintainer-guide}.
These policies often require build-script patches, as vcpkg maintainer
Robert Schumacher recounts~\cite{schumacher2019herd}. In \sys, user preferences enter
through distinct, manifest-checked relations that author rules cannot silently override
(\cref{sec:design:inputs}).

\section{The Design of \sys}
\label{sec:design}

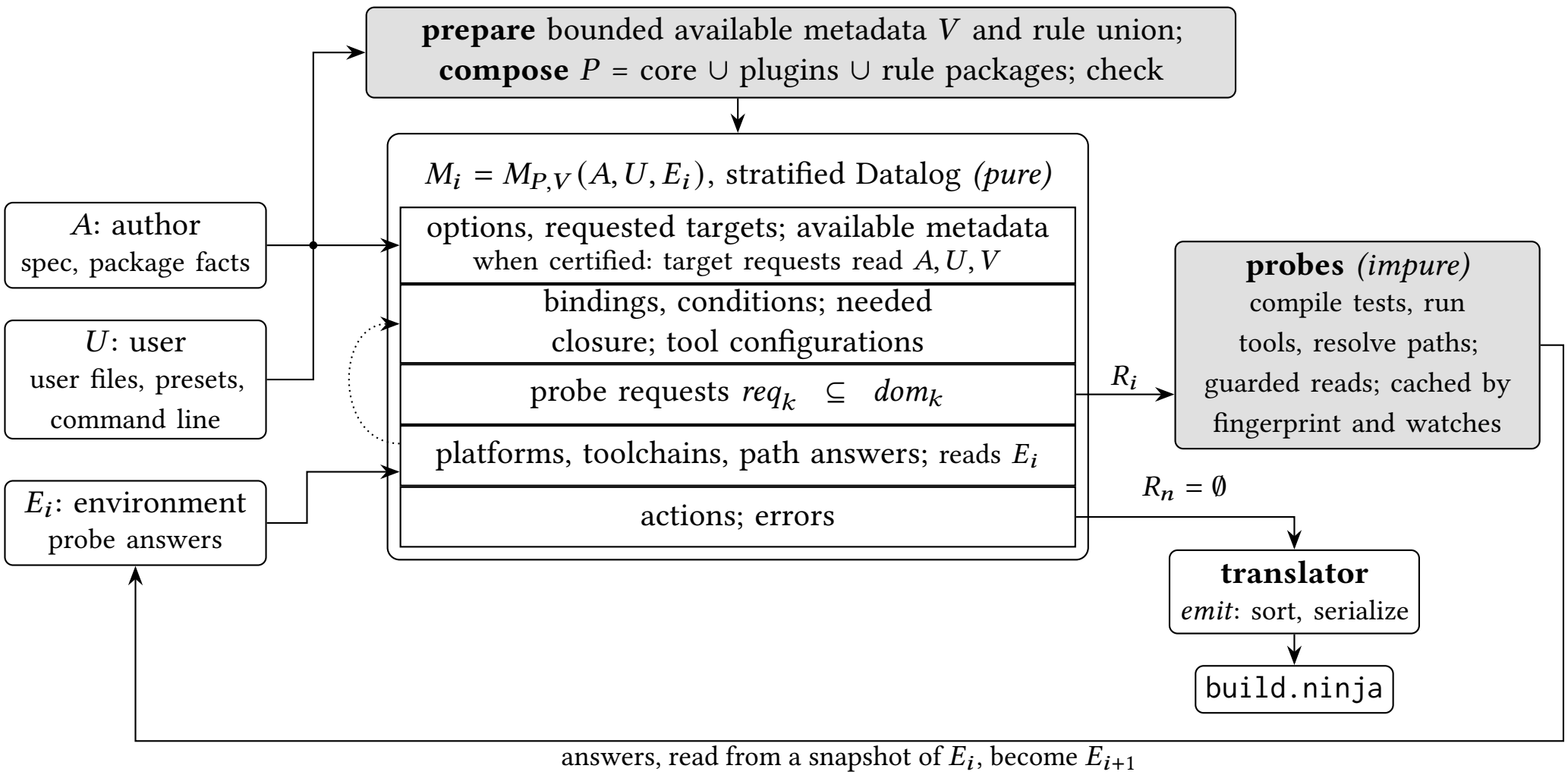
\begin{figure}[t]
\centering
\begin{tikzpicture}[
  font=\footnotesize,
  >=Stealth,
  box/.style={draw, rounded corners=2pt, align=center, inner sep=2.5pt},
  pure/.style={box, fill=white},
  impure/.style={box, fill=black!12},
  stratum/.style={draw, fill=white, text width=5.6cm, minimum height=0.52cm, align=center,
    inner xsep=4pt, inner ysep=1pt},
  lab/.style={font=\scriptsize, align=center, inner sep=1pt},
]
\node[pure, text width=2.1cm] (A) at (1.1, 2.62) {$A$: author\\[-1pt]\scriptsize spec, package facts};
\node[pure, text width=2.1cm] (U) at (1.1, 1.45) {$U$: user\\[-1pt]\scriptsize user files, presets, command line};
\node[pure, text width=2.1cm] (E) at (1.1, 0.2) {$E_i$: environment\\[-1pt]\scriptsize probe answers};

\node[stratum, draw=none, fill=none] (H) at (6.35, 3.22) {\textbf{$M_i = M_{P,V}(A, U, E_i)$}, stratified Datalog \emph{(pure)}};
\node[stratum, minimum height=0.66cm] (S1) at (6.35, 2.62) {options, requested targets; available metadata\\[-2pt]{\scriptsize when certified: target requests read $A,U,V$}};
\node[stratum, below=0pt of S1] (S2) {bindings, conditions; needed closure; tool configurations};
\node[stratum, below=0pt of S2] (S3) {probe requests $\mathit{req}_k \subseteq \mathit{dom}_k$};
\node[stratum, below=0pt of S3] (S4) {platforms, toolchains, path answers; {\scriptsize reads $E_i$}};
\node[stratum, below=0pt of S4] (S5) {actions; errors};
\node[draw, rounded corners=3pt, fit=(H)(S5), inner sep=3pt] (M) {};

\node[impure, text width=7.4cm] (L) at (6.9, 4.3)
  {\textbf{prepare} bounded available metadata $V$ and rule union;
   \textbf{compose} $P$ = core $\cup$ plugins $\cup$ rule packages; check};

\node[impure, text width=3.0cm] (probe) at (11.75, 1.75)
  {\textbf{probes} \emph{(impure)}\\[-1pt]\scriptsize compile tests, run tools, resolve paths;
   guarded reads; cached by fingerprint and watches};
\node[pure, text width=2.0cm] (emit) at (11.2, -0.4) {\textbf{translator}\\[-1pt]\scriptsize $\mathit{emit}$: sort, serialize};
\node[pure] (ninja) at (11.2, -1.25) {\texttt{build.ninja}};

\coordinate (AU) at (2.65, 2.62);
\draw (A.east) -- (AU);
\draw (U.east) -| (AU);
\fill (AU) circle[radius=1pt];
\draw[->] (AU) |- (L.west);
\draw[->] (AU) -- (S1.west);
\draw[->] (L.south -| S1.north) -- (M.north -| S1.north);
\draw[->] (E.east) -- ++(0.35,0) |- ([yshift=-4pt]S4.west);
\draw[->, densely dotted] ([yshift=3pt]S4.west) to[out=180, in=180, looseness=1.4] (S2.west);
\draw[->] (S3.east) -- node[lab, above] {$R_i$} (S3.east -| probe.west);
\draw[->] (probe.east) -- ++(0.2,0) |- node[lab, pos=0.75, below] {answers, read from a snapshot of $E_i$, become $E_{i+1}$} (1.1, -1.7) -- (E.south);
\draw[->] (S5.east) -- node[lab, above=3pt] {$R_n = \emptyset$}
  (S5.east -| emit.north) -- (emit.north);
\draw[->] (emit) -- (ninja);
\end{tikzpicture}
\caption{How \sys configures a build. Preparation fixes the composed program $P$ and available metadata $V$
before inference, with no package-driven back edge. When full-program decisions are answer-independent,
$A,U,V$ fix explicit target requests (package-use guards can otherwise depend on discovery), while discovered
facts extend the needed closure (dotted). Each round evaluates $P$ to $M_i$ and answers new probe requests
$R_i$ against a snapshot to form $E_{i+1}$. Shaded boxes perform environment I/O.}
\Description{A flow diagram. Inputs A (author), U (user) and E (environment) feed a stack of
strata inside a box labeled stratified Datalog (pure): options and requested targets (reading
A, U and fixed V when answer independence is certified), the needed closure, which a dotted edge shows is
elaborated by discovered facts, probe requests, platforms and toolchains and packages, and actions. A loader box above
prepares bounded available metadata and composes and checks the program once. Probe requests
flow to a shaded impure probes box, whose answers return to E without changing the program.
When no new probe request arises, a translator writes build.ninja. No edge returns to preparation.}
\label{fig:overview}
\end{figure}

\Cref{fig:overview} shows the three stages of a \emph{configure run}, which computes the command
graph for one or more requested \emph{configurations} (each a target platform paired with settings
such as optimization and linkage).
\emph{Preparation} first catalogs package metadata and rules within declared dependency roots.
\emph{Discovery} is impure: \emph{probes} ask the environment questions (does this compiler
accept C++20? which library exists under this sysroot?) and return the answers as facts.
\emph{Derivation} is pure: a stratified Datalog program over the author's specification, the
user's files and command line, and the probe answers, whose model contains the build graph that a
thin translator emits as Ninja.

A \emph{layer} pairs Datalog rules with a manifest declaring which relations it reads,
extends or accepts as input; layers perform no environment I/O.
The \emph{core} and installed \emph{plugins} supply layers plus Python
\emph{probe kinds} (families of environmental questions) and \emph{metadata providers}, which
enumerate package layouts within declared roots and use \emph{importers} to decode
pkg-config, CPS or CMake exports.
Libraries can also ship \emph{rule packages}---a manifest, layers and facts, with no impure
code---and the loader guards each package's \emph{module} to separate its declarations from
its effects.

\subsection{Configuring Is Deriving}
\label{sec:design:loop}

A \emph{target request} asks for a build target in a configuration. A target is \emph{needed}
if requested, or if a needed target links, loads or runs it; this transitive dependency
relation is the \emph{needed closure}.
In \cref{fig:surface}, for example, requesting \texttt{libfribidi} requires its generated
tables, whose action runs \texttt{gen-bidi-type-tab} and makes that program needed in the
build-machine configuration; Ninja later executes these derived actions, whereas probes run
only during configuration.

Packages follow a distinct three-stage progression: \emph{available} definitions participate
in name resolution, \emph{binding} selects a definition and exposes its declarations, and
\emph{use} enables build effects for needed consumers, leaving available but unused modules
\emph{dormant} rather than invisible.
Suppose the user's prefix contains Qt: preparation makes QtCore's definitions available and
composes its rules before probing, so a condition can ask whether QtCore exists without
triggering a build. Binding then resolves each target name (source declarations and aliases
take precedence, and a demanded ambiguous binding is an error), while only a needed consumer
enables effects such as \texttt{moc} actions (\cref{sec:design:generators}); unused modules
contribute neither actions nor target requests, and environmental probes never add new package
rules.

Each discovery \emph{round} evaluates the composed program and reads \emph{probe requests} from
its model (\cref{fig:overview}). Unlike a target request, a probe request asks an
environmental question, such as whether the selected compiler supports a needed language
standard. The driver checks that each request lies in its declared finite domain, runs the
round's probes against a single answer snapshot, and adds their results for the next round.
All rounds evaluate the same prepared program and metadata, and only preparation and probes
perform environment I/O (\cref{sec:implementation}).

Authored target-request and target-kind conditions may test options, not probe results, though
package-use guards remain a separate source of activation (\cref{sec:design:options}).
Within discovery rules, a \emph{guarded} answer read requires the matching probe request in the
same rule body, making any retained answer whose probe request disappears inert in that rule.
By convention, \texttt{req\_} names probe requests and \texttt{res\_} their answers; these
core rules request imported files' architectures and guard their reads:
\begin{lstlisting}[language=Datalog,xleftmargin=0.08\linewidth]
// c: config; f: file; p: platform; x: format
req_file_archs(f)  :- import_file_in(_, f).
covers(f, p)       :- res_file_arch(f, a), req_file_archs(f), platform_cpu(p, a).
archs_known(f)     :- res_file_format(f, x), x != "unknown", req_file_archs(f).
error(c, f) :- import_file_in(c, f), config_platform(c, p), archs_known(f), !covers(f, p).
\end{lstlisting}
The error rule negates \texttt{covers} only after \texttt{archs\_known} confirms the probe has answered.

The loader inserts binding guards on package declarations and use guards on package effects,
including fact payloads and disjuncts (\supp{sec:formal-supp:module-guards}).
Because errors propagate to dependents only after the final round completes, discovery rules
chain on positive ``not found'' facts rather than on failure.
Given the probe answers, the stratified model is unique; within-round probe-order independence
holds under the hypotheses of \cref{sec:formal:discovery}, and the \texttt{why} command
reconstructs the proof of any derived fact from source and probe leaves (\cref{sec:design:why}).

\subsection{The Core and Its Layers}
\label{sec:design:core}

The model holds the whole configured build, down to the commands. The bottom of the core is a
set of action relations, the only relations the translator reads: an action's outputs, typed
inputs, dependency file, arguments and Ninja flags. Arguments whose order matters form a set with
\emph{before} edges, sorted topologically with a lexical tie-break, so even a link line's order is
derived. The translator (about 330 lines of Python) derives nothing; the core checks one action per
output, that every input exists or is produced, and that there are no cycles.
Any layer or project can add actions directly; an action's \emph{words} are its command-line argument tokens.
A raw action's words are taken as written, checked
for integrity but not inspected for tools. Everything above is a lowering into them, a design
inspired by MLIR's dialects~\cite{lattner2021mlir}. A Halide generator lowers to libraries,
libraries to objects and link items, and those to actions, so a provenance explanation follows
a derived flag through every lowering level (\cref{sec:design:why}).

The next layer knows no language. It holds targets, dependency edges with a visibility, and
properties that propagate along edges as in \cref{sec:background:closure}.
Edges and properties may have conditions. A condition can compare a discovered version,
such as \texttt{cxx.version >= 13}; reads of discovered values must be guarded.
\label{sec:design:lattice}
Every property key declares an algebra, most of them sets. A \emph{lattice} key's value is the
join of the values a target states and inherits: needing C++20 and depending on a C++17 library
gives C++20, and a join the selected compiler does not support (probed) is an error blaming the
target, configuration and compiler.
Each \emph{setting} key is declared by the layer that owns it, with finite values, a default
and a \emph{tool policy}, which says how the key changes for tools. The
core declares optimization, sanitizer and linkage, and plugins add more (e.g., the Python plugin
adds the interpreter), so release, debug and an address-sanitized build are declarations, none
privileged over other configurations.
Every configuration $c$ has a \emph{tool configuration}, in which programs that run during a
build of $c$ are built: the build machine's platform, with each setting rewritten by its key's
policy (tools are optimized, unsanitized, and inherit the linkage). In a native build whose product settings match the tool policy, the two configurations
coincide and one target instance serves both roles, so the tool is built only once
(Invariant~\ref{prop:exec}).

Platforms and toolchains are facts that plugins derive from probes. For each language, build
platform and target platform, two candidate toolchains without a \emph{pin} (an explicit selection) are an error; a pin specifies the
whole toolchain key, since in a cross build ``use gcc for C++'' could mean the product's compiler or the
tools'.
Native code (objects, libraries, executables, link closure and order) is language-neutral and
in the core. The C/C++ plugin, \texttt{buildlog-cpp}, adds clang toolchain inventory, C and
C++ layers, C configure checks (header, symbol and compile tests), instruction-set knowledge, install rules and importers for
pkg-config, CPS~\cite{cps-spec} and CMake's export files. A Python plugin builds
extension modules.

\paragraph{A surface that desugars}
\label{sec:design:surface}
Authors write per-subject tables in TOML (\cref{fig:surface}). The loader desugars them into
rows of the core's relations and records the line that wrote each row, which explanations cite
(\cref{sec:design:why}); the tables have no semantics of their own. The surface does not hide
the fact schema, which comprises every relation that the core and the loaded plugins declare.
Any table can be replaced by the rows it stands for, and a project can add Datalog rules that
write these relations, subject to the manifests' ownership checks. The schema is the final
lowering and the escape hatch, and project rules are how larger builds such as zlib-ng and
Halide get abstraction.
\label{sec:design:templates}
A \emph{target template} also desugars to rows: those of a target with named parameters in its
values (\texttt{"src/arch/\{isa\}/impl.c"}). Its instances are rows that rules could write; what
the template adds is that rules can derive the instances, as ggml derives one per ISA level from
the platform's features, and each instance's flags have provenance through the template line.
Parameters may not depend on the needed closure, because they decide edges and sources.

\begin{figure}[t]
\begin{lstlisting}
config_decl = [["debug", "arm64-macos"]]   # the configurations the author supports
request_all = [["libfribidi"]]             # built in every configuration

[option.TOOLS]                         # an option: default off
help = "build the command-line tools"

[target.libfribidi]
kind = "library"                       # linkage is chosen, not fixed by the author
sources = ["lib/fribidi.c", "lib/fribidi-bidi.c"]
public = { "cc.include_dir" = "src:lib" }
gen_include = { private = [""] }       # its generated files are on its own path

[target.gen-bidi-type-tab]             # a program; nothing says where it runs
kind = "executable"
sources = "gen.tab/gen-bidi-type-tab.c"

[generator.bidi_type]                  # runs the program, captures its output
owner = "libfribidi"                   # "2" below: fribidi's table compression depth
run = ["gen-bidi-type-tab", "2", "${src:gen.tab/unidata/UnicodeData.txt}"]
stdout = "${out:bidi-type.tab.i}"

[target.fribidi-cli]
kind = "executable"
sources = "bin/fribidi-main.c"
deps = { private = ["libfribidi"] }
request_when = "TOOLS"                 # decides what is built; may test only options
\end{lstlisting}
\caption{A specification modeled on the fribidi study (the \texttt{TOOLS} option is ours); every TOML
table desugars to core relations. The \texttt{src:} prefix resolves relative to the source root, and
\texttt{owner} attaches the generator's action to \texttt{libfribidi}, causing \texttt{gen-bidi-type-tab}
to be built for the build machine.}
\label{fig:surface}
\Description{A TOML listing with a declared configuration and a request, an option, a library,
a table-generating program, a generator that runs it and captures its output into the library's
generated directory, and a command-line program requested only when the option is on.}
\end{figure}

\subsection{Authors and Users}
\label{sec:design:inputs}

Each layer's manifest separates \emph{author inputs}, such as targets, dependencies and
defaults, from \emph{user inputs}, such as toolchain pins, package locations, linkage and
configurations. Authors and packages may supply the former; user files, user-selected
presets and the command line supply the latter. Dependencies' user facts are not inherited.
An author-supplied user relation is an error unless listed under \texttt{author\_decides}
with a reason, which diagnostics quote. These checks do not ensure that arbitrary compiler
or linker flags respect user preferences.

\Cref{fig:walkthrough} contrasts an ordinary preference with a legitimate author constraint.
The user overrides \texttt{codec}'s default linkage in one line, which the explanation cites
(\cref{sec:design:why}). By contrast, \texttt{filters}' static initializers require static
linkage. The author states that reason explicitly, and a conflicting user choice is an error
that cites both sources and gives the fix.

\begin{figure}[t]
\begin{lstlisting}
default_linkage = [["codec", "shared"]]  # a preference: the user may change it
linkage_prop = [["codec", "static", "cc.define", "CODEC_STATIC", "public"]]

[target.filters]
kind = "library"
sources = "src/filters.c"
linkage = { require = "static", why = "registers filters from static initializers" }
\end{lstlisting}
\vspace{-5pt}\noindent\rule{\linewidth}{0.4pt}\vspace{1pt}
\begin{lstlisting}[language={},basicstyle=\ttfamily\footnotesize,breaklines=true,literate={<-}{{$\leftarrow$}}2]
$ W=walkthrough
$ buildlog configure -C build-default $W    # libcodec.dylib, libfilters.a
$ cat $W/user-static.toml
library_linkage = [["codec", "static"]]
$ buildlog configure --user $W/user-static.toml -C build-static $W  # both .a
$ buildlog why -C build-static --flag -DCODEC_STATIC \
    --output viewer.c.o --leaves $W
  library_linkage("codec", "static")  <- W/user-static.toml:1
  linkage_prop("codec", "static", "cc.define", "CODEC_STATIC", "public")  <- W/spec.toml:4
  target_deps("viewer", "private", "codec")  <- W/spec.toml:18  ... (8 more leaves)
$ cat $W/user-shared.toml
library_linkage = [["filters", "shared"]]
$ buildlog configure --user $W/user-shared.toml -C build-shared $W
error: user linkage contradicts the library's requirement: filters: the user's library_linkage
  asks for shared, but its author requires static (registers filters from static initializers);
  fix: drop library_linkage(filters, shared) or state static
\end{lstlisting}
\caption{The author/user split on linkage, excerpted from a runnable project. Above the rule:
the author's statements. Below: the user's files, three configures and one explanation; comments
give the libraries built. \texttt{W} abbreviates \texttt{walkthrough} in source citations.}
\label{fig:walkthrough}
\Description{A TOML excerpt with an author default linkage and a reasoned linkage requirement,
followed by a terminal session: a default configure, a user override that makes codec static,
a why query whose leaves cite the user file and the spec, and a conflicting user choice that
fails with an error citing both sources.}
\end{figure}

\begin{samepage}
Three decisions that project authors commonly hard-code in build scripts show how this
manifest split works in practice: library linkage, compiler warnings and header include order.

\paragraph{Linkage}
\label{sec:design:linkage}
A built library's linkage is a parameter, derived per configuration from, in order, a reasoned
requirement, the user's per-library and global choices (as CMake's \texttt{BUILD\_SHARED\_LIBS}),
the author's default, and static; a configuration may ask for both. A usage requirement may hold
for one linkage only (\texttt{CODEC\_STATIC} in \cref{fig:walkthrough}), applied by a core rule, so no author inspects
linkage. An imported library's linkage is a fact, not a choice.
\end{samepage}

\paragraph{Warnings}
\texttt{-Werror} is the user's decision wherever the user chooses the compiler, because which
code warns depends on the compiler. An author's \texttt{-Werror} is an error unless the author
decides it with a reason or the author's facts fix the compiler.

\paragraph{Include order}
\label{sec:design:order}
Include order can change the output, since the first directory with a header wins. A compile's include directories fall into three classes: the target's own (\texttt{-I}), every
dependency's (\texttt{-isystem}, so a dependency's warnings are its author's), and other system
directories. Within a class they are a set, emitted in lexical order; where two hold a header of the same
name, the author can state a priority (a cycle is an error). \texttt{buildlog-cpp} adds the compiler's
\texttt{-Wshadow-header} warning by default where the toolchain accepts it; the core does not
diagnose a header in two directories or a symbol in two archives.

\subsection{Options That Decide What Is Built}
\label{sec:design:options}

Options select target requests (\texttt{TOOLS} in \cref{fig:surface}), target kinds or dependency
edges. Because policies such as ``build the CUDA backend if CUDA is found'' (Meson's \texttt{auto})
make the product vary by machine, \sys permits only option tests in authored target-request and
target-kind conditions, resolved from the user's value or the author's default: missing CUDA is
an error, not a changed option, as is requesting or needing an option-disabled target (including
from a package).

Whereas the loader checks authored decision rules before inserting module guards, those guards
can introduce probe-answer dependencies into packages' target requests. Full-program answer
independence for target requests and kinds therefore requires a further dependency restriction
(\cref{sec:formal:consequences}) that loader acceptance does not certify; conversely, failing that
sufficient audit implies neither a changed request nor an incorrect build.

\subsection{Generators and Rule Packages}
\label{sec:design:generators}

A generator's run is a fact from which the core derives its outputs' edges, owners and
consumers. A generator runs a program that writes files. Its argument vector is a template whose words are
literal text or typed placeholders: an output, a source, a file of an imported package, another
program, or a list of words (\cref{fig:surface}). The types tell the core which words are inputs
and outputs, so edges, output ownership and provenance are derived; a program a word mentions
untyped, or a script runs, is invisible to the core. The first word is a
program, and that use alone makes it a build-time tool, built in the tool configuration; neither
the program nor the generator is marked as a tool. An imported program's platform comes from its
package's facts or a probe. A generator may itself be a template.
Libraries can likewise ship and install the rules their consumers need (\cref{sec:design:import}):
Halide's rules lower a generator into a library and header for C++ targets,
a small DSL fixture exports a \texttt{.dsl}-to-C++ rule that downstream builds run (which
CPS alone cannot carry),
and two installed generator libraries compose in one consumer through ordinary
runtime dependencies while a third available package in the prefix stays dormant (\supp{sec:eval-supp:composition}).

Our Qt~6.11.3 packages import Qt's exports and could be installed beside its libraries
(116 lines total, 41 of Datalog). The rules below derive \texttt{moc} actions for listed headers of QtCore
consumers, including through usage-passing edges, with their compile flags (\texttt{...} elides atoms):
\begin{lstlisting}[language=Datalog]
qt_passes(d, c) :- eff_dep(d, c, "Qt6::Core", v), v != "private".
qt_passes(d, c) :- eff_dep(d, c, x, v), v != "private", qt_passes(x, c).
qt_user(t, c) :- eff_dep(t, c, "Qt6::Core", _).
qt_user(t, c) :- eff_dep(t, c, x, _), qt_passes(x, c).
qt_input(t, f, "moc", "moc_") :- target_sources(t, f), path_ext(f, e),
  (e = ".h" ; e = ".hpp" ; e = ".hxx"), !qt_moc_exclude(t, f).
qt_gen(t, c, f, k, cat(g, "/", d, x, s, ".cpp")) :- qt_user(t, c), qt_input(t, f, k, x), ...
action(o, "MOC"), action_run(o, 0, "", "Qt6::moc", "run"), ... :- qt_gen(_, _, _, "moc", o).
action_arg(o, 20, "-D", v), ... :- qt_gen(t, c, _, "moc", o), use(t, c, "cc.define", v, _).
action_arg(o, 30, "-I", v), ... :- qt_gen(t, c, _, "moc", o), use(t, c, "cc.include_dir", v, _).
\end{lstlisting}
Further rules compile each generated source into the target and run \texttt{rcc} on its
resource lists. A rule lies in a higher stratum than the relations it negates and aggregates, so
``after the target is fully described'' (\cref{sec:background:qt}) holds by construction.

Because dependency rules affect what is built, the \emph{lock} (\texttt{buildlog.lock}) records
content hashes for explicit packages, rule modules used in the final model and their plugins,
while preparation observations track dormant modules (\cref{sec:implementation}). Configure
rejects hash changes until approved; unused modules stay out of the lock, and the lock itself
never activates a package.

\subsection{Imported and Built Libraries}
\label{sec:design:import}

A model records that a fact holds, not how the fact was obtained, so a library is consumed the
same way whether it is built or installed: through facts about its targets, such as files,
dependencies and usage requirements. A built library's facts are derived; an installed library's
are loaded, and its targets are \emph{imported}. The link closure and the standard's join read
the same facts either way. An installed library is a rule package: its manifest, its facts and
optional rules. The facts can be Datalog, or an importer can
decode them from a pkg-config, CPS or CMake export file, found by search or listed in the
manifest's \texttt{imports}. An upstream that ships only CMake exports can still ship a manifest and
rules beside them, as our LLVM and Qt packages do. Unreadable metadata is an error when demanded. A package may assert only about
targets in namespaces it owns, and sources that disagree are an error, never resolved by
precedence.
Installing reverses this: it writes down the facts that a consumer of the built library would
derive. Datalog is the canonical export format. In \sys's own format the export is a file of
ground facts, which Soufflé reads as it is, plus the project's rule layers if the specification
exports them; a downstream loads it like any rule package, pinned in its lock.
Exports to CMake, CPS and pkg-config serve consumers of other tools. Their \emph{exporter} spells each key in the format or declares it
unspelled with a reason, and each lost key is surfaced to the user as a warning (pkg-config has no field for a C++
standard). Under \cref{sec:formal:export}'s conditions, a lossless export preserves the
supported usage facts as a set; this is \emph{set-level usage preservation}.
\label{sec:design:runtime}
Libraries loaded with \texttt{dlopen} are \emph{run-time dependencies}: built in the calling program's
configuration, placed beside that program or each executable that links it, and passing no usage
requirement or link item; only the Datalog export carries the requirement.

\subsection{Explanations}
\label{sec:design:why}

The \texttt{why} command explains a fact by its proof relative to the configured model and the
recorded probe answers; an explanation is not a log. It reruns the composed program on the final
round's input in Soufflé's provenance mode, maps leaves to origins and prints the proof as a DAG
that shares repeated subproofs.
Negations, constraints and aggregate results are leaves checked against the model; other leaves
cite their origin: a specification line, a user file, a package or a probe answer with its
probe request (\texttt{-{}-demand} explains that request). Because the loader records the line that
wrote each desugared row, a leaf cites the TOML line, not the row, as the user file does in
\cref{fig:walkthrough}. Because every layer lowers into the one below, an explanation can
follow a flag from an action through the objects and libraries that entail it.
A query may spell a flag as the Ninja file does.
On the questions the studies' authors asked, default answers take 16 to 52 lines
(\supp{sec:eval-supp:why}); the longest shows how zlib-ng's AVX2 requires SSE4.2 down to SSE2,
each step with its check, and a test asserts that no unrelated premise appears.
\Cref{sec:evaluation:why} excerpts an answer through a lattice join and gives the behavior of
\texttt{why-not}, which explains a missing fact.

\subsection{Implementation}
\label{sec:implementation}

\sys is a Python driver around Soufflé: 12,232 lines of Python and 2,989
of Datalog in the core, and 3,330 and 1,084 in the C/C++ plugin, excluding
tests (Python by \texttt{cloc}; Datalog non-blank, non-comment).
\sys uses 23 Soufflé patches (about 1,500 added lines, tests excluded): 15 for provenance,
2 for serving multiple rounds in one process and 6 for per-run overhead.
Timings depend on the last eight; each patch has a regression test.
The patches pass Soufflé's full regression suite (3,841 tests) and change no output of our
programs.

\texttt{configure} emits the Ninja file, \texttt{check} validates the model without emitting it,
and \texttt{test} runs configured tests; \texttt{why}, \texttt{why-not} and \texttt{query} inspect
the model without running probes. Determinism tests shuffle input facts and within-round probe
order and require identical Ninja output.

\paragraph{Preparation}
Before probing, metadata providers follow includes and rule pointers within declared roots;
their \emph{preparation certificate} records the resulting filesystem observations (rather than
provider correctness) and is rechecked before publishing a successful run.
The loader composes the reachable Datalog union once, rejecting invalid or conflicting rules
even if unused, whereas ordinary metadata errors wait until demanded; probes cannot extend
this prepared scope (\supp{sec:formal-supp:cache}).

\paragraph{Loader}
The loader rejects private-relation violations, I/O, foreign functions and unstratified
rules, reporting offending cycles. It verifies guarded answer reads and authored decision
dependencies (\cref{sec:formal:discovery,sec:formal:consequences}), while checking inserted
module guards for declaration/use stratification rather than full-program answer independence.

\paragraph{Rounds and cache}
A single Soufflé process evaluates every round, and a guarded path probe resolves symbolic
sysroot paths with ordered negative lookups. Cached outputs are reused when inputs repeat,
and unchanged runs validate preparation, probe keys and watches without inference.
Unfingerprinted probe kinds run on every configure unless the user states their answers.

\paragraph{Refresh and regeneration}
Compile checks key their arguments and validate \texttt{stat} watches on depfile inputs and
compiler-reported search roots; directory watches are nonrecursive, not a complete record of
negative or transitive lookups. Cache validity therefore assumes unobserved external and
toolchain inputs stay stable until \texttt{-{}-fresh}, which bypasses warm skipping, retained
answers and cached results (\supp{sec:formal-supp:cache}).
Ninja regenerates on specification, user-file or package edits; after probe-only changes,
re-running \texttt{configure} revalidates keys and watches.

\paragraph{Emission}
The translator reads only the action and test relations. A model with errors deletes the Ninja
file and fails, unless \texttt{-{}-keep-going} turns errors of a target or action into failing
edges; a loader, lock or probe failure stops earlier and leaves the old file.
Copies, installs and configured files are emitted as built-in driver commands.

\section{Formalization}
\label{sec:formal}

Discovery can withdraw probe requests, so accumulating answers need not produce growing models.
Under the stated hypotheses, guarded reads make irrelevant retained answers inert,
supporting least closed answers and agreement between fresh and validated cached runs.
Finite progress across declared domains yields termination, and per-round snapshots yield
within-round probe-order independence.
\Cref{tab:guarantees} separates checks from author obligations; \supp{sec:formal-supp} gives the proofs.

\newtheorem{coreinvariant}{Invariant}
\crefname{coreinvariant}{Invariant}{Invariants}
\Crefname{coreinvariant}{Invariant}{Invariants}
\newtheorem{exportremark}{Remark}
\crefname{exportremark}{Remark}{Remarks}
\Crefname{exportremark}{Remark}{Remarks}

\begin{table}[tp]
\caption{Assurance boundary. Runtime checks cover reached rounds, not all hypothetical
answer sets. Detailed conditions and mechanization status are in \cref{tab:guarantees-coded}.}
\label{tab:guarantees}
\small
\begin{tabularx}{\textwidth}{@{}l>{\RaggedRight}X@{}}
\toprule
Responsibility & What it establishes or assumes \\
\midrule
Loader & Origins/visibility, stratification, guarded reads, authored decision dependencies and
error isolation. \\
Driver & Domain membership, unique argument tuples, answer shape, cache confirmation,
rule-union and observation checks. \\
Plugin/provider author & Finite evaluation and probe-request universes; returning functional probes;
complete finite enumeration; sufficient cache keys and acyclic prerequisites. \\
Lean & Abstract model/driver results and finite-graph closure; core toolchain,
tool-configuration and export-classification invariants. \\
\bottomrule
\end{tabularx}
\end{table}

\paragraph{Mechanization boundary}
Lean~4.35.0 (with bundled Std) mechanizes a finite abstraction: it proves answer inertness for
finite ground guarded rules rather than verifying the loader's translation, and its
cache-agreement and answer-free-decision theorems take completed-phase and dependency
certificates as hypotheses. A 201-entry source-correspondence test checks the core-rule
transcription against the shipped Datalog, and an axiom audit confirms that the development
uses only standard Lean axioms (including classical choice), with no project axioms or
admitted proofs.
The loader, certificate extraction, provider traversal, module-guard insertion, Souffl\'e,
seed-loop termination, the general (F) reduction and shipped-kind finiteness/fixed-answer
arguments are not mechanized.

\subsection{Programs, Inputs and One Derivation}
\label{sec:formal:programs}
\label{sec:formal:derivation}

Let $P$ be the union of the prepared Datalog layers and $I=(A,U,V,E)$ contain author facts $A$,
user facts $U$, certified available metadata $V$ and probe answers $E$;
$D(I)=A\cup U\cup V\cup E$. Manifests classify inputs by origin and relations by visibility
(\cref{sec:design:inputs}); the loader enforces both and rejects predicate cycles through
negation or aggregation.

\paragraph{Unique model (standard fact)}
\label{prop:model}
Stratified $P$ with finite stratum evaluations has a unique perfect model $M_P(I)$
\citep{apt1988declarative}. Term construction makes finiteness an obligation.
Below $M_P(A,U,E)$ fixes $V$; neither $P$ nor $V$ changes between rounds.
A run's observable \emph{outcome} comprises its sorted error report, emitted build graph and exit status.

\begin{coreinvariant}[Toolchain uniqueness]
\label{lem:toolchain}
For each required (language, build platform, target platform) key, multiple unpinned candidates,
conflicting pins or a non-candidate pin derive an error. Otherwise at most one toolchain is selected; no selection derives
``no toolchain'' at every target with an action in the configuration.
\end{coreinvariant}

\subsection{Discovery}
\label{sec:formal:discovery}

Here ``request'' means a probe request, not a target request.
Each probe kind $k$ declares requests $\mathit{req}_k$, domains $\mathit{dom}_k$ and results
with a primary row, keyed by $(k,\mathit{id}(r))$; $\mathit{req}$ collects all kinds' requests.
A \emph{guarded} read, including negation or aggregation, entails a positive
identifier-matching request in the same disjunct; the loader checks a sufficient syntactic
condition (W3). Package selection reads $V$; all shipped environmental kinds are guarded.
Writing $R_i$ for the round-$i$ probe requests still lacking a primary answer in $E_i$, discovery iterates
\begin{align*}
E_0 &= \emptyset, \qquad M_i = M_P(A, U, E_i), \\
E_{i+1} &= E_i \cup \textstyle\bigcup_{(k,r) \in R_i} \mathit{probe}_k(r, E_i).
\end{align*}
It stops when $R_n=\emptyset$. Answers grow, but negation means models need not.
The driver checks domain membership, unique argument tuples (W6) and primary-row totality
(P1), including negative answers. Obligations are: (P0) probes/fingerprints return;
(F) every finite answer set has a finite model; (P2) fixed finite universes $\mathcal U_k$
contain every round's domains; and (P3) answers depend only on arguments, fingerprinted
answers, watched filesystem observations and the environment.
Fresh means no answers and no cache (\texttt{-{}-fresh}).

\begin{theorem}[Fresh termination]
\label{thm:termination}
For stratified $P$ satisfying (F), (P0) and (P2), a fresh run stops or fails a check within
$1+\sum_k|\mathcal U_k|$ rounds: every continuing round answers a new pair.
\end{theorem}

\begin{theorem}[Fresh determinism]
\label{thm:determinism}
For fixed stratified $P$ satisfying (F), (P0), (P2) and (P3), and fixed driver options,
the fresh outcome and stopping round depend only on $A$, $U$ and the environment through
the snapshot-dependent answer function $\beta_k(r,E)$.
Fact ingestion order and within-round probe execution order do not affect the outcome.
\end{theorem}
This is within-round order independence, not arbitrary rescheduling under (P3).

\paragraph{Guarded answers have a least closed set}
When a later round withdraws a request, its earlier answer stays stored, but no guarded rule
can use it without the request; pruning these inert answers proves leastness without requiring
requests to grow monotonically.
Condition~(A) strengthens~(P3) so each request has one answer $\alpha_k(r)$ across runs and
rounds: a \emph{correct} set unions such answers and is \emph{closed} if it answers every
request of its model. We also require that (RL) every result row carry its request identifier,
(I) identifiers determine arguments across runs, and $\alpha$ be defined on finite universes;
(P2) and (CK), passing domain/tuple/primary-row/declaration checks, must hold over
\emph{every correct-answer subset}, not merely reached rounds.

\begin{theorem}[Least closed guarded answers]
\label{thm:lfp}
For fixed stratified $P$ guarding every kind and satisfying (F), (A), (RL) and the universe
hypotheses above, correct closed sets exist and agree off probe-result relations, giving $M_*$
there. The demanded answers $E^*=\alpha(\mathit{req}(M_*))$ form the unique least correct closed answer set.
\end{theorem}

\paragraph{Reusing answers}
Whole-run replay validates a recorded run, whereas seeding confirms retained answers against
current requests and keys from confirmed prerequisites, dropping stale seeds without probing
and retrying fresh if a check fails.
Cache soundness assumes (FC) that equal keys and unchanged watches determine answers (for
shipped compiler probes, unobserved external and toolchain inputs must stay stable until
\texttt{-{}-fresh}, \cref{sec:implementation}; observation completeness is not checked) and
(FL) that keys and probes read only acyclic prerequisite answers demanded by every derivation of
the probe request.

\begin{corollary}[Cached and fresh runs agree]
\label{cor:seeding}
Under the determinism and least-closure hypotheses, (FC) and (FL), completed replay,
seeded and fresh runs have the same outcome and agree off probe-result relations.
Retained inert answers need not agree.
\end{corollary}

\paragraph{Preparation before inference}
\begin{proposition}[Bounded preparation]
\label{prop:preparation}
Given complete finite provider enumeration and fixed, answer/demand-independent links,
$L_{j+1}=L_j\cup\mathit{links}(L_j)$ from declared roots terminates at the least reachable
closure. Composition consumes its valid executable union once; environmental rounds cannot
change that union or its metadata.
\end{proposition}

\paragraph{Coverage}
The shipped-kind argument establishes (A) from (P3) except for CUDA, where it remains an
obligation. The (F) audit covers core, C/C++, Halide and Qt, not arbitrary projects or the
Python, Emscripten, NDK and CUDA plugins
(\supp{sec:shipped-checks} and \supp{sec:formal-supp:complexity}).
The supplement also generalizes to unguarded plugins; their fixed answer set is empty here.

\subsection{Consequences of the Stratification}
\label{sec:formal:consequences}

The \emph{decision relations} are explicit $\mathit{request}$, $\mathit{request\_all}$ and
declared kind; they are \emph{answer-free} when no positive, negative or aggregate path
reaches probe results (W5). Because the loader checks only authored clauses whereas
module-use guards can add paths (including to packages' target requests), full-program W5
remains a hypothesis: all 13 audited study/SDK configurations pass (Qt was not included in
that audit), while our two-library generating consumer builds even though its
dormant-package control fails W5 (\supp{sec:eval-supp:composition}).

\begin{proposition}[Explicit target requests are answer-free]
\label{prop:options}
For fixed stratified $P$ satisfying (W5), explicit target requests and their kinds depend only
on $A$, $U$ and fixed metadata $V$, not answers.
\end{proposition}

Error propagation waits for $\mathit{complete}\leftarrow\neg\mathit{pending}(\_)$.
\begin{proposition}[Errors do not steer discovery]
\label{prop:errors}
For well-formed $P$ ((W1)--(W6), Definition~\ref{def:wellformed}), let $P'$ remove the
error-propagation gate. Under (F) for both and (P0), (P3), their requests, answers and
driver checks agree each round; completed models agree on errors and attributions.
\end{proposition}
(W4) protects requests, domains and discovery checks from completion-gated heads.

\begin{coreinvariant}[Tool configurations]
\label{prop:exec}
With no setting/tool-configuration error and exactly one fact identifying the build machine's
platform, every configuration $c$
has a unique $\mathit{tool}(c)$. A typed tool reference to $g$ in an action of $c$ derives
$\mathit{needed}(g,\mathit{tool}(c))$. Direct tool-induced edges arise only from such
references or built-toolchain tools.
\end{coreinvariant}

\subsection{Built and Imported Artifacts}
\label{sec:formal:export}

The export result is conditional and set-level, not a command or binary equivalence claim.
An export $(F,R)$ pairs usage facts $F$ with hash-pinned rules $R$. \sys's native format loses
neither, whereas foreign formats classify each own fact exclusively as spelled, lost with a
warning or an error, and warn whenever rules $R$ are dropped (Lemma~\ref{lem:export-exporter}).

\begin{exportremark}[Set-level downstream use]
\label{prop:export}
Given an error-free export with no loss (E1), inverted spelling/relocation (E2), the same
resolved components/configurations (E3), unchanged other consumer inputs/dependencies (E4),
and rules adding no usage facts about their own targets (E5), downstream use preserves the
built target's supported usage facts as a set, relocated to the install prefix.
In \sys's format the consumer also receives $R$.
\end{exportremark}

This remark concerns supported dependency, link and setting facts rather than emitted commands
or binaries; order, multiplicity, version acceptance, external absolute paths and hand-written
package files are excluded (\supp{sec:formal-supp:export}).

\section{Case Studies}
\label{sec:case-studies}

Eleven studies of ten projects port pinned open-source builds, each isolating a capability
and comparing with the project's own build, called \emph{upstream}
(\cref{sec:evaluation:method}). Meson and Bazel supply two
non-CMake engines; nanobind uses CMake through scikit-build-core.
The supplement lists releases and scope (\supp{sec:eval-supp:portfolio}) and further
capabilities, including C++20 modules and CUDA (\supp{sec:eval-supp:capabilities}).
The studies ran on an arm64 Mac with Apple clang; unless another destination is specified,
``cross'' means arm64-to-x86\_64 macOS. The studies also build on arm64 Linux with
GCC~13 and Clang~18; scoped execution records are separate (\supp{sec:eval-supp:linux}).

\subsection{Halide}
\label{sec:cs:first}
\label{sec:cs:halide}

Halide~\cite{ragankelley2013halide} is the stress test: an LLVM-based compiler whose generators,
runtimes and autoschedulers are rule strata that entail library facts, run during the build.

\paragraph{The full build} \sys builds libHalide and runs 941 of the 950 tests defined by
upstream's CMake build, including Python bindings, tutorials and autoschedulers: native
execution yields 848 passes, 93 skips and no failures across 941 executions. The nine excluded tests either
invoke a nested build whose outputs are ordinary targets here, drive a debugger, or check a
Windows-only filter (\supp{sec:eval-supp:fidelity}). Because native tools inherit the Python
setting, generators share the product's libHalide build; release, debug and ASan are declared
configurations.

\paragraph{The cross build} From arm64 macOS to Linux on AArch64, \sys builds libHalide, its
runtime and a 32-test validation subset (20 correctness and 12 generator/AOT tests),
in one tree from one configure, the generators and build-time tools
for macOS and the rest for Linux; the tests give 30 passes and 2 skips, as under CMake, whose
build takes two trees and two configures. Rules relocate run paths and prefix imported
paths with the sysroot. The tests cover mechanisms such as C sources, image I/O, multitarget
output and separate runtimes, not a statistical sample. The user pins an otherwise ambiguous build-machine compiler.

\paragraph{Specification size} The current port uses 2,642 specification and rule lines
against CMake's 5,361, excluding blank lines, comments and source-only lists.
The count includes the Python and tutorial rules as well as reusable generator helpers
(\supp{sec:eval-supp:size}).

\paragraph{Cross-built apps} Upstream builds Halide's HelloWasm and HelloAndroid apps with a
super-build and a nested configure; in \sys each is one configure and one graph. Generator runs
are identical to CMake's (15 and 4), all four WebAssembly variants run 64 frames under Node.js
with identical final frames, and the Android libraries export all eight JNI entry points (not
run). Two plugins stand in for Emscripten's and the NDK's toolchain files
(\supp{sec:eval-supp:apps}).

\subsection{Build-Time Tools: flatbuffers and fribidi}
\label{sec:cs:flatbuffers}

\paragraph{flatbuffers in CMake and Bazel} flatbuffers' \texttt{flatc} compiles schemas into C++ headers during the build, so in a cross
build it must run on the build machine. A CMake cross build requires two build trees and a hand-spelled
tool path (google/flatbuffers\#5525), and silently runs a \emph{stale} build-machine binary whenever \texttt{flatc}'s sources change.

\sys's project uses \texttt{flatc} as the program an action runs, and nothing else says where it
runs. Natively every compile pairs with CMake's (\cref{tab:action-fidelity}). With \texttt{-{}-config x86\_64}
(140 commands), \texttt{flatc} is built once for arm64 and run 11 times, the generated headers
are byte-identical to CMake's, and the x86\_64 tests pass under Rosetta. A universal2
configuration (210 commands) builds fat products and both slices' tests pass.

In a slice of flatbuffers' Bazel~8.8.1 cross build, every \sys action matches by configuration
role, type and unit; the generated header is byte-identical. Bazel's \texttt{genrule} names
\texttt{flatc} twice, once in \texttt{tools} with its exec transition; \sys names it once.
Misplacing \texttt{flatc} among Bazel's product-configured sources silently builds and runs
an x86\_64 tool under Rosetta; the corresponding \sys mistake fails with both machines named.

\paragraph{fribidi and Meson's native flag}
\label{sec:cs:fribidi}
fribidi~1.0.17 generates its Unicode tables with seven programs that must run on the build
machine. Meson asks the author to write \texttt{native: true} on each and, separately, to ask
for the build machine's compiler for their header checks; fribidi's ask the cross compiler, right
here only because both platforms share one SDK. In \sys no row marks a build-machine tool, and a check is
answered in the configuration of the target that uses it. Cross-compiled for x86\_64 from a user
file, \sys builds the seven programs once for arm64 and splits the 42 objects as Meson does.

\subsection{Further Studies}
\label{sec:cs:zlibng}
\label{sec:cs:ggml}
\label{sec:cs:spirv}
\label{sec:cs:nanobind}
\label{sec:cs:zlib}
\label{sec:cs:qt}

\paragraph{Checks and exports}
zlib~1.3.2's five checks agree with CMake's cache and its \texttt{zconf.h} is byte-identical,
without CMake's generator-expression bug (\supp{sec:eval-supp:further}).
nlohmann/json~3.12.0's downstream usage facts agree in-tree, through CPS and through CMake
exports from either system, also after relocation. Installed headers/package files agree;
pkg-config's lost language standard is reported.

\paragraph{ISA variants}
zlib-ng~2.3.3's universal2 CMake build silently omits SIMD in both slices (zlib-ng\#1016).
\sys derives per-source ISA flags for both slices in one configure, matching separate
single-architecture CMake builds (\supp{sec:eval-supp:further}).
ggml's run-time CPU modules instantiate one template per ISA level
(\cref{sec:design:templates,sec:design:runtime}); module sets, flags and symbols match CMake's.
On ARM, upstream accepts \texttt{-march=native} tuning with all variants but emits modules without the
scoring entry point ggml's backend loader requires, so it selects none. \sys rejects that combination.

\paragraph{Imported data and interpreter configurations}
SPIRV-Tools' six generated files from imported SPIRV-Headers grammars are byte-identical to
CMake's; \texttt{why} traces their inputs to the user's prefix.
nanobind makes the Python interpreter a configuration setting and derives the stable ABI, definition,
file name and wheel tags together. Symbols and wheel metadata match scikit-build-core/CMake
across both tested Python interpreters (3.11 and 3.13).

\paragraph{Qt: rules a library ships}
A Qt~6.11.3 consumer uses the rules of \cref{sec:design:generators}, never naming \texttt{moc}.
Its 15-line specification compares with CMake's 13; both tests pass on macOS and Linux.
This compares consumer specifications, not coverage of CMake's Qt implementation:
the package handles listed headers and resources, not \texttt{uic},
\texttt{Q\_OBJECT} in \texttt{.cpp} files or compiler-predefine generation.
\texttt{why} traces a generated source to the QtCore dependency.

\paragraph{Further capabilities}
TinyCC bootstraps three compiler stages in one graph: a compiler built by one stage serves
as the next stage's compiler, linker and archiver. Stages two and three are byte-identical.
C++20 named modules exercise build-time dependency discovery, as Fortran modules do:
source scans feed a rule stratum that derives Ninja dyndep files and shared or specialized
module interfaces. A \texttt{fmt} module consumer builds and runs without naming its module
in the specification. CUDA and stdgpu compile and link with a real toolkit.
These are capability checks beyond the eleven portfolio studies
(\supp{sec:eval-supp:capabilities}); no NVIDIA GPU was available for CUDA execution.

\section{Evaluation}
\label{sec:evaluation}

We assess whether \sys produces correct builds in practical configuration times, not superiority
to mature systems. Execution checks and command comparisons remain separate evidence, while
provenance queries and pin ablations evaluate explainability and user control (\supp{sec:eval-supp}).

\subsection{Method}
\label{sec:evaluation:method}

We tokenize actions with machine paths as placeholders, attributed through CMake's File API,
\texttt{meson introspect} or \texttt{bazel aquery}.
Pairs match unit (target/source for compiles, output for libHalide), action type and configuration
role, checking tool versus product configuration on both sides (\supp{sec:eval-supp:fidelity}).
Output checks cover file hashes, symbols, archives, load commands, installed trees and tests,
as applicable.
Artifact scripts configure \sys against recorded upstream baselines at matching pinned revisions.
The full native Halide graph comparison and execution checks cover the current port;
execution results are reported separately from command comparisons.

Native fidelity records and configure API measurements use an Apple M3 Max (16 cores,
48\,GiB), macOS~26.5.2, Xcode~26.6 (Apple clang~21.0.0), CMake~4.4.3, Meson~1.12.1, Ninja~1.13.2,
and Soufflé at a fixed revision with the artifact's local changes (\cref{sec:implementation}).
Configure API measurements use Python~3.14.0 and serial runs at a one-minute load of 3.4--5.3;
the separate portfolio CLI comparison also uses Python~3.14.0, with its own recorded load.
Their whole-process CLI and in-process API timing boundaries are distinguished in \cref{sec:evaluation:cost}.

\subsection{Fidelity}
\label{sec:evaluation:fidelity}

We first check whether the configured graphs build usable artifacts and pass the scoped
tests (\cref{tab:fidelity}). We then compare commands to locate differences that those tests
may miss. Neither passing tests nor matching normalized commands proves full build equivalence,
equal runtime performance or equal code size.

\begin{table}[t]
\caption{Build validation for the ported configurations. Halide counts are executions
(941 distinct macOS tests); exclusions and Linux failure controls appear in the text.}
\label{tab:fidelity}
\begin{tabularx}{\textwidth}{@{}l>{\RaggedRight\arraybackslash}X@{}}
\toprule
Study & Tests and artifact checks \\
\midrule
zlib / zlib-ng & Variants: 2/2 and 7/7 tests pass; symbols, headers and packages checked \\
flatbuffers & CMake port: 1/1 test passes; Bazel slice generates the same header and runs it \\
fribidi & Native library/tools: 9/9 tests pass; generated tables and symbols checked \\
ggml & Selected CPU variants pass; loader selects the predicted backend \\
SPIRV-Tools & Ported library/tools: 4/4 tests pass; six generated files identical \\
nlohmann/json & Installed headers identical; in-tree and installed consumers build \\
nanobind & Python 3.13 and 3.11 tests pass; ABI modes and wheel contents checked \\
Qt consumer & 2/2 tests pass; corresponding CMake consumer builds with \texttt{moc}/\texttt{rcc} \\
libHalide & macOS: 848 pass, 93 skip, 0 fail; Linux (877 tests): 783 pass, 92 skip, 2 fail \\
\bottomrule
\end{tabularx}
\end{table}

\paragraph{Command differences}
The detailed pairing table and taxonomy are in \supp{sec:eval-supp:fidelity}.
Spelling normalization removes compiler paths and equivalent platform spellings, not
optimization or visibility flags. Rebuilding with baseline flags changes no zlib or fribidi
output; flatbuffers' C++11 versus CMake's C++14 changes 3 of 85 objects. These experiments
support only the pairs rebuilt, not a general claim that flag differences are harmless.
Bazel applies a different optimization and hardening policy, not replicated by the slice.

Of 4,748 Halide pairs, 2,814 differ at most in spelling, 538 only in warnings/includes,
and 1,396 carry potentially consequential differences (PCH, wrappers, export/LTO,
custom commands and links; \supp{sec:eval-supp:fidelity}).
Current-graph replay gives identical objects for 438/438 wrappers and 728/730 PCH
compiles; two runtime-buffer tests gain code with PCH.
Four tutorial regressions check Release optimization, normal visibility and C++17.
End-to-end runtime performance and the code-size effects of remaining flag differences are not measured.

Unpaired work includes fixtures, staging and exports CMake writes during configuration;
counts alone do not measure missing functionality (\supp{sec:eval-supp:fidelity}).

\paragraph{Scope and failures}
The full Linux GCC Halide run gives 783 passes, 92 skips and two
failures; a focused rerun passes one and reproduces the other's speed assertion against both
the \sys build and the independent CMake-built library. We do not count these as all tests
passing (\supp{sec:eval-supp:linux}). The Linux scope excludes Python bindings.

\paragraph{Normalization}
Except libHalide's, the comparisons treat flags as sets and sort link inputs. An order-keeping
comparison of the other studies finds only independent flag groups interleaved,
with no potentially consequential order differences (\supp{sec:eval-supp:order}).

\paragraph{Pins}
Removing pins one at a time from the user files of TinyCC and every study but Qt's,
one of 43 rows is necessary with Apple clang alone, versus 36 when Homebrew's LLVM is also
offered. On Linux with GCC and Clang, 21 of 34 are necessary.
A user-global file with one row per toolchain role (C, C++ and the linker driver) replaces the
repeated toolchain choices; Halide's separate LLVM-linkage pin remains.
Thus machine-wide toolchain policy can be stated once in a user file rather than per project
(\supp{sec:eval-supp:pins}).

\paragraph{Specification size}
Excluding shared infrastructure but charging project-specific rules, ggml uses 501 lines against
CMake's 644 and zlib-ng 523 against 1,223. SPIRV-Tools' port is also shorter; fribidi's is longer
(301 against Meson's 263), largely for \texttt{config.h}. \Cref{sec:cs:first} gives Halide's
count, and \supp{sec:eval-supp:size} details the counting rules and per-study breakdown.

\subsection{Provenance Answers}
\label{sec:evaluation:why}

For \texttt{app}, \texttt{util}'s public C++20 requirement dominates its dependency
\texttt{core}'s C++17. This excerpt traces the winning value:
\par\noindent\begin{minipage}{\linewidth}
\begin{lstlisting}[language={},basicstyle=\ttfamily\footnotesize,breaklines=true,literate={<-}{{$\leftarrow$}}2]
action_arg("<object>", 10, "-std=c++", "20")  [cpp:L4_cxx.dl:33]
  ...
  use_join("app", <x86>, "cxx.std", "20")  [L1:760]
    use_cand("app", <x86>, "cxx.std", "20", 1)  [L1:742]
      eff_dep("app", <x86>, "util", "private")  [L1:596]
        ...
      eff_prop("util", <x86>, "cxx.std", "20", "public")  [L1:718]  (via iface)
        ...
        prop("util", "cxx.std", "20", "public")  <- cross/spec.toml:70
    lattice_key("cxx.std")  [L1:748]
\end{lstlisting}
\end{minipage}
Here \texttt{<object>} abbreviates the object path and \texttt{L1} is \texttt{L1\_targets.dl}.
The answer shows the winning C++20 requirement,
not the losing C++17 candidate; negation and joins are checked against the model
(\supp{sec:eval-supp:why-join}).
\texttt{why-not} explains blocking premises using overridable witnesses.
For a complete generator explanation, see \supp{sec:eval-supp:why}.

\subsection{Configure Cost}
\label{sec:evaluation:cost}
\paragraph{End-to-end comparison}
Ten serial CLI trials alternate tool order with private sources, builds and caches
(\cref{tab:single-composition}). Timing includes startup, probes and graph generation,
not builds/tests. Cold empties tool state, not filesystem caches; warm invokes configure
unchanged; edit adds a corresponding private definition.
zlib is shared-only; Halide includes Python, tutorials and tests on both sides.
Port omissions remain: CMake includes extra ggml tests/examples, but its SPIRV-Tools tree lacks
the GoogleTest checkout and omits tests.
Nanobind's upstream time sums two direct CMake configures, one per interpreter.
\supp{sec:eval-supp:cli-cost} details scopes and SDK-app costs, whose upstream outer
configures do different work.

\begin{table}[t]
\caption{Whole-process configure seconds, \sys{} / upstream, each a
\textbf{mean}$\pm$relative 95\% confidence half-width from ten trials.
Upstream is CMake except fribidi (Meson) and the Bazel slice.
$^\dagger$Ported test and example scope differs from upstream; see text.}
\label{tab:single-composition}
\setlength{\tabcolsep}{3pt}
\begin{tabular*}{\textwidth}{@{\extracolsep{\fill}}lrrr@{}}
\toprule
Project & Cold & Warm & Edit \\
\midrule
zlib & \textbf{2.47}$\pm$4.8\% / \textbf{1.43}$\pm$8.7\% & \textbf{0.14}$\pm$5.9\% / \textbf{0.11}$\pm$9.6\% & \textbf{0.87}$\pm$9.4\% / \textbf{0.11}$\pm$13.6\% \\
nlohmann/json & \textbf{2.03}$\pm$4.4\% / \textbf{0.33}$\pm$7.2\% & \textbf{0.12}$\pm$3.1\% / \textbf{0.10}$\pm$7.9\% & \textbf{0.70}$\pm$5.0\% / \textbf{0.10}$\pm$10.8\% \\
zlib-ng & \textbf{3.49}$\pm$4.0\% / \textbf{4.56}$\pm$6.5\% & \textbf{0.15}$\pm$6.7\% / \textbf{0.29}$\pm$6.1\% & \textbf{0.97}$\pm$17.2\% / \textbf{0.29}$\pm$11.7\% \\
ggml$^\dagger$ & \textbf{4.30}$\pm$4.1\% / \textbf{3.73}$\pm$4.6\% & \textbf{0.19}$\pm$5.7\% / \textbf{0.42}$\pm$6.6\% & \textbf{1.08}$\pm$1.6\% / \textbf{0.42}$\pm$6.8\% \\
fribidi & \textbf{2.60}$\pm$2.8\% / \textbf{1.41}$\pm$8.0\% & \textbf{0.14}$\pm$4.6\% / \textbf{0.36}$\pm$4.8\% & \textbf{0.87}$\pm$2.7\% / \textbf{0.36}$\pm$6.1\% \\
flatbuffers & \textbf{2.35}$\pm$4.1\% / \textbf{0.71}$\pm$7.2\% & \textbf{0.14}$\pm$10.8\% / \textbf{0.12}$\pm$8.2\% & \textbf{0.83}$\pm$2.6\% / \textbf{0.12}$\pm$12.2\% \\
SPIRV-Tools$^\dagger$ & \textbf{2.36}$\pm$2.9\% / \textbf{0.86}$\pm$5.3\% & \textbf{0.14}$\pm$6.4\% / \textbf{0.37}$\pm$7.0\% & \textbf{0.97}$\pm$2.3\% / \textbf{0.37}$\pm$7.6\% \\
nanobind & \textbf{2.51}$\pm$4.2\% / \textbf{0.96}$\pm$7.0\% & \textbf{0.13}$\pm$9.8\% / \textbf{0.25}$\pm$8.6\% & \textbf{0.81}$\pm$3.0\% / \textbf{0.23}$\pm$5.6\% \\
Halide & \textbf{6.74}$\pm$2.3\% / \textbf{6.12}$\pm$2.7\% & \textbf{0.43}$\pm$5.2\% / \textbf{2.97}$\pm$3.3\% & \textbf{4.08}$\pm$3.5\% / \textbf{2.91}$\pm$3.1\% \\
Bazel slice & \textbf{2.17}$\pm$2.9\% / \textbf{6.22}$\pm$6.3\% & \textbf{0.13}$\pm$8.6\% / \textbf{0.19}$\pm$3.3\% & \textbf{0.78}$\pm$1.9\% / \textbf{0.14}$\pm$3.9\% \\
Qt consumer & \textbf{2.61}$\pm$1.6\% / \textbf{0.79}$\pm$1.5\% & \textbf{0.24}$\pm$2.3\% / \textbf{0.18}$\pm$1.8\% & \textbf{1.13}$\pm$4.0\% / \textbf{0.18}$\pm$2.4\% \\
\bottomrule
\end{tabular*}
\end{table}

Cold configuration completes in seconds, though fixed startup overhead remains noticeable on
small projects, demonstrating practical feasibility rather than uniform speed parity with CMake.

\paragraph{One composition}
In-process API instrumentation shows one composition and evaluator start cold or after an
edit, with four evaluations cold and one after edit; unchanged runs only validate
certificates, where repeated scans dominate warm Halide (\supp{sec:eval-supp:single-composition}).

\paragraph{Cached against fresh}
Tests compare seeded/fresh facts and graphs after configuration, answer, dependency, pin,
prefix and metadata changes, checking probe order, invalidation, fallback and zero-evaluation
reuse (\supp{sec:formal-supp:cache}). They complement, not discharge, probe-author
obligations (Corollary~\ref{cor:seeding}).

\subsection{Threats to Validity}
\label{sec:discussion:threats}

Feature-selected projects validate mechanisms, not typical-build coverage.
C/C++ run checks expose only build-machine exit status, not printed values, compiler output matching
or chained answers; general probe kinds admit acyclic prerequisites (\cref{sec:formal:discovery}).
Source content cannot select files: \texttt{moc} runs on every listed header.
Normalization affects fidelity; functional checks do not establish equal performance or code size.
Timings use one machine, warm filesystem caches and stated port omissions, not representative
workloads. Finally, no user study evaluates authoring specifications, diagnosing errors or
reading provenance explanations.

\section{Related Work}
\label{sec:related}

\paragraph{Build systems}
\citet{mokhov2018bsalc,mokhov2020bsalcjfp} model Make, Shake~\cite{mitchell2012shake}, Bazel and
others as a scheduler and a rebuilder; Pluto~\cite{erdweg2015pluto},
Rattle~\cite{spall2020rattle,spall2022forward} and CloudBuild~\cite{esfahani2016cloudbuild} prove
dynamic dependencies sound, trace them, and distribute builds. They \emph{execute} graphs;
\sys derives them.

\paragraph{Pants}
Pants~v2's typed \texttt{@rule}s form a static rule graph, reject ambiguous inputs and memoize
results~\cite{pants-architecture}, close to a Horn program over types. Rules are Python
coroutines, however, without proof explanations; Pants~2.33's C/C++ backend has no
library targets~\cite{pants-backend-source}.

\paragraph{Bazel and Buck2}
Bazel's Skyframe~\cite{bazel-skyframe} and Buck2's DICE~\cite{buck2-dice} memoize host-language
computations, and Bazel selects compatible toolchains by registration priority where \sys
requires uniqueness.
Like Shake's oracles~\cite{shake-oracle} and Bazel's repository rules and module
extensions~\cite{bazel-repo-rules,bazel-module-extensions}, \sys separates impure answers
from pure evaluation, while also deriving the probe requests as facts.
Where Bazel and Buck2 declare configuration transitions~\cite{bazel-config,buck2-configurations}
and \texttt{aquery} lists configured commands~\cite{bazel-aquery}, \sys infers tool
configurations (\cref{sec:cs:flatbuffers}) and explains flags by proofs.

\paragraph{qbs}
qbs's \texttt{Probe} items cache imperative script results in properties~\cite{qbs-probe},
whereas \sys's probes have rule-derived requests, finite domains, cache keys and provenance.
Like \sys, qbs activates \texttt{moc} when a product depends on
\texttt{Qt.core}~\cite{qbs-module,qbs-qt-core}, but through scripted module items rather
than library-supplied Datalog rules composed into a single checked stratified program.

\paragraph{Gradle}
Gradle's configuration cache reuses graphs while observed inputs match~\cite{gradle-config-cache};
\sys also caches individual probe answers.
Gradle's lazy properties derive edges from producing tasks' output-to-input
wiring~\cite{gradle-lazy-config}, like \sys's typed placeholders but without proof explanations.

\paragraph{Packages and configuration}
Version resolution is NP-complete~\cite{mancinelli2006edos}; solvers use SAT, ASP or
optimization~\cite{abate2020depsolving,pinckney2023pacsolve}.
Spack~\cite{gamblin2015spack} resolves choices with answer-set
programming~\cite{gamblin2022asp,gebser2019clingo} and could supply them to \sys.
Deriving a graph from fixed choices needs a unique model, not optimization.
Nix and Guix~\cite{dolstra2004nix,courtes2013guix} derive package-level steps;
Nix can evaluate purely~\cite{nix-pure-eval} or wait on builds~\cite{nix-ifd}, whereas
\sys confines effects after preparation to between-round probes.

\paragraph{Datalog and provenance}
\sys uses Soufflé and stratified negation's unique perfect model~\cite{apt1988declarative},
as in Datalog program analysis~\cite{whaley2004bddbddb,bravenboer2009doop,avgustinov2016ql}.
Its rules also derive which environmental facts to collect.
\sys encodes lattices with ranks and aggregates; Flix~\cite{madsen2016flix} and
Datafun~\cite{arntzenius2016datafun} support them natively.
Probes resemble HEX external atoms~\cite{eiter2005hex,calimeri2007value}, but are answered between rounds.
Provenance semirings~\cite{green2007semirings}, why-not provenance~\cite{chapman2009whynot},
PUG~\cite{lee2019pug} and Soufflé's proof reconstruction~\cite{zhao2020provenance} explain queries.
\sys uses the latter to trace flags to specification, user, package and probe-request leaves,
separating proofs from demand (cf.\ magic sets~\cite{bancilhon1986magic}).

\paragraph{Studies and analyses of build code}
Build maintenance is costly~\cite{mcintosh2011buildeffort,mcintosh2012javabuild,mcintosh2015buildtech};
prior work migrates, analyzes and traces build scripts~\cite{gligoric2014migration,tamrawi2012symake,licker2019buildrules},
and translates CMake to formulae~\cite{nguyen2022cmake}, traces affected
commands~\cite{nejati2024buildchanges} and types scripts~\cite{han2026cmakesonar}. None derives the
configured graph as a logic program's model, nor does Biomake~\cite{biomake}, a
Make-compatible executor whose Prolog rules are order-dependent.

\section{Conclusion}
\label{sec:discussion}

Expressing configuration as relational inference makes a build's command graph the model of
an explicit logic program rather than the side effect of imperative scripts.
In \sys, plugins answer only the probe requests derived by that program, while stratified
rules compute a unique model that rejects ambiguous toolchain or package candidates and gives
every derived fact a provenance proof.
Authors state constraints and users state preferences through distinct, statically checked
input relations; libraries ship domain-specific consumer rules using the same facts for
installed and built code; and running a compiled tool in a build step entails its
build-machine configuration.
Across eleven case studies of ten projects, \sys configures C and C++ builds in practical
times while reporting remaining command differences, and derives Halide's build-machine
generators and target-platform products in one graph rather than nested build trees.

\paragraph{Limitations and future work}
A build currently admits one rule-package version per library, and edits re-evaluate the full
Datalog program (\cref{sec:evaluation:cost}).
Future work pairs incremental Datalog evaluation~\cite{szabo2018inca} with Windows/MSVC
support, additional plugins, and a Python ABI lattice with wheel exports---extending a small,
explainable model in which build configuration follows by logical entailment from discovered
environmental facts.

\section*{AI Use}
The authors used generative AI throughout this work, under their direction and with their
review. \emph{Implementation:} \sys, its tests and the case-study specifications were written in
tight collaboration with AI coding agents (GitHub Copilot CLI driving Claude and GPT models); the
authors designed the system, specified each change, and reviewed and verified every change
against the full test suite before merging. \emph{Proofs:} the Lean development was written with
AI assistance; Lean's kernel checks every proof. \emph{Evaluation:} AI agents performed reproduction
runs and baseline comparisons that the authors checked.
The authors take full responsibility for all content.

\clearpage
\section*{Data-Availability Statement}
\sys, its test suite, and the case-study baselines and deviation tables will be
made available as an artifact. The upstream sources of the case studies are
pinned git submodules of the artifact's repository. Prerequisites that are built or installed
separately (the patched Soufflé, a Halide install, SPIRV-Headers, Qt, Python environments) are
located by environment variables, listed in an example file, and the
tests that need an unset one are skipped.
The artifact will be archived with a DOI for the Available badge, with the hardware and
software needed to reproduce the case studies.

\clearpage
\appendix
\section*{Supplementary Material}
\setcounter{table}{0}
\setcounter{figure}{0}
\renewcommand{\thetable}{S\arabic{table}}
\renewcommand{\thefigure}{S\arabic{figure}}
\renewcommand{\theHtable}{S.\arabic{table}}
\renewcommand{\theHfigure}{S.\arabic{figure}}
\section{Proofs and Details for the Formalization}
\label{sec:formal-supp}
\sloppy
\crefname{exportremark}{Remark}{Remarks}
\Crefname{exportremark}{Remark}{Remarks}

This section gives the proofs of the results stated in \cref{sec:formal}, the full
cache-reuse algorithm with its guarantee, the finite-construction audit, and the
export conditions with the complete table of round trips. Unless stated here, the statements,
definitions and hypotheses are those of the paper; we refer to them by their numbers there. The
export lemma and downstream-use remark, summarized in \cref{sec:formal:export}, are stated in
\cref{sec:formal-supp:export}. \Cref{tab:glossary} maps the notation to relations,
\cref{tab:guarantees-full} maps guarantees to conditions in full, and \cref{tab:conditions}
defines each condition, grouped by role, with its discharge. The guarantee-indexed summary, with
the assurance boundary, is \cref{tab:guarantees}; \cref{tab:guarantees-coded} gives each result's
conditions and Lean status.

\begin{definition}[Terms]
Values are symbols, numbers and constructed terms:
a \emph{setting} $\langle k, v\rangle$; a \emph{configuration} $\langle p, s\rangle$ of a
platform $p$ and a finite list $s$ of settings, sorted by key; and a \emph{condition}, built
from $\mathsf{eq}(k,v)$, $\mathsf{defined}(t)$ and $\mathsf{named}(n)$ with $\neg$, $\wedge$
and $\vee$. Symbols may also be built by concatenation.
\end{definition}
The general condition evaluator reads effective settings, target-name existence for
$\mathsf{defined}(t)$ and the condition expression declared under $n$ for $\mathsf{named}(n)$.
Request and kind conditions instead admit only option tests
(\cref{sec:formal-supp:consequences}).

\begin{table}
\caption{Notation of \cref{sec:formal} and the relations of the submitted core that realize it.}
\label{tab:glossary}
\small
\begin{tabular}{@{}p{0.43\linewidth}p{0.50\linewidth}@{}}
\toprule
Concept & Relation(s) \\
\midrule
needed closure: target $t$ is built in config.\ $c$ & \texttt{needed}$(t, c)$ \\
tool configuration $\mathit{tool}(c)$ & \texttt{exec\_config}$(c, e)$ \\
decision relations & \texttt{request}, \texttt{request\_all}, \texttt{native\_kind} \\
effective option value (user value, else author default) & \texttt{option\_effective} \\
probe kind $k$: $\mathit{req}_k$, $\mathit{dom}_k$, results & \texttt{req\_}$k$, \texttt{domain\_}$k$, \texttt{res\_}$k$\ldots,
  declared in a layer manifest \\
$\mathit{pending}$, $\mathit{complete}$ & \texttt{probe\_pending}, \texttt{discovery\_complete} \\
toolchain selection; export spelling and loss & \texttt{tc\_select}; \texttt{export\_spelled}, \texttt{export\_lossy} \\
\bottomrule
\end{tabular}
\end{table}

\begin{table}[tp]
\caption{The guarantees of \cref{sec:formal} by condition: the conditions each adds to those of the rows it cites, and their discharge:
L loader check, D driver check, S argument for the shipped rules and probe kinds, O plugin-author
obligation. Checks cover reached rounds only. Lean: A abstract engine model, T transcription of
the shipped core rules (drift-checked). Conditions are defined in \cref{tab:conditions}.}
\label{tab:guarantees-coded}
\scriptsize
\begin{tabularx}{\textwidth}{@{}>{\RaggedRight}p{3.0cm}@{\hspace{5pt}}>{\RaggedRight}p{3.1cm}@{\hspace{5pt}}>{\RaggedRight}X@{\hspace{5pt}}>{\RaggedRight}p{2.6cm}@{}}
\toprule
Guarantee & Needs & Discharged by & Lean \\
\midrule
Termination (Thm.~\ref{thm:termination}) & (W1), (F), (P0), (P2) & (W1) L; (F) S/O; (P0) O; (P2) D membership, S/O finiteness & A \\
Order independence of a fresh run (Thm.~\ref{thm:determinism}) & + (P3) & O & A; fresh = \texttt{-{}-fresh} \\
Least closed guarded answers (Thm.~\ref{thm:lfp}) & + (W3), (RL), (A), (I), (CK) & (W3) L; (RL) O; (A), (I), (CK) S/O; (CK) D on reached rounds & A; (A)'s shipped argument not mechanized \\
Cache agreement (Cor.~\ref{cor:seeding}) & + (FC), (FL); guarded-first schedule & (FC) O; (FL) S/O; schedule and fallbacks D & A, given confirmation certificates; seed loop not mechanized \\
Bounded preparation (Prop.~\ref{prop:preparation}) & finite complete provider graph; fixed links & enumeration O; code checks and observations D & A: finite graph only \\
Answer-free target requests (Prop.~\ref{prop:options}) & (W1), (W5) & L on authored clauses; full guarded dependency certificate O & A, given dependency certificates \\
Errors do not steer (Prop.~\ref{prop:errors}) & (W1)--(W6), (F), (P0), (P3) & L; (W6) and domains D & A, (W4) gate discharged \\
Toolchain uniqueness (Inv.~\ref{lem:toolchain}) & shipped core rules & S & T \\
Tool configurations (Inv.~\ref{prop:exec}) & + one build-machine platform fact; no setting or tool-configuration error & S; errors reported & T, functional form \\
Export fidelity (Rem.~\ref{prop:export}) & error-free export; (E1)--(E5) & (E1) S in \sys's format, else O; (E2), (E3) O, fixture-tested; (E4) S in part; (E5) O & T: exclusive classification, (E1) premise \\
\bottomrule
\end{tabularx}
\end{table}

\begin{table}[tp]
\caption{Full guarantee conditions and discharge; cited results' hypotheses also apply.
Shipped rules: core, C/C++, Halide and Qt. Shipped probe kinds: core and production/case-study
plugins, excluding fixtures.}
\label{tab:guarantees-full}
\label{tab:enforced}
\scriptsize
\begin{tabularx}{\textwidth}{@{}>{\RaggedRight}p{3.3cm}@{\hspace{6pt}}>{\RaggedRight}p{4.4cm}@{\hspace{6pt}}>{\RaggedRight}X@{}}
\toprule
Guarantee & Conditions & Checked, or obligation \\
\midrule
A unique model (\S\ref{prop:model}) & $P$ stratified; its model on $D(I)$ finite &
  Stratification of the composed program checked by the loader. Finiteness an obligation,
  argued only for the shipped rules (Lemma~\ref{lem:shipped-finite}). \\ \addlinespace[2pt]
No silent toolchain choice for a needed target (Invariant~\ref{lem:toolchain}) & as
  \S\ref{prop:model}; the core's selection and error rules &
  Not a loader check: the core's rules derive an error for an open or inconsistent choice, and
  ``no toolchain'' when none is selected. \\ \addlinespace[2pt]
Discovery stops, with nothing pending or at a failed check, within $1 + \sum_k
  |\mathcal{U}_k|$ rounds (Thm.~\ref{thm:termination}) & $P$ stratified; (F); (P0) and (P2)
  for every probe kind &
  The driver checks domain membership, argument tuples and missing primary answers in each
  round, so (P1) is a check, not a hypothesis. (P0) an obligation: the driver sets no time limit.
  (F) an obligation, argued for the shipped rules (Lemma~\ref{lem:shipped-finite}). Finiteness of each
  $\mathcal{U}_k$, which may depend on the environment, an obligation. \\ \addlinespace[2pt]
The outcome, and which check fails, depend only on $A$, $U$ and the environment, not on
  order (Thm.~\ref{thm:determinism}) & as Thm.~\ref{thm:termination}; (P3) for
  every probe kind; a fixed prepared program and available metadata &
  (P3) an obligation. The driver reports all of a round's check failures in canonical order;
  implemented and tested under shuffled probe orders. Shuffle tests also compare emitted graphs.
  \\ \addlinespace[2pt]
Guarded answers have a unique least correct closed set; all correct closed sets give $M_F$
  (generalization of Thm.~\ref{thm:lfp}) & fixed stratified $P$; (RL), (F), (A), (I); finite request universes;
  (P2), (CK) over every correct-answer subset; correct unguarded $F$ &
  (W3) checked by the loader. The universal request-domain and answer hypotheses are
  obligations. (A) proved for the shipped kinds from (P3)
  (Lemma~\ref{lem:fixed}). \\ \addlinespace[2pt]
A replay and a seeded configure have a fresh configure's outcome (Cor.~\ref{cor:seeding}) &
  as Thm.~\ref{thm:determinism} and Thm.~\ref{thm:lfp}, including (RL);
  (FC), (FL) &
  The driver answers nothing in a round that drops a stale seed, reruns seeded when a replay deviates, and reruns unseeded
  when a seeded run fails a check. (FC), (FL), (I) obligations; the shipped request
  rules preserve domain membership, and shipped identifiers are injective by construction
  (\cref{sec:shipped-checks}). \\ \addlinespace[2pt]
Preparation reaches the least include/rule closure (Prop.~\ref{prop:preparation}) &
  finite complete enumeration; fixed links independent of answers and demand &
  Enumeration is a provider obligation. The driver validates the executable union before
  inference and confirms observations; no environmental round admits code. \\ \addlinespace[2pt]
Requested targets and their declared kinds are decided before discovery
  (Prop.~\ref{prop:options}) & fixed $P$, $A$, $U$, $V$; $P$ stratified; (W5) &
  W5 checked on authored clauses; full guarded W5 is an additional obligation.
  Stratification checked on the guarded union. \\ \addlinespace[2pt]
Errors do not steer discovery (Prop.~\ref{prop:errors}) & $P$ well-formed: (W1)--(W6),
  including (W2)'s finite domains and (W3) for kinds seeded beyond replay;
  (F) for $P$ and $P'$; (P0), (P3) &
  (W1), (W3), (W4) checked by the loader; W5 has the qualification above.
  (W4) protects request, domain and
  discovery-time check relations from completion-gated rule heads.
  (W6) and domain membership checked by the driver in each round. Finiteness of the
  domains across rounds, in (W2), an obligation. \\ \addlinespace[2pt]
A target an action runs is needed in its unique tool configuration (Invariant~\ref{prop:exec}) &
  no setting or tool-configuration error; one fact identifying the build machine's platform &
  By the core's rules, which derive the tool configuration and the need; an action that runs
  something unrunnable derives an error. Mechanized against the transcribed rules
  (\texttt{base\_exec\_exists\_unique}); ``the model exists'' alone does not give a total
  $\mathit{tool}(c)$. \\ \addlinespace[2pt]
For an exported configuration, each own fact is spelled, lost with a warning, or an export
  error, exclusively on error-free models;
  each requirement is stated or an error; foreign formats warn that shipped rules are lost
  (Lemma~\ref{lem:export-exporter}) & the model exists; an active export format &
  By the export layer's rules. Tests cover examples of loss and warning, not every exporter
  case. \\ \addlinespace[2pt]
Depending on an exported target gives the built target's relocated usage facts, as a set
  (Remark~\ref{lem:export-downstream}) & an error-free export; (E1)--(E5), for the component and its requirements.
  Excludes order and multiplicity, version acceptance, absolute paths outside the package, and
  hand-written packages with no built counterpart &
  (E1) by construction only in \sys's format; the other formats report losses. (E2) and (E3)
  tested on fixtures, not proved. Ownership partly discharges (E4) for shipped rules, except
  their actions. (E5) an unchecked assumption on shipped rule layers. \\ \addlinespace[2pt]
Every fact in $M_n$ has a proof tree (\cref{sec:formal:export}) & the final model exists and
  is finite &
  Souffl\'e's provenance; \texttt{why} prints it. Negated premises, constraints and aggregate
  results are leaves. An installed fact's proof stops at its installed facts file, not at its
  producer's derivation. \\
\bottomrule
\end{tabularx}
\end{table}

\begin{table}[tp]
\caption{Conditions by role and their discharge: L loader check, D driver check, S
shipped-kind/rule argument, O plugin-author obligation. Checks cover reached rounds only; S has
the audit boundaries of \cref{sec:shipped-checks,sec:formal-supp:complexity}.}
\label{tab:conditions}
\scriptsize
\begin{tabularx}{\textwidth}{@{}>{\RaggedRight}p{1.9cm}@{\hspace{6pt}}>{\RaggedRight}X@{\hspace{6pt}}>{\RaggedRight}p{4.6cm}@{}}
\toprule
Condition & Meaning & Discharged by; where \\
\midrule
\multicolumn{3}{@{}l}{\emph{Probe obligations}} \\
(P0) Return & Probe and fingerprint calls return; no driver timeout. & O; \cref{sec:formal:discovery} \\
(P3) Functional & Answers depend only on arguments, fingerprinted answers, watches and environment. & O; \cref{sec:formal:discovery} \\
(RL) Row ownership & All answer rows carry their request's identifier. & O; \cref{sec:formal-supp:cache} \\
(FC) Sufficient keys & Equal keys and unchanged watches imply equal answers across configures. & O; \cref{sec:formal-supp:cache} \\
(FL) Prerequisites & Probes/keys read only acyclic prerequisite answers used by every request derivation. & S inventory argument/O; Lemma~\ref{lem:fixed} \\
\multicolumn{3}{@{}l}{\emph{Rule and domain conditions}} \\
(F) Finite evaluation & Every finite answer set has a finite model. & S/O; \cref{sec:formal-supp:complexity} \\
(P2), (W2) & Domains lie in fixed finite request universes; requests lie in domains. & D membership; S/O finiteness, \cref{sec:shipped-checks} \\
(A) Fixed answers & Each request has one answer across runs and rounds in an environment. & S from (P3), Lemma~\ref{lem:fixed}; otherwise O \\
(I) Stable identifiers & An identifier determines its arguments across configures. & S length-prefixing; O, \cref{sec:shipped-checks} \\
(CK) Passing checks & Domain, tuple, primary-row and declaration checks pass over every correct-answer subset. & S/O; D checks reached rounds only, \cref{sec:shipped-checks} \\
Preparation finiteness & Providers enumerate a finite complete graph within declared roots. & O; Prop.~\ref{prop:preparation} \\
Fixed links & Metadata includes and rule pointers do not depend on demand or probe answers. & Provider contract; D bounded-decoding and observation checks, Prop.~\ref{prop:preparation} \\
\multicolumn{3}{@{}l}{\emph{Loader and driver checks}} \\
(P1) Totality & Every answer has a matching primary row, even if negative. & D at each call; \cref{sec:formal-supp:discovery} \\
(W1) Stratified & No negative/aggregate cycle, including generated pending rules. & L at composition; \cref{sec:formal:programs} \\
(W3) Guarded & Each result read entails its matching positive request. & L syntactic sufficient check; \cref{sec:formal-supp:cache} \\
(W4) Error isolation & Requests, domains and discovery checks cannot depend on completion-gated heads. & L dependency check; \cref{sec:formal-supp:consequences} \\
(W5) Answer-free & Explicit target requests and declared kinds cannot depend on answers. & L on authored clauses; O for full guarded union; \cref{sec:formal:consequences} \\
(W6) Unique tuples & One argument tuple per kind/identifier in a round. & D each round; \cref{sec:formal-supp:discovery} \\
\multicolumn{3}{@{}l}{\emph{Export conditions} (for an error-free export)} \\
(E1) No loss & No export loss fact for the component or its requirements. & S native format; O otherwise, \cref{sec:formal-supp:export} \\
(E2) Inversion & Import spelling inverts export spelling, with relocated paths. & O, fixture-tested; \cref{tab:formats} \\
(E3) Resolution & Requirements resolve to the same components/configurations. & O, fixture-tested; \cref{tab:formats} \\
(E4) Isolation & Consumer inputs and other dependencies remain unchanged. & S ownership in part/O; \cref{sec:formal-supp:export} \\
(E5) Neutral rules & Exported rules add no usage facts about their package's targets. & O; \cref{sec:formal-supp:export} \\
\bottomrule
\end{tabularx}
\end{table}

\subsection{One Derivation and Discovery}
\label{sec:formal-supp:discovery}

\paragraph{General schedule}
The main paper specializes to shipped guarded kinds. For the generic plugin interface,
let $G$ be the guarded kinds and let $Q_i$ be all requests of $M_i$ lacking a primary
answer in $E_i$. The driver selects
\[
R_i=\begin{cases}
\{(k,r)\in Q_i\mid k\in G\},&\text{if this set is nonempty},\\
Q_i,&\text{otherwise}.
\end{cases}
\]
Thus $R_i=Q_i$ when all kinds are guarded; the general loop stops when $Q_i=\emptyset$.
The finite-progress and round-induction proofs below apply to either schedule.

\paragraph{Rule inspection for Invariant~\ref{lem:toolchain} (toolchain uniq{}ueness)}
Here a layer requires $\ell$ in $c$ (\texttt{tc\_required}), so $(\ell, e, p)$ is used
(\texttt{tc\_used}). Without a pin every candidate is selected; with one, only the pinned
candidate is. An error naming $(\ell, e, p)$ is derived if two candidates are selected, if two
pins disagree, or if a pin names a non-candidate. The cases are exhaustive. If no toolchain is
selected, \texttt{has\_tc}$(\ell, c)$ fails, and the rule for ``no toolchain'' derives the
error at every target with an action in $c$. The error rules use only negation of
lower-stratum relations.
These are rule-level invariants of the submitted core, not properties of arbitrary
extensions. The selection and rejection rules in \texttt{L2\_toolchains.dl}, abbreviating
diagnostic strings and the two pin-conflict diagnostics, are:
\begin{verbatim}
tc_select(l,e,p,t) :- tc_candidate(l,e,p,t), !pinned(l,e,p).
tc_select(l,e,p,t) :- tc_candidate(l,e,p,t), toolchain_pin(l,e,p,t).
error("ambiguous",...) :- tc_used(l,e,p),
  tc_select(l,e,p,a), tc_select(l,e,p,b), a < b.
error("non-candidate pin",...) :- tc_used(l,e,p),
  toolchain_pin(l,e,p,t), !tc_candidate(l,e,p,t).
error("conflicting pins",...) :- toolchain_pin(l,e,p,a),
  toolchain_pin(l,e,p,b), a < b.
error_at("no toolchain",...,t,c) :- tc_required(l,c),
  !has_tc(l,c), cfg_name(c,n), target_action(t,c,_).
\end{verbatim}
Here \texttt{pinned(l,e,p) :- toolchain\_pin(l,e,p,\_)}.
The missing-selection diagnostic uses \texttt{!has\_tc(l,c)}, where
\texttt{has\_tc(l,c) :- tc\_for(l,c,\_)}.
Both invariants are mechanized in Lean against a transcription of the shipped rules, kept in
step with the sources by a drift test (\cref{sec:formal}); the inspection above and the
property tests below are secondary evidence.
Property-based tests in \texttt{tests/test\_core\_invariants\_prop.py} check ambiguity,
pin conflicts, invalid pins and no silent choice, as well as the two directions of the
tool-induced configuration invariant below.
They generate small projects from a fixed seed (20261012): 130 for toolchain selection and 110
for the tool configuration, compared with an independent least fixpoint, plus a check that the
generator covers pins, zero to three toolchains and a tool also used as a library. All 241 cases
pass.

\begin{lemma}[Finite rounds]
\label{lem:finite}
If $P$ is stratified and satisfies (F), and every probe call returns (P0), every round reached
has a finite $E_i$ and a finite, unique $M_i$.
\end{lemma}
\begin{proof}
By induction on $i$. $E_0$ is empty. If $E_i$ is finite, then by (F) $M_i$ is finite, and
the standard model fact (\S\ref{prop:model}) makes it unique, so $R_i$ is finite. By (P0) each of the round's
probe calls returns, and a returned answer is a finite set of rows, so $E_{i+1}$ is finite.
\end{proof}

\begin{proof}[Proof of Theorem~\ref{thm:termination} (fresh termination)]
By Lemma~\ref{lem:finite}, each round's evaluation and probe calls terminate. Let $\mathit{Ans}_i$ be the
set of pairs $(k, \mathit{id})$ with a primary result in $E_i$. It grows with $i$. In a round
with $R_i \neq \emptyset$, either a check fails (a request outside its domain, two argument
tuples for one pair, an empty primary answer) and configure stops, or every request in $R_i$
gains a primary result, so $\mathit{Ans}_{i+1} \supsetneq \mathit{Ans}_i$. The per-round domain
check places every pair keyed by $R_i$ in $\{(k, \mathit{id}(r)) \mid r \in
\mathit{dom}_k(M_i)\}$, and (P2), whose across-round part is not checked, places that in the
finite set $\{(k, \mathit{id}(r)) \mid r \in \mathcal{U}_k\}$. $\mathcal{U}_k$ may depend on the
environment, as \texttt{cxx\_std}'s does through the inventory's answers, but the environment does not change during configure, so
$\mathcal{U}_k$ is one fixed set for the run. A strictly increasing chain of subsets of a finite
set has at most $\sum_k |\mathcal{U}_k|$ strict steps.
\end{proof}

\begin{proof}[Proof of Theorem~\ref{thm:determinism} (fresh determinism)]
By Theorem~\ref{thm:termination}, there is a last round $n$. By induction on $i \le n$, $E_i$
is a function of $(A, U)$ and the environment's snapshot observations $\beta_k(r,E_i)$:
$M_i$ is unique by Lemma~\ref{lem:finite},
so $R_i$ and the checks of round $i$, which read only $M_i$ and $E_i$, are determined; every
probe of the round reads $E_i$ and no other probe's answer of the same round, so by (P3) each
answer is determined, and the union defining $E_{i+1}$ does not depend on the order in which
they are computed. Hence the round at which configure stops is determined. If a check fails in
that round, the set of violations reads only $M_i$, $E_i$ and the round's answers, so it is
determined too, and the driver reports all of them in a canonical order (below), so the report
does not depend on the order of probes. On success
or failure in the model, the outcome reads only $M_n$: $\mathit{emit}$ depends only on the model
as a set, and the errors are the extension of the error relations in $M_n$.
\end{proof}

The driver checks a round in two steps. It first checks every request against its domain and
for a second argument tuple. If either check fails, it reports every such violation of the round
and runs no probe. Otherwise it runs every probe of the round, collects every empty answer and
every undeclared result relation, and, if there is any, reports them all. In both steps the
report lists the violations sorted by check, probe kind and request arguments, so it is the same
for every order of probes. A test runs a round in which two probes leave three requests without
an answer and a third returns an undeclared relation, under reversed and shuffled probe orders,
and checks that all four violations are reported, in the same order.

The determinism observation is also tested: the driver can shuffle the input facts and the order of probes in
each round with a seed, and the integration tests configure the cross-compilation example and
the Halide case study under shuffles and compare the emitted graphs. Within a round, every probe
reads a read-only snapshot of the answers confirmed at the start of the round, which is
$E_i$ in the definition above, so no probe can observe the order in which the others ran. The
driver's tests check that reversed and shuffled probe orders give the same result.

\subsection{The Order of Discovery and Reusing Answers}
\label{sec:formal-supp:cache}

\paragraph{A withdrawing request}
\label{sec:formal-supp:withdraw}
Consider an illustrative guarded layer over \texttt{cc\_check}, abbreviating two fixed
argument tuples and their identifiers by \texttt{a}, \texttt{b}, and its request/result
relations by \texttt{req}, \texttt{res}. Assume the prerequisite inventory is already fixed:
\begin{lstlisting}[language=Datalog]
dom("a"). dom("b").
req("a") :- dom("a").
req("b") :- dom("b"), req("a"), !res("a","yes").
use(X) :- req(X), res(X,"yes").
\end{lstlisting}
With no answers, both requests hold. Suppose both checks answer ``yes'' in that round.
The next model withdraws \texttt{req("b")} and derives only \texttt{use("a")};
\texttt{b}'s retained answer is inert. With unchanged program, non-answer input, keys and
watches, replay starts with both cached answers, confirms \texttt{b} from the record's
withdrawn list and confirms \texttt{a} from its request, calling neither check.
If replay is unavailable, guarded seeding leaves \texttt{b} unconfirmed and prunes it.
Both have the fresh outcome, although guarded seeding need not retain the fresh run's
probe-result facts.

A configure may start from the previous configure's answers instead of from
$E_0 = \emptyset$. We state what the driver does and then what it guarantees. A probe kind
may declare a \emph{fingerprint}, a function of the request's arguments and of the answers
given so far; the answer's \emph{key} is the kind, the arguments and the fingerprint. Answers
are cached under their keys. A probe may also record a \emph{watch set} of filesystem
observations; the answer is reused only while those observations remain unchanged.
Shipped compile checks watch depfile-recorded files, compiler-reported header/framework
search roots and linker search paths, and selected library files.
File identities record resolved path, size, modification time, inode and device, not
content hashes; directory identities omit size and are nonrecursive. A recorded missing
path is invalidated by its appearance. These observations do not establish complete
negative or transitive lookup coverage: a nested earlier candidate, an unrecorded
\texttt{\_\_has\_include} input, or dependencies omitted by a failed compilation may
change without invalidating the retained answer.

The shipped probes therefore assume that inputs outside their keys and recorded watches
remain stable while an answer is reused. After installing, removing, replacing or
restructuring external/toolchain include or library inputs, users explicitly refresh:
\begin{quote}
\texttt{buildlog configure -{}-fresh -C build project}
\end{quote}
with the intended user files, presets and configuration selections.
This bypasses the unchanged-configure fast path, seeds and cached probe answers, and
executes the reached requests. It refreshes those answers, not every old cache entry;
other affected configurations or newly enabled checks also need a fresh configure.
Disabling seeding alone does not bypass answer-cache reuse. For these probes, (FC)
includes the stable-unobserved-environment assumption; validating observations does not
validate their completeness. Changes must be visible through the recorded identity fields.

A kind without a fingerprint has no key: its answers are never
cached or seeded, and it runs in every configure. At the end of a configure, the driver records
every answer of the run with its key, a digest of the composed program, and a digest of the
round-0 input other than answers ($A$, $U$, the packages' facts).

A kind $k$ is \emph{guarded} if every rule that reads one of
its result relations, positively, negatively or inside an aggregate, binds the result's identifier, in the same
disjunct, in a positive atom of $\mathit{req}_k$ or of a relation all of whose rules bind their
first argument that way. The loader checks this syntactically for every kind that does not
declare itself \emph{unguarded}, with a stated reason (W3). The shipped prepared union has
no unguarded package kinds: selection reads fixed available metadata instead.
The generated $\mathit{pending}$ rules read a result negatively, joined with its request, so
they are guarded too.

After bounded preparation and one composition, the driver proceeds as follows
(\texttt{run\_fixpoint}). The generic schedule still accommodates explicitly justified
unguarded plugins; with shipped guarded kinds it simply answers all pending requests.
\begin{enumerate}
\item \emph{Seed.} Take the record for the current program, else the latest one, and keep the
  answers whose kind exists, whose key is in the cache, whose watch set is unchanged and whose
  key, recomputed against the kept answers, is unchanged. The run is a \emph{replay} if the program and the round-0 input
  other than answers are unchanged and every recorded answer is kept. The record lists
  withdrawn requests (step~4); records without that list use their replayable flag;
  otherwise it is \emph{seeded}, and the answers of unguarded kinds are dropped.
\item \emph{Rounds.} The kept answers are added to $E_0$, unconfirmed, except that replay
  confirms the recorded withdrawn answers, including unguarded ones. In each round, after
  the checks, a seeded answer is \emph{confirmed} if its request is in $M_i$ and its key,
  recomputed against the confirmed answers only, is unchanged; this is repeated until nothing
  changes. A seeded answer whose request is in $M_i$ but is not confirmed is \emph{stale} and
  is dropped, and a round that drops one answers no request: the next round computes the
  requests again without it. Otherwise, if a request of a guarded kind is pending, the round
  answers only the pending guarded requests; otherwise it answers the pending unguarded ones.
  A request that waits is not kept: the next round computes the requests again from its own
  model. Requests are answered from the cache when their key is there and its watch set holds, and otherwise by the
  probe, against a snapshot of the confirmed answers. The run stops when nothing is pending.
\item \emph{Fallback.} If a replay meets a pending request, a stale answer, or a seeded answer
  left unconfirmed whose request did not arise, it is abandoned, and the configure runs
  again as a seeded run from the same record. If a run that started with seeds fails a check,
  it is abandoned, and the configure runs again without seeds and reports that run's outcome.
\item \emph{End.} When nothing is pending, the seeded answers that were never confirmed are
  removed from the final input facts, and the record is written. It holds every answer of the
  run that remain after pruning, including confirmed answers whose requests were later
  withdrawn, and lists those withdrawn requests. Records are kept separately for each
  composed program. A fallback re-evaluates the same program; it never composes a larger one.
\end{enumerate}
Before publishing success, the driver confirms the preparation observations and rejects any
used file, root or rule module outside the certified union. Final locks include explicit
packages and finally used modules only.

What is written when a configure fails depends on where it fails. A failure before emission (the
loader rejects the composed program, the lock does not hold, a driver check or a probe fails,
or preparation confirmation fails) stops the configure and leaves the build directory's previous
Ninja file in place, stale with respect to the inputs. If the final model has errors,
\texttt{configure} still writes the fact files, metrics and lock, deletes the Ninja file and exits
with failure. With \texttt{-{}-keep-going}, errors attributed to a target or action instead become
edges that print the error and fail when built, the rest of the graph is emitted and
\texttt{configure} succeeds; errors not attributed to a target or action still delete the Ninja file.
The regeneration edge of an emitted Ninja file depends on the files the configure read, not on
probe answers, so a change that only a probe would see takes effect at the next explicit configure.

The results below use (RL), (A), (FC), (FL), (I) and (CK) as stated in
\cref{sec:formal:discovery}. Throughout, an \emph{answer} is the set of rows that one call
gives for one pair $(k, \mathit{id})$; every result row carries that pair's identifier,
including rows retained from an earlier configure. This answer-row ownership (RL) is an obligation
on probe authors: the driver checks for a matching primary row and declared relations, but
does not check every secondary row's identifier. We write $M(E)$ for
$M_P(A, U, E)$, $G$ and $N$ for the guarded and unguarded kinds, and $M \approx M'$ if two
models agree on every relation other than the probe results. Emission and the error report
read only relations other than the probe results, so models that agree in this sense have the
same outcome.
This agreement excludes retained probe-result rows and the final answer-input files:
guarded seeding can skip a transient request that a fresh run answers. The example above
exhibits this difference. Equality of the previous and fresh final input facts is not a
premise or conclusion of Corollary~\ref{cor:seeding}.

The \emph{confirmation invariant} concerns provenance and correctness, not persistence
of requests. An executed answer has a request in the model of its probing round. A
normally confirmed seed has a request in the confirming round and an unchanged key
computed from confirmed prerequisite answers, with valid watches. A replay's recorded
withdrawn answer has the recorded run's certificate instead; it need not have a request
in the replay model. Subsequent rounds retain these certificates when requests withdraw.
Unconfirmed guarded seeds remain tentative and are removed if never confirmed.

\paragraph{What (W3) checks}
\texttt{layers.\_check\_answer\_reads} checks each result atom's identifier variable against
a positive, top-level request atom with exactly that variable in the same body disjunct.
It also admits auxiliary guards: starting from requests, it certifies a relation only if
every clause entails an already certified guard with the head's identifier variable.
Ground facts and accepted, user or driver input relations cannot certify auxiliary guards;
aggregate-local atoms and atoms inside parentheses cannot certify positive guards.
Thus nested disjunctions cannot borrow a guard from another branch.
The check also traverses Souffl\'e's expanded initial AST, so macros and includes cannot
hide result reads or bypass auxiliary-guard certification.
Induction on this certification establishes request entailment, including for negative and
aggregate result reads. This is a sound, conservative syntactic sufficient condition for
semantic guardedness: it may reject semantically valid joins with identifier equalities,
parenthesized guards or more involved implications. Unguarded declarations are explicit
opt-outs and receive no speculative seeding.

\paragraph{Identifier stability and fingerprints}
(I) is not enforced across configures. The cache key is exactly the kind, the entire
argument tuple and the fingerprint (\texttt{discovery.cache\_key}), so changing arguments
prevents reusing the old answer even for the same identifier. A simultaneous collision
with two tuples is rejected in that round. Neither behavior proves the cross-configure
stability required by Theorem~\ref{thm:lfp}; a violating plugin cannot invoke that theorem.
Secondary-row ownership (RL), (FC) and (FL) remain plugin obligations; violating them
can invalidate seeding equivalence without being detected. In (FL), ``earlier answers''
are prerequisite answers in the acyclic request-dependency order: every derivation of
the request uses their rows. Confirmation follows this dependency order, not a chosen
ordering of independent probes.
Under (A), the fixed-answer function $\alpha_k(r)$ replaces the snapshot-dependent
$\beta_k(r,E)$ used by fresh determinism; these are distinct functions with distinct premises.

\begin{lemma}[Fixed answers for the shipped kinds]
\label{lem:fixed}
If the shipped probe kinds satisfy (P3), they satisfy (A).
\end{lemma}
\begin{proof}
The inventory kinds of the C/C++, Emscripten and NDK plugins read no earlier answer, so by (P3)
each answers its one constant request as a function of the environment. Every other shipped
probe reads, among the earlier answers, only inventory rows about the toolchain and platforms
that its request names (a toolchain's tools, a platform's triple, SDK and constraints). Only an
inventory answer states a toolchain or platform (by the shipped domain rules), so every
derivation of the request uses those rows. A round that answers requests has dropped no stale
answer, so its unconfirmed seeds are inert (Lemma~\ref{lem:guard} below) and its requests are
those of the model over the confirmed and probed answers, which form the snapshot. The rows a
probe reads are thus in its snapshot, and they are the inventory's fixed answer. By (P3), each
other probe's answer is then a function of its request and the environment.
\end{proof}

\paragraph{Checks over every correct-answer subset}
\label{sec:shipped-checks}
For the shipped rules, domain membership is structural, not an induction limited to the
fresh schedule. The arguments below fix $A$, $U$, the environment and the finite package
answers allowed by the theorem. Let $V$ contain all their input values, the shipped rule
constants and all correct inventory values. Finiteness of configurations, path derivations
and template expansions follows from Lemma~\ref{lem:shipped-finite}. Deleting correct answers
may expose negated branches, but cannot introduce values outside those finite sources.
Each domain is therefore contained in a fixed finite universe even over arbitrary subsets.
\begin{description}
\item[Inventory, Emscripten inventory, NDK inventory.]
Each request and domain uses the same constant identifier and the same
\texttt{config\_decl} premise. Their universes are singletons, there are no arguments,
and their probes return a primary row: the build-machine row for C/C++, a status row for
Emscripten and NDK even when no toolchain is found.
\item[\texttt{cc\_capability}.]
The request copies its domain, formed from the finite toolchain, language and platform
sets. Negated cross-mismatch and built-toolchain tests only filter this product.
The probe returns a yes/no primary row for every tuple.
\item[\texttt{cxx\_std}.]
The request copies its domain, a subset of inventory toolchains times the finitely stated
\texttt{lattice\_rank} standards. Its yes/no answer always has a primary row.
\item[\texttt{cc\_warning}.]
The request copies its domain, a subset of toolchains times the two languages and the
finite \texttt{cc\_default\_warning} flags; its probe always returns a yes/no primary row.
\item[\texttt{cc\_implicit\_include}.]
The request copies its domain, toolchains times the two languages. The primary result
includes an empty-directory row when there are no include directories.
\item[\texttt{cc\_check}.]
The request copies its domain of finite \texttt{cc\_check\_want} tuples; negating
\texttt{cc\_check\_stated} only removes tuples. Subjects, headers and argument words are
drawn from the fixed finite input and derived configurations. Every completed probe
returns a primary yes/no row; the optional size row uses a declared secondary relation.
\item[\texttt{file\_archs}.]
The request copies its domain, one tuple per derived imported file path. These paths come
from finitely many package answers and inputs. The primary format row is present even for
an unknown format; architecture rows belong to its declared secondary relation.
\item[\texttt{python\_interp}.]
The domain uses the finite input interpreter paths; the request only filters it by
\texttt{python\_wanted}. A missing interpreter returns a primary ``found=no'' row.
\item[\texttt{python\_package}.]
The domain is the finite product of interpreter paths and stated distribution/module/
specifier tuples; wanted and found-interpreter premises only filter it. A primary
``found'' row is returned even when no package is found.
\end{description}
All these implementations return only their declared result relations. Thus domain
membership, answer totality and declaration checks hold over all correct subsets.
The tuple-uniqueness part of (CK), and (I), require collision-free identifiers. Every
shipped multi-argument kind builds its identifier by length-prefixing each field but the
last (\texttt{len:value:}), so the identifier is injective in the argument tuple by
construction, whatever characters the arguments hold: subject \texttt{a|b} with headers
\texttt{c} and subject \texttt{a} with headers \texttt{b|c} get distinct identifiers.
Single-argument kinds are injective trivially. Third-party kinds still carry the obligation:
they must be injective and preserve domain coverage to use the theorem.
These per-kind arguments discharge all-subsets (P2) for the shipped composition, not
arbitrary extensions of its open relations.

\begin{lemma}[Inert answers]
\label{lem:guard}
Let $P$ be stratified and satisfy (F) and (RL), let $E$ be a finite set of answers, and let $X$ be a finite set of answers of guarded
kinds. If no request answered in $X$ holds in $M(E)$, or none holds in $M(E \cup X)$, then
$M(E) \approx M(E \cup X)$.
\end{lemma}
\begin{proof}
Write $M = M(E)$ and $M' = M(E \cup X)$, and suppose that no request of $X$ holds in $M^\circ$,
which is one of the two. We show by induction on strata that $M$ and $M'$ agree on every
relation other than the probe results. Let the strata below $s$ agree. A rule instance of
stratum $s$ that reads a row of $a \in X$, positively, negatively or inside an aggregate, has in
the same disjunct a positive premise $g(\mathit{id}(a))$, where $g$ is $\mathit{req}_k$ or a
relation all of whose rules bind their first argument in $\mathit{req}_k$; so $g(\mathit{id}(a))$
holds in a model only if $\mathit{req}_k(\mathit{id}(a))$ does. If $g$ lies below $s$, it agrees
in $M$ and $M'$, and it is false in $M^\circ$, hence in both. If $g$ lies in stratum $s$,
consider the iterates $J_0 \subseteq J_1 \subseteq \cdots$ of the stratum's
immediate-consequence operator over $M^\circ$'s lower strata, which reach $M^\circ$'s stratum
$s$. Each $J_m$ lies inside $M^\circ$, so $g(\mathit{id}(a))$ is false in it and the instances that
read $X$ have a false positive guard (aggregate contributions are unchanged); the operators
for $E$ and for $E \cup X$ therefore agree on every $J_m$, and the
two least fixpoints coincide. In either case stratum $s$ agrees, and the induction closes.
\end{proof}

In the shipped rules, a guard relation other than the request is always a lower one, but the
proof does not need this.

\paragraph{Guarded discovery}
Throughout this characterization, fix a composed $P$, $A$, $U$ and environment and finite
universes $\mathcal{U}_k$ of request tuples. Each identifier determines one argument tuple
across these universes (I). Every $\alpha_k(r)$ is a correct, finite answer with a primary
row and only declared result relations; (RL), (F), (A), (P2) and (CK) hold for every model on
subsets of these answers. In particular, all requests of those models belong to
$\mathcal{U}_k$, not merely the requests of a fresh run. Quantification over answer sets
below is only over subsets of these fixed answers, not arbitrary result rows or stale seeds.
Fix a correct set $F$ of answers of unguarded kinds. A correct set $E$ of answers of guarded kinds is
\emph{closed over $F$} if every answer in $E$ is $\alpha_k(r)$ for its request $r$, and every
request of a guarded kind that holds in $M(E \cup F)$ is answered in $E$.

\paragraph{Generalized least-closure statement}
Let $G$ be the guarded kinds. Under the hypotheses above, sets closed over $F$ exist
and all give the same model
off probe-result relations, denoted $M_F$.
Then $E^*_F=\alpha(\mathit{req}_G(M_F))$ is the unique least correct closed set over $F$.
This proves Theorem~\ref{thm:lfp} by taking $F=\emptyset$ and all kinds guarded.

\begin{proof}[Proof of the generalized Theorem~\ref{thm:lfp}]
\emph{Existence.} Let $E_0 = \emptyset$ and $E_{i+1} = E_i \cup \alpha(\mathit{req}_G(M(E_i \cup
F)))$. Each $M(E_i \cup F)$ is finite by (F), and by (P2) the identifiers answered lie in a finite
set, so the chain stops at a set closed over $F$.
\emph{Uniqueness.} Let $E_1$ and $E_2$ be closed over $F$, and let $X = E_2 \setminus E_1$. If
the request $r$ of some $a \in X$ held in $M(E_1 \cup F)$, then $E_1$ would answer it with
$\alpha_k(r)$, which is $a$ by (A); so no request of $X$ holds in $M(E_1 \cup F)$, and
Lemma~\ref{lem:guard} gives $M(E_1 \cup F) \approx M(E_1 \cup E_2 \cup F)$. Symmetrically
$M(E_2 \cup F) \approx M(E_1 \cup E_2 \cup F)$. Call the common model, up to $\approx$, $M_F$.
\emph{Least closed set.} Let $E^*_F = \alpha(\mathit{req}_G(M_F))$, which is well defined, since
requests are not probe results. For any $E$ closed over $F$, $E^*_F \subseteq E$, and the requests
of $E \setminus E^*_F$ do not hold in $M(E \cup F) \approx M_F$, so by Lemma~\ref{lem:guard}
$M(E^*_F \cup F) \approx M_F$; hence $E^*_F$ is closed over $F$ and
$\alpha(\mathit{req}_G(M(E^*_F \cup F))) = E^*_F$. A correct set $E$ of answers with
$\alpha(\mathit{req}_G(M(E \cup F))) \subseteq E$ is closed over $F$, so it contains $E^*_F$.
\end{proof}

The operator $E\mapsto\alpha(\mathit{req}_G(M(E\cup F)))$ need not be monotone: an answered
request can withdraw another request. The proof obtains leastness among correct closed sets
by inertness and model agreement, not by applying Knaster--Tarski to this operator.
Guarded reads entail positive request dependencies. Within a stratum a cycle of such guards
cannot justify itself: every request needs a finite least-derivation tree rooted outside the
cycle. Negative and aggregate dependencies on derived relations must go to lower strata.
Together these facts prevent two guarded, self-consistent closed sets from sustaining
mutually withdrawing requests; Lemma~\ref{lem:guard} is the precise stratum-induction argument.

\begin{lemma}[What a round's model is]
\label{lem:phase}
Under the generalized least-closure hypotheses above, additionally assume (FC) and (FL),
and let a run of the driver reach a round $i$ that drops no
stale answer and in which no request of a guarded kind is pending after confirmation.
Let $F$ be the unguarded answers of $E_i$. Then $M_i \approx M_F$, and the confirmed and
probed guarded answers of $E_i$ are closed over $F$.
\end{lemma}
\begin{proof}
Every answer that the run gave by calling a probe is $\alpha_k(r)$ by (A), and every answer
taken from the cache is too, by (FC). A confirmed seed is as well. By (FL) and the guard, the
answers on which its key depends answer requests that hold whenever its own does, so, by
induction along derivations, they are confirmed before it, and its key, recomputed against
answers equal to $\alpha$, is the key of $\alpha_k(r)$; by (FC) and (I) the seeded rows are
$\alpha_k(r)$. Split $E_i$'s guarded answers into the confirmed and probed ones, $C$, and the
unconfirmed seeds, $X$. A seed whose request holds in $M_i$ is confirmed or stale, and the round
drops no stale answer, so no request of $X$ holds in $M_i =
M(C \cup X \cup F)$, and Lemma~\ref{lem:guard} gives $M_i \approx M(C \cup F)$. Every guarded
request of that model holds in $M_i$, is not pending and is not answered in $X$, so $C$ answers
it: $C$ is closed over $F$, and $M_i \approx M_F$ by the generalized least-closure result.
\end{proof}

\paragraph{Optional unguarded plugins}
For the unguarded kinds, define $F_0 = \emptyset$ and $F_{j+1} = F_j \cup \alpha(Q_j)$, where
$Q_j$ is the set of requests of unguarded kinds that hold in $M_{F_j}$ and are not answered in
$F_j$; let $m$ be the first $j$ with $Q_j = \emptyset$. The sequence is defined by $A$, $U$ and
$\alpha$ alone, and stops by (P2). It is a schedule, not a fixpoint characterization. Consider
two unguarded requests with $\mathit{req}(a) \leftarrow \neg\,\mathit{found}(b)$ and
$\mathit{req}(b) \leftarrow \neg\,\mathit{found}(a)$, where each answer is $\mathit{found}$.
Both $\{\alpha(a)\}$ and $\{\alpha(b)\}$ are sets whose model asks for nothing else, and
neither is contained in the other, so there is no least one; the schedule answers both at
once and ends with $F_1 = \{\alpha(a), \alpha(b)\}$, in whose model neither request holds.
This illustrates the generic API's limit, not shipped package selection, which reads fixed
available metadata and uses no unguarded probe.

Answering unguarded kinds before guarded closure also loses schedule independence.
For example, a constant request $\mathit{req}_g$ of a guarded kind and
$\mathit{req}_n\leftarrow\mathit{req}_g,\neg\mathit{res}_g$ of an unguarded kind satisfy the
guard on the negative read of $\mathit{res}_g$. Before $g$ is answered, $n$ is pending;
afterwards it is absent. If $n$ was answered early, its unguarded result remains observable;
if $g$ was answered first, $n$ never runs. Waiting for guarded closure chooses the latter
consistently, including when $g$ is a confirmed seed.

\begin{proof}[Proof of Corollary~\ref{cor:seeding} (cached and fresh configures agree)]
\emph{Checks and progress.} A fresh run contains only correct fixed answers, so (CK) makes
its checks pass and Theorem~\ref{thm:termination} completes it.
Tentative seeds need not be correct: a seeded run may fail a check despite (CK) for correct
answer subsets. The driver then reruns unseeded, obtaining the fresh outcome. It remains to
consider runs whose checks pass. A round that drops a stale seed probes nothing. In every
other probing round, confirmation makes the active seeds equal to $\alpha$ by (FC), (FL)
and (I); the unconfirmed seeds have absent requests and are inert. Requests therefore lie
in the finite universes of the theorem. A round that defers unguarded requests answers a
guarded one, so a run makes only finitely many probing rounds and seed-removal rounds.
\emph{Fresh and seeded runs.} Neither seeds an unguarded answer, so a run's unguarded answers
change only in rounds that answer unguarded requests, and in such a round no guarded request
is pending. By induction on these rounds, the $j$th of them starts with unguarded answers $F_j$:
by Lemma~\ref{lem:phase} its model is $M_{F_j}$, so its pending unguarded requests are $Q_j$, it
answers all of them, by (A) and (FC) with $\alpha$, and its unguarded answers become
$F_{j+1}$. The last round has nothing pending, so by Lemma~\ref{lem:phase} its model is
$M_{F_m}$. The removal of unconfirmed seeds at the end changes the model only up to
$\approx$, by Lemma~\ref{lem:guard}. Probe order and the snapshot do not enter, as in
Theorem~\ref{thm:determinism}.
\emph{Replays.} Induct over completed records for each fixed composed program.
The invariant is stronger than outcome agreement: a record contains the canonical
unguarded answer set $F_m$, correct guarded answers $C$ closed over $F_m$, and the
answers to every prerequisite used to obtain or confirm those retained answers.
It need not contain the fresh run's transient guarded answers.
Fresh and guarded-seeded runs establish this invariant by the phase argument above.
Pruning removes only never-confirmed seeds, not an answer used by a probing or confirming
round. Thus the record retains every guarded phase certificate needed for its unguarded
schedule.

Replay keeps the recorded program, non-answer input and all answers, and validates every
key and watch against the retained answer snapshot. Induct on the acyclic prerequisite
order from (FL): each recorded prerequisite remains present, its answer is correct by
the induction hypothesis, and (FC) makes the equal-key cached answer equal to the current
$\alpha$. This argument includes withdrawn answers of both guarded and unguarded kinds;
it uses their historical request certificates, not a request in the replay model.
The recorded phase certificates therefore reproduce the canonical package schedule
$F_0,F_1,\ldots,F_m$ under the current oracle.
A completing replay has nothing pending, so its correct guarded answers $C$ are closed
over $F_m$. Its model is $M(C\cup F_m)\approx M_{F_m}$ and its record preserves the invariant.
In particular, withdrawn unguarded answers are retained as part of $F_m$, not discarded
by the inert-answer lemma, which applies only to guarded kinds.
A replay that deviates is abandoned before it calls a probe or writes a record. The seeded run
that replaces it keeps no confirmation and no unguarded answer of the replay, and starts from
the record's guarded answers alone, so it is a seeded run as above.
\emph{Without (CK).} A run with seeds that fails a check, whether a seeded run or one that
replaces a replay, is replaced by a run without seeds, and only the retry's check failures
are reported. This says nothing about outcome equivalence for cached runs that never
trigger fallback, or about which fresh intermediate check they would reach.
\end{proof}

A completed replay reconstructs the recorded final input, not necessarily the fresh final
input. The record invariant supplies the missing link: canonical unguarded answers and
correct guarded closure imply agreement off probe results. The same argument applies
to records written by guarded-seeded runs and, by induction, to replayed records.

\begin{proof}[Proof of Proposition~\ref{prop:preparation} (bounded preparation)]
Fix the finite provider graph, including metadata includes, manifest requirements and rule
pointers. Its reachability operator $T(L)=L\cup\mathit{links}(L)$ is inflationary and
monotone because its edges are fixed. Every nonterminal iteration adds a vertex, so at most
$|V_{\mathrm{graph}}|-|L_0|$ strict growth steps are possible. By induction every iterate is
contained in every link-closed superset of $L_0$; the terminal iterate is therefore the least
one. These are \texttt{Preparation.terminates}, \texttt{closed} and \texttt{least} in Lean.
Complete enumeration and the relation between the graph and actual filesystem observations
remain provider/implementation obligations. The driver rejects conflicting executable code
and never adds a vertex in response to an inference request.
\end{proof}

With every environmental kind guarded, \texttt{Stratified.fixed\_union\_closure} specializes
least guarded closure to $F=\emptyset$; \texttt{CachePhase.fixed\_union\_agreement} derives
the same canonical model from a completed confirmation certificate. These do not verify
Python certificate extraction or module-guard insertion. Available declarations may affect
conditions; dormancy suppresses effects, not availability itself.

The seeding tests compare seeded and unseeded configures of the cross-compilation example after
narrowing and widening its configurations and after invalidating one recorded answer, and find
the derived facts, the final input facts and the emitted graphs identical. Further tests compare
a seeded and a fresh configure after a dependency appears, after one disappears and after a
package pin, prefix or file changes, under shuffled probe orders, and test that a check failure
in a seeded run falls back, that a round with a stale answer probes nothing, that a replay that
meets a pending request reruns seeded, that inactive answers are pruned from final provenance,
and that uncertified generated roots or files fail without recomposition. Additional tests
cover dormant modules, strict unused-code conflicts, metadata-only instances, relative path
spellings and late sysroots with negative lookup invalidation. An unchanged warm configure
can validate its certificate and skip inference altogether. (A), (FC), (FL) and (I) are obligations on probe authors,
like (P3).

\subsubsection{Module-guard rewriting}
\label{sec:formal-supp:module-guards}

The loader makes the availability/binding/use separation explicit in the rules.
For example, Qt's authored clause
\texttt{source\_claimed(t,f) :- qt\_input(t,f,"rcc",\_).} becomes
\begin{center}
\begin{minipage}{0.72\linewidth}
\begin{lstlisting}[language=Datalog]
source_claimed(t,f) :-
  module_used("QtCore-id",_), qt_input(t,f,"rcc",_).
\end{lstlisting}
\end{minipage}
\end{center}
Here \texttt{"QtCore-id"} abbreviates the stable manifest identity; private qualification
is omitted. Declaration clauses instead receive the corresponding \texttt{module\_bound}
guard. Helpers inherit their consumers' stratum, and every head, fact payload and disjunct
is guarded. Thus availability does not activate a package, and binding does not wait on
its actions.

Typed tool references in the core derive the tool configuration used by the fribidi
example of \cref{sec:design:loop}:
\begin{lstlisting}[language=Datalog]
tool_dep(c, g) :- target_action(t, c, a), action_runs(a, r), tool_resolves(t, r, g).
needed(g, e)   :- tool_dep(c, g), exec_config(c, e).
\end{lstlisting}

\subsection{Consequences of the Stratification}
\label{sec:formal-supp:consequences}

\begin{definition}[Well-formed composition]
\label{def:wellformed}
A composed program $P$ is \emph{well-formed} if
(W1) $P$, with the generated $\mathit{pending}$ rules, is stratified;
(W2) every probe kind's domain is finite in the sense of (P2);
(W3) every probe kind whose answers are seeded beyond full replay is guarded;
(W4) no request, domain or discovery-time check relation depends on a completion-gated
rule head or on $\mathit{complete}$;
(W5) the decision relations are answer-free; and
(W6) each pair $(k, \mathit{id})$ has one argument tuple.
The loader generates $\mathit{pending}(r)\leftarrow\mathit{req}_k(r),
\neg\mathit{answered}_k(\mathit{id}(r))$ for each kind.
\end{definition}
The consolidated enforcement and obligation table is \cref{tab:conditions}.

\begin{proof}[Proof of Proposition~\ref{prop:options} (req{}uests are decided before discovery)]
In a stratified program, the extension of a relation in $M_P(D)$ is determined by the facts of
$D$ in the relations it reaches in the dependency graph, stratum by stratum. A decision relation
reaches no probe-result relation, so its extension in $M_P(A, U, E_i)$ equals its extension in
$M_P(A, U, \emptyset)$ for every $i$, fixing $V$ throughout. This uses W5 for the full $P$.
For the authored core, its
clauses for the decision relations read author relations (requests, request conditions, declared
and conditional kinds, declared configurations), user relations (option values, the selected
configurations) and two derived relations: an option's effective value, defined from the user's
value and the author's default, and the evaluator of request and kind conditions, which
descends into the subterms of conditions written in $A$ and reads only effective option values.
This evaluator is separate from the per-configuration one that general conditions use, and a
request or kind condition that tests anything other than an option, such as whether a target is
defined, is an error when the specification is loaded. None of these relations reaches a
probe-result relation.
\end{proof}

Plugins and rule packages may add clauses to $\mathit{request}$ and $\mathit{request\_all}$,
which are open. After the engine checks the guarded union's types and stratification,
the loader checks authored clauses without their implicit module guards for the W5 dependency
condition: an edge runs from a rule's head to
every relation in its body, positive, negated, or inside an aggregate, and a path from a
decision relation to a probe-result relation is rejected with the chain of rules that forms it.
The needed closure, the pending-request relation and the end of discovery all reach probe
results, so a clause that makes a request depend on any of them is rejected, even when it reads
it positively, which the engine alone accepts. This source-level check is not a certificate
of W5 for the guarded union. The core, the C/C++ plugin, the Halide and Qt
packages and every case study pass the check, and the 16 reference projects' graphs are
byte-identical with and without it.

Preparation fixes $P$ before inference, including every available executable package.
Module guards prevent unused packages from adding requests merely through availability.
They can make package-derived requests depend on use, which can depend on answers.
Such effects are outside the answer-free guarantee unless the full dependency graph
independently satisfies W5; merely passing the loader does not establish that premise.
Package-derived requests are included in these decision relations. A separate audit
preprocesses each complete executable union, including scoped fact-payload bridges,
generated source declarations and I/O rules, and applies the loader's checker with all
declared relations and probe results. All 13 audited study/SDK configurations pass,
including both Halide scopes; Qt was not included in that audit. These are configuration-specific
certificates, not a loader guarantee. The composition example
(\cref{sec:eval-supp:composition}) deliberately includes an accepted exception:
its use-gated self-request introduces the path
$\mathit{request\_all}\to\mathit{module\_used}\to\mathit{build\_platform}
\to\mathit{host\_platform}\to\mathit{res\_host\_platform}$.
The predicate-level check fails even while the package is dormant; this is not a
demonstrated wrong build. Audit inputs, diagnostics and reproduction commands are in
\texttt{notes/data/review-followup}.
A core error rule reports a request for a target whose option gates all fail (an answer-free
rule, so (W5) is unaffected).
A second rule reports a target that is needed, not requested, while its gates all fail; it
reads needed targets, which may depend on answers, but derives only an error, never a request.
Under that full W5 premise, target kinds and explicit requests remain answer-free throughout
the environmental rounds. Authored option values do not change when module use changes.

\paragraph{Full statement of Proposition~\ref{prop:errors}}
Let a fixed $P$ be well-formed, let $P'$ remove $\mathit{complete}$ from error-propagation
rules, let both satisfy (F), and let probes satisfy (P0) and (P3). Under (W4), including
domains and every discovery-time check relation, discovery over $P$ and $P'$ makes the same
requests, answers and driver-check outcome in every round. If discovery completes with
answers $E_n$, then $M_P(A,U,E_n)$ and $M_{P'}(A,U,E_n)$ have the same errors and attributions.

\begin{proof}[Proof of Proposition~\ref{prop:errors} (errors do not steer discovery)]
$P$ and $P'$ differ only in the propagating rules. By (W4), neither requests nor domains nor
any discovery-time check relation reaches a changed rule head or $\mathit{complete}$.
Their reachable rules are identical, so their extensions and all relation-based checks
agree in each round. The loader checks the expanded AST's dependency reachability through positive, negative
and aggregate reads, protecting every completion-gated rule head conservatively, not just
the core's error heads. The protected roots include every manifest request and domain.
Primary-answer and
declared-result checks read returned Python rows, not additional model relations.
By induction on rounds, the two discoveries compute the same $R_i$
and, by (P3), the same $E_{i+1}$; they stop in the same round $n$ with the same answers $E_n$
(and fail the same check, if one fails). The stated (F) for both programs and (P0) ensure
finite evaluation and returning calls; (W2) ensures finite progress.
If discovery completes, in $M_P(A, U, E_n)$ no request is pending, so
$\mathit{complete}$ holds and the gated rules fire as the ungated ones of $P'$ do; the two models
agree on every error relation, and on the targets and actions the errors are attributed to.
\end{proof}

\paragraph{Rule inspection for Invariant~\ref{prop:exec} (tool configurations)}
The relevant rules in \texttt{L1\_targets.dl} and \texttt{L2\_toolchains.dl} are:
\begin{verbatim}
exec_config([p,s],[h,e]) :- base_config([p,s]),
  host_platform(h), exec_settings(s,e).
exec_config(e,e) :- exec_config(_,e).
tool_ref(t,c,a,r,u) :- action_run(a,_,_,r,u), target_action(t,c,a).
tool_ref(t,c,a,r,u) :- action_run_item(a,_,_,r,u), target_action(t,c,a).
tool_dep(t,c,g) :- tool_ref(t,c,_,r,_), tool_resolves(t,r,g).
tool_dep(t,c,g) :- needed(t,c), target_toolchain(t,tc),
  toolchain_tool_target(tc,_,g).
needed(g,e) :- tool_dep(_,c,g), exec_config(c,e).
\end{verbatim}
The settings rules implement the per-key tool policy. Resolving an action's tool reference
therefore supplies exactly its tool configuration to the last rule. Conversely, that last
rule requires a \texttt{tool\_dep} premise, whose only two sources in the core are the
action references and built-toolchain association printed above. This is minimality of
the \emph{direct tool-induced edge}, not ``needed if and only if run'': an explicit request,
a test or an ordinary dependency can independently need the same target, and the built
toolchain rule demands every associated tool before individual actions are known.
The latter can include a tool no emitted action actually runs. The property tests in
\texttt{tests/test\_core\_invariants\_prop.py} check both directions of the scoped invariant.

\paragraph{Staging and raw actions}
Negation and aggregation read complete lower-stratum extensions; Qt's finalization therefore
sees every lower fact. Adding facts back from above is unstratifiable and rejected.
Input checks read relations. The \texttt{-Werror} check reads words of user-selected programs'
actions, not inside words; other rule-action flags remain the author's obligation.

\subsection{Finite Evaluation for the Shipped Rules}
\label{sec:formal-supp:complexity}

Polynomial data complexity for fixed-program Datalog~\citep{vardi1982complexity} does not
bound constructed terms: $n$ recursive $\texttt{cat}(s,s)$ applications yield length $2^n$,
and $h$ two-valued template parameters yield $2^h$ expansions. Each round's fixed stratified
program has polynomial data complexity only without such construction; \cref{sec:evaluation:cost}
measures cost. The argument below establishes only (F), not an arbitrary-package bound.
(F) is not checked, and term size has no run-time limit; it is an obligation on whoever writes
rules. The loader's gated-recursion check rejects some unbounded
constructions, but the argument below does not rely on it. A \emph{constructed} term is a
symbol, record or number that a string functor, constructor, arithmetic, \texttt{to\_number} or
an aggregate produces. A constructed term inside a recursive component needs variables bound
outside it, a strictly progressing column, or the core's template expansion to pass that check.

\begin{lemma}[Shipped rules have finite evaluation]
\label{lem:shipped-finite}
The core, the C/C++ plugin, and the Halide and Qt rule packages satisfy (F), and hence the
finiteness hypothesis of the standard model fact (\S\ref{prop:model}).
\end{lemma}

The lemma does not cover a project's own rules (zlib-ng's, for instance), nor the Emscripten,
NDK and Python plugins, which we have not audited.

\begin{proof}
Fix a finite set of answers. Souffl\'e's numbers have a fixed width, so only symbols and records
can grow without bound. A non-recursive constructing rule builds finitely many terms from
finitely many body tuples, so along the fixed stratum order it keeps every relation finite. It
remains to bound the components in which a constructed value flows back, through a dependency
cycle, into a column of the same component. A column-level scan of the shipped rules
(\texttt{scripts/construct\_sites.py}) over-approximates these cycles. Each is finite:
\begin{itemize}
\item \emph{Ordered joins} (the core's, and Halide's feature string) build one suffix per rank
  of a group. The recursion is gated on the group being well-formed: a tie in rank, an item with
  two ranks, an unranked item, or two separators is reported as an error and derives no suffix.
\item \emph{Template expansion} (two copies in the core) starts only for a row whose
  placeholders are all closed, declared and bound for the instance, which a walk over the row's
  positions checks. It carries an injective, length-prefixed encoding of the values chosen so
  far, so every partial expansion is a prefix of one of finitely many complete choices over the
  finitely many bound values. A template that is itself an instance, directly or along a chain,
  does not expand, so an expanded value is never expanded again.
\item \emph{Configurations} have one name per suffix of a declared settings list. A setting key
  with two tool-configuration policies or values is gated, so the tool map is a partial
  function; it is idempotent (also checked), so it adds at most one configuration per declared
  configuration and build-machine platform.
\item \emph{Second linkage instances} append a fixed suffix along a chain bounded by the
  linkage names that facts state.
\item \emph{Generated-file paths} join a directory and a relative path; the cycle is between
  relations only, since paths into a package's data feed action arguments and inputs, never
  the properties they are read from.
\end{itemize}
Every gate that a recursion negates is evaluated in a stratum below it, except the template
walk, which walks positions, not choices. Each ill-formed input is still reported as an error.
The remaining components build lists of bounded length or count up to a bound stated in the rule.
\end{proof}

\subsection{Built and Imported Artifacts}
\label{sec:formal-supp:export}

An export is derived at the exporter, and an import at the consumer from the files the
exporter wrote; the two share no facts. The loss of \cref{sec:formal:export} is visible only at
the exporter. A foreign format $f$ has a \emph{spelling table}, stated by its plugin, that spells
each key as a field or declares it unspelled with a reason. For an exported component $m$, built
as $t$ in configuration $c$, let $\mathit{own}(t, c)$ be the properties $t$ gives its consumers
itself, excluding build-internal keys, and $\mathit{req}(t, c)$ its exported requirements on
sibling components or on imported packages by name and minimum version. Let $\mathit{loc}$ map
build- and source-tree paths to paths relative to the install prefix, and $\mathit{usage}(t,
c)$ be the usage facts a consumer derives from a dependency on $t$ in the build, through $t$'s
own properties and transitively through its requirements.

\begin{lemma}[What an export states]
\label{lem:export-exporter}
In $M_n$, for every exported component $m$, configuration $c$ in which it is exported, and active
format $f$, at least one of the following holds for each fact $(k, v) \in \mathit{own}(t, c)$,
and exactly one if $M_n$ derives no error:
the export spells it in $f$, with a path value mapped by $\mathit{loc}$; or $f$ declares $k$
unspelled, and the exporter derives a loss fact for $(m, k, f)$ and a configure warning; or the
exporter derives an error. Every requirement in $\mathit{req}(t, c)$ is stated, or is an error
when it names a built target outside the package or an imported target whose package has no
name. In each foreign format, shipped rules $R$ are reported as a loss, with a
warning.
\end{lemma}
Exclusivity needs the error-free premise: a key both spelled and declared unspelled derives
an error, but the spelling and loss facts still coexist
in that erroneous model.

The downstream-use remark has five conditions, for $t$ and the components it requires.
\begin{description}
\item[(E1) No loss.] In $f$, no loss fact is derived; in \sys's format this holds by
  construction.
\item[(E2) Inversion.] $\mathit{parse}_f(\mathit{spell}_f(u)) = u$ for every fact $u$ whose key
  $f$ spells, with paths resolved relative to the file the importer reads.
\item[(E3) Resolution.] Every requirement resolves at the consumer to the export of the same
  component or imported package for the same configuration.
\item[(E4) Isolation.] The consumer's own properties, other dependencies and user facts are
  unchanged.
\item[(E5) Rule neutrality.] At the consumer, $R$ derives no usage fact about the package's
  targets.
\end{description}

\begin{exportremark}[Set-level downstream use; Remark~\ref{prop:export} in full]
\label{lem:export-downstream}
(i)~\emph{Facts.} For an error-free export under (E1)--(E5), a consumer that imports the export of $t$ in format $f$ for
configuration $c$ derives, from its dependency on $t$, the set of usage facts
$\mathit{loc}(\mathit{usage}(t, c))$. Without (E1), under (E2), the property rows the importer
reads for each exported component are its spelled rows, that is, $\mathit{loc}(\mathit{own}(t,
c))$ less the facts with lost keys, and the consumer derives no loss fact; what the consumer
derives from those rows is not characterized. (ii)~\emph{Rules.} In \sys's format the
consumer's program contains $R$, as in-tree; a foreign format carries none of it.
\end{exportremark}

Ownership discharges (E4) for shipped rules, except their actions. (E5) is an assumption that
nothing checks: ownership lets a package's rules add usage facts to its own targets.
Hand-written packages (LLVM, Qt) have no built counterpart and are outside the remark. The remark
equates \emph{sets}, so order and multiplicity are outside it. (E2) and (E3) are tested, not
proved (\cref{tab:formats} and \sys's own format below).

\begin{proof}[Proof of Lemma~\ref{lem:export-exporter} (what an export states)]
By the rules of the export layer: a property is spelled when the table spells its key (and, for
keys with enumerated values, its value); the loss fact is derived from an exported fact, an
active format and the table's unspelled entry, and the warning from the loss fact; a key that is
neither spelled nor declared unspelled, a value with no spelling, a rooted path under no install
location, and an exported edge to a non-exported built target are each an error. A static
library's link-only requirements, and native kinds other than static and shared libraries, are
errors as well. The loss fact for shipped rules is derived from an exported rule package and an
active foreign format (\texttt{export\_rules\_lossy}).
\end{proof}

\paragraph{Why the conditional set-level observation holds}
By induction on the requirement graph of the export. For a component, by
Lemma~\ref{lem:export-exporter}, the export being error-free, and (E1), each fact of
$\mathit{own}(t, c)$ is spelled (no loss alone leaves the error alternative open), and by
(E2) the importer turns it back into $\mathit{loc}$ of that fact. By (E3) each requirement
resolves to an export for which the induction hypothesis holds. The consumer's propagation rules
are shared by built and imported targets, so, with (E4), the facts it derives from the
dependency are those of the database it reads, and by (E5) the shipped rules add no usage fact
to them. Without (E1), (E2) still applies to every spelled row, and lost rows are absent; how the
consumer's rules combine the remaining rows is not characterized. Part (ii) holds by construction: \sys's format installs $R$ unchanged, and the loader loads
it as it would in-tree; a foreign format carries no rules.
Two configurations that would install different exports to one place are a configure error;
version acceptance and absolute paths outside the package (kept, not relocatable) are outside
the remark, and metadata the CMake importer cannot read is an error.

\begin{table}[t]
\caption{Round trips tested in the submission. ``Facts'': the downstream's derived usage facts
are compared, as sets, with those it derives in-tree. ``Consumer'': a foreign consumer builds and
runs against the install.}
\label{tab:formats}
\small
\begin{tabular}{@{}p{0.27\linewidth}p{0.17\linewidth}p{0.48\linewidth}@{}}
\toprule
Package & Format & Result \\
\midrule
nlohmann/json, two variants & CPS, CMake & facts equal, also after moving the prefix \\
nlohmann/json & pkg-config & facts equal except the language standard, a loss the exporter
  derives \\
nlohmann/json, CMake's install & CMake export & facts equal (CMake's \texttt{.pc} also lacks the
  definition) \\
two-component fixture & CPS, CMake & facts equal, including an inter-component requirement \\
zlib, shared and static & CMake package & consumer: upstream's \texttt{find\_package} test builds
  and passes; a request for the next major version is refused \\
Halide, CMake's install & CMake export & the import states every fact that a hand-written binding
  states \\
three fixtures (14, 10 and 8 facts) & \sys & the importer's rows equal the exported facts as
  multisets up to the prefix, also after copying the install \\
DSL fixture & \sys & its shipped rule builds and runs a downstream; read through CPS alone
  (a CPS-only search) it does not configure \\
\bottomrule
\end{tabular}
\end{table}

The CMake importer reads \texttt{FRAMEWORK} targets and the generator expressions that
installed export files use; metadata it cannot read (object libraries, unknown compile features,
other generator expressions) is an error of that target, not a silent drop. It reads Qt's own
\texttt{Qt6*Targets.cmake}, which the Qt study's packages list in their manifests.

\paragraph{\sys's own format}
An install writes \texttt{<prefix>/share/buildlog/<Pkg>/}: a manifest, a facts file
\texttt{<Pkg>.facts.dl} of ground facts in a \texttt{pkg\_*} vocabulary (packages, components,
locations, requirements and properties by \sys's own keys, with paths relative to the prefix,
which is found from the file's location), and the project's rule layers if its specification
names them (\texttt{export\_rules}). The facts are derived from the same export relations as the
CPS, pkg-config and CMake files, so the format has no spelling table and nothing unspelled. The
file is Soufflé input as it stands: a test runs Soufflé on it with a generated header of
declarations and gets the same tuples back. The round trip of \cref{tab:formats} compares tuples
as multisets; it excludes absolute paths outside the package (\texttt{pkg\_prop\_abs}, stated
but not relocated) and metadata (compatibility, description, required versions), which the CPS
importer skips too. It is a tested property, not a proof. The format has no conditions: a package
installed from several configurations must state the same facts in each.

Preparation enumerates the manifest, its facts, imports and executable rules within the declared
roots. The rules enter the checked union before environmental inference; available facts are
bound and used under the module guards. The final lock pins explicit packages and finally used
modules by content, while the observation certificate also covers dormant code and data.
A relocated package has the same content hash; an edited locked package is an error until the
user accepts it. A decision relation reading a probe answer is a W5
error, and the visibility and layering checks apply as to any package. \texttt{why} on an
imported fact ends at the line of the installed facts file; it does not continue into the
upstream project's derivation, which the exporter does not keep.

A manifest may also list nearby export files of other formats (\texttt{imports}), which are
handed to their importers, so an upstream whose install has only CMake exports can ship a
hand-written manifest and rules beside them; the files must stay inside the prefix, and the lock
covers finally used modules. Preparation enumerates and decodes these imports without probe
answers; their includes and rule pointers are part of its finite closure.
Binding and module use select their effects within the checked stratified union.
Preparation enumerates every candidate in every declared search directory; two candidates for one name are an error
until the user pins one, never a first match, and \texttt{package\_file} names an exact manifest,
the interface by which a resolver hands \sys its choice. When one install writes both a CPS and a
\sys export, the CPS file names its sibling manifest (\texttt{x\_buildlog\_manifest}, a
prefix-relative path). The CPS importer verifies the marker and states a format-neutral row that
the two files are one package, so the core sees one candidate, and a plain prefix search binds
the \sys route, facts and shipped rules. This is identity, not precedence: no format outranks
another. A third party's unmarked CPS file beside a manifest is ambiguous until pinned, and a
dangling or mismatched marker is an error, not a fallback. (A CPS file can still name a rule
package, \texttt{x\_buildlog\_rules}, for hand-written or third-party exports.)

Halide's LLVM, Clang and LLD and Qt's CMake packages are found this way: each has one
hand-written manifest that imports the package's own export file
(\texttt{LLVMExports.cmake}, \texttt{ClangTargets.cmake}, \texttt{LLDTargets.cmake},
\texttt{Qt6*Targets.cmake}). Package-level facts, such as LLVM's version, definitions and RTTI
setting, are published through the open \texttt{package\_version} and \texttt{package\_attr}, and
LLVM's 31-line rule layer derives the C++ ABI properties of every target the import defines;
Clang and LLD require the LLVM version they were built with. The loader rejects a package whose
rules assert about a target outside the namespaces its own imports define
(\texttt{import\_ns\_target}), and a property the export and the rules state differently is an
error, not resolved by precedence. For macOS's LLVM~22 the three packages are 132 lines, 37 of
them Datalog, and no Python.

\paragraph{Provenance}
An inner node of a proof tree is a rule instance; its positive premises are its children and
are explained in turn. Its negated premises, built-in constraints and aggregate results are
leaves, so a proof does not show why a negated fact is absent or which tuples an aggregate
counted; \texttt{why-not} explains an absent fact one rule at a time. The other leaves are
input facts, with their origins (a file and line, or the probe request that produced it), and
fact clauses of layers. An
installed fact's proof stops at its installed facts file. A tree deeper than Souffl\'e's limit
is cut and marked as truncated.

\section{A Survey of Meta Build Systems}
\label{sec:survey}

\Cref{tab:survey} describes twelve build and packaging systems along eleven axes, from how they
discover the environment~(A) to how closely they resemble a logic program~(K). It is
descriptive: each cell summarizes the system's documentation, and the artifact gives the
source of every cell. Every system with usage requirements or providers computes a monotone
closure of the kind \cref{sec:background:closure} shows (column~B), although none describes
itself as a logic program. The columns also show how the systems differ around the closure:
most interleave discovery with declaration~(A), several resolve choices by an order the user
does not see~(G), and few model compatibility~(F).
\begin{table}[p]
\centering
\rotatebox{90}{\begin{minipage}{0.97\textheight}
\caption{Twelve build and packaging systems along eleven axes. Each cell summarizes the
system's documentation; the sources for every cell are in the artifact. Twenty cells were
checked against primary documentation. Cells marked~\unv{} could not
be confirmed against primary documentation. Spack's column~F follows \citet{gamblin2022asp}.}
\Description{A table with one row per system (CMake, GN, Meson, build2, Bazel, Buck2, Pants v2,
SCons, Autotools, Cargo, Nix/Guix, Spack) and one column per axis of the survey.}
\label{tab:survey}
\fontsize{7}{8.2}\selectfont
\setlength{\tabcolsep}{2.5pt}
\renewcommand{\arraystretch}{1.0}
\begin{tabularx}{\linewidth}{@{}>{\bfseries}l*{11}{>{\RaggedRight\arraybackslash}X}@{}}
\toprule
& A.~Discovery & B.~Propagation & C.~Imported vs.\ built & D.~Toolchains, configs
& E.~Codegen to library & F.~Compati\-bility & G.~Hidden choices & H.~Meta-level
incrementality & I.~Extension & J.~Backend & K.~Resemblance to LP \\
\midrule
CMake & interleaved, cached & closure of \texttt{INTERFACE\_*} & \texttt{IMPORTED}, same surface
& one toolchain per tree & custom command, wrapped by hand & \texttt{COMPATIBLE\_\allowbreak
INTERFACE\_*} & package search order & full re-configure & modules, toolchain files & Make,
Ninja, IDEs & closure only \\
GN & none (\texttt{exec\_script}) & public configs, second phase & no imported kind & many
toolchains & \texttt{action} & none found & none & full regeneration & \texttt{template} & Ninja
& closure only \\
Meson & one non-Turing-complete pass & dependency objects\unv & inter\-change\-able & one config per
directory; three machines & \texttt{generator} & none & pkg-config, CMake, frameworks, in order
& regeneration rule & built-in modules & Ninja, IDEs & closest to a fact base \\
build2 & \texttt{configure} meta-operation & metadata protocol & \texttt{sys:} packages, same
surface & many toolchains & \texttt{in} module, recipes & avoided\unv & three-tier import search
& persistent configuration & modules & own executor & named closure \\
Bazel & phased\unv & \texttt{CcInfo}, depsets & \texttt{cc\_import} gives \texttt{CcInfo} &
platforms, exec/target, transitions & rules, \texttt{genrule} & constraint matching &
first-match toolchains & Skyframe & Starlark & own, remote & closure, ordered search \\
Buck2 & one DICE graph & transitive sets & same providers\unv & per-node exec platform & prelude
rules & ALL/ANY\unv & most refined \texttt{select} & DICE & Starlark, BXL & own, remote-first &
finer than Bazel \\
Pants v2 & memoized rules & typed parameters, inference & distinct types\unv &
environments\unv & sandboxed & no validation & ambiguity is an error & memoized rule graph &
plugins & local, Docker, remote & static Horn-like, no backtracking \\
SCons & read all, then build & none (environments) & no distinction & one environment per
config & custom builders & none found & first-found tools & dependency tracking & programmable
& own executor & mostly imperative \\
Autotools & shell script & none & no distinction & one build/\allowbreak host/\allowbreak target
per run & none & checks only & \texttt{PATH} order & cache off by default & m4 macros & Make & no
fact base \\
Cargo & \texttt{build.rs} & feature union & n/a & one target; proc-macros on host\unv &
\texttt{build.rs} & SemVer & backtracking order & lockfile & crates & \texttt{rustc} &
backtracking search \\
Nix/Guix & one lazy evaluation & splicing & no distinct kind & many platforms & derivations &
user code & none in engine & none; flake cache\unv & overlays & store builder & pure functional
\\
Spack & facts, then solve & fixpoint, optimization & external or built, one spec & solved
jointly & n/a & NP-complete & solver objectives & reused concretizations &
\texttt{.lp} files, recipes & own installer & answer-set program \\
\bottomrule
\end{tabularx}
\end{minipage}}
\end{table}

\section{Further Evaluation Data}
\label{sec:eval-supp}
\sloppy

This section gives the evaluation data that \cref{sec:evaluation} summarizes: the portfolio, the spec-size
accounting and table, the categories of difference per study, the full account of libHalide's differences, which
comparisons keep order and what an order-keeping comparison finds, further capabilities (a built
toolchain, C++ modules and CUDA), the pins needed on Linux, the sizes of \texttt{why}'s answers,
and configure API measurements. The machine, versions and method are those of
\cref{sec:evaluation:method}.

\subsection{The Portfolio}
\label{sec:eval-supp:portfolio}

\Cref{tab:studies} lists the case studies with their pinned releases, their baselines and the
capability each isolates.

\begin{table}[ht]
  \caption{The case-study portfolio. Every study compares against the upstream project's own
  build at the pinned release, whose sources are a pinned git submodule.}
  \label{tab:studies}
  \small
  \begin{tabular}{@{}l l l p{0.44\linewidth}@{}}
    \toprule
    Project & Release & Baseline & Capability isolated \\
    \midrule
    zlib & 1.3.2 & CMake & configure checks, configured headers, install and export \\
    Halide & pinned main & CMake & a DSL whose generators run during the build; configurations \\
    nlohmann/json & 3.12.0 & CMake & header-only library; install/export round trip \\
    flatbuffers & 25.12.19 & CMake & a code generator that must run on the build machine \\
    flatbuffers (slice) & 25.12.19 & Bazel 8.8.1 & tool configuration against \texttt{cfg = "exec"} \\
    fribidi & 1.0.17 & Meson 1.12.1 & \texttt{native: true} derived; a project-owned config header \\
    zlib-ng & 2.3.3 & CMake & per-source ISA flags; universal2 from one configure \\
    ggml & 0.25.3 & CMake & per-ISA variants as run-time-loaded modules \\
    SPIRV-Tools & 2026.3 & CMake & code generated from an imported package's data \\
    nanobind\_example & pinned commit & scikit-build-core & Python extension ABI, abi3, wheels \\
    Qt consumer & Qt 6.11.3 & CMake & rules a library ships for its consumers (\texttt{moc}, \texttt{rcc}) \\
    \bottomrule
  \end{tabular}
\end{table}

\subsection{Spec Size}
\label{sec:eval-supp:size}

Line counts compare unlike things unless one decides what each side is billed for. We decide by
asking whether comparable tools ship a piece of logic in their core. Logic that CMake, Meson or
Bazel provide built in (path manipulation, a module that an executable loads without linking it,
joining flags, argument-vector templates, one configured-file formatter) lives in \sys's core and
is not billed to projects. Project-specific conventions are billed: test conventions, ISA tables
and score weights, a generator's argument vector. A port ports intent, not another tool's
features; fribidi's \texttt{config.h}, written by Meson without a template, becomes a template
that the project owns and pays for. Rule packages are reusable library code, like a CMake module
or a \texttt{.bzl} file: they are reported separately and not counted in project size. The same
rules apply to the baselines, whose helper modules and \texttt{.bzl} files are counted as library
code where upstream ships them as such. We report per study, not a pooled ratio.
Sizes are counted by a script in the artifact: non-blank, non-comment lines of every TOML,
Datalog and template file of the project, skipping lines that hold nothing but source paths
(argument words, flags and names count) and skipping presets, which are user input.

\Cref{tab:size} gives the project-billed size of each study's specification. Upstream
builds usually cover more (more platforms, install, packaging, more
tests); where only part of an upstream build is in scope we count that part, or give the
study's estimate, marked $\sim$. For zlib-ng and ggml a script (\texttt{scripts/upstream\_lines.py})
counts the in-scope part by rule: from the files that hold the ported logic it drops the
branches, helper functions and commands of the platforms, backends and features the port leaves
out, with the same line measure, which reproduces flatbuffers' 481. Some 20 to 30 lines it keeps
may still be out of scope.
\begin{table}[t]
  \caption{Project-billed specification size in lines, under the accounting policy of
  \cref{sec:eval-supp:size}. ``DL'' is the Datalog part of ``Lines''; rule packages are
  library code, reported separately.}
  \label{tab:size}
  \small
  \begin{tabular}{@{}l r r r l p{0.34\linewidth}@{}}
    \toprule
    Study & Lines & DL & Rules & Upstream & Scope of the upstream count \\
    \midrule
    nlohmann/json & 37  & 0   & --  & -- & not counted \\
    zlib          & 88  & 2   & --  & -- & not counted \\
    flatbuffers   & 217 & 10  & --  & CMake 481 & also options, shared library, install, packaging \\
    flatbuffers/Bazel & 49 & 1 & -- & 19 & use site; the \texttt{.bzl} rule path is $\sim$130 \\
    fribidi       & 301 & 47  & --  & Meson 263 & \texttt{config.h} template billed to the port \\
    zlib-ng       & 523 & 91 & --  & CMake 1,223 & in scope, of 2,241 in the same five files; upstream has 5 more architectures, MSVC, install, fuzzers \\
    ggml          & 501 & 115 & --  & CMake 644 & in scope, of 1,728 in the same five files; ISA tables upstream keeps in C are billed here \\
    SPIRV-Tools   & 220 & 8   & --  & CMake $\sim$590 & also install, other libraries, fuzzers, all tests \\
    nanobind      & 17  & 0   & 99  & 16 & CMake 12 + \texttt{pyproject} 4; the rule package matches nanobind's CMake module \\
    Qt consumer   & 26  & 0   & 116 & CMake 13 & one configuration there, four here (15 with one); the packages are Qt's \\
    libHalide & 2,293 & 502 & 690 & -- & not counted; the rule package serves the apps \\
    \bottomrule
  \end{tabular}
\end{table}
Where the upstream count covers a comparable scope
(SPIRV-Tools by the study's estimate, ggml and zlib-ng by count), the specification is shorter.
fribidi's is longer than its Meson build, largely because the configured header's template,
which Meson generates from a dictionary, is billed to the project. The Bazel slice's 49 lines
compare with 19 at Bazel's use site plus about 130 of rule library, and nanobind's 17 with
upstream's 16. What remains in Datalog is project abstraction (zlib-ng's source sets and
directory-wide properties, ggml's variant scoring); architecture and extension knowledge is
library code in the C++ plugin (about 220 lines), not billed to the projects.
For the whole-build comparison in \cref{sec:cs:first},
\texttt{scripts/spec\_size.py} counts all top-level TOML and Datalog files in
\texttt{libhalide} and the reusable generator rules, including Python and tutorial rules.
It additionally removes fact rows that name only a source file with a derived name and
constant columns, which CMake leaves implicit. The totals are 1,573 TOML lines, 502
project-rule lines and 567 reusable-rule lines, against CMake's 3,797 project lines
and 1,564 helper lines. The table uses the source-list policy above without this
fact-row normalization. Fresh per-file counts and source hashes for both measures are in
\texttt{notes/data/spec-size-refresh}.
Like for like, the CMake generator-helper lines that
implement what \sys's generator rules model number 495, against 292 lines of those rules; the
count of modeled lines includes 317 lines of scaffolding we did not assign to one feature, which
favors CMake.

\subsection{Fidelity in Detail}
\label{sec:eval-supp:fidelity}

\begin{table}[ht]
\caption{Recorded action pairing; the full Halide row uses the current native validation.
Scope counts ported upstream targets of all targets, not a denominator for unmatched actions.
Paired units use their worst difference: spelling, warnings/includes, or potentially
behavior-changing flags (distinct differences in parentheses). ``Out.'' counts changed
outputs in baseline-flag replay; n.r.\ means not rebuilt.}
\label{tab:action-fidelity}
\footnotesize
\setlength{\tabcolsep}{2.5pt}
\begin{tabular*}{\textwidth}{@{\extracolsep{\fill}}>{\RaggedRight}p{3.7cm}rrrrrrrr@{}}
\toprule
& & \multicolumn{5}{c}{Units in both builds} & \multicolumn{2}{c}{In one only} \\
\cmidrule(lr){3-7}\cmidrule(lr){8-9}
Study (baseline) & scope & paired & spell. & warn./incl. & flag & out. & base & \sys \\
\midrule
\multicolumn{9}{@{}l}{\emph{Compiles, paired by target and source}} \\
zlib (CMake) & 6/6 & 34 & 0 & 4 & 30 (1) & 0 & 0 & 0 \\
flatbuffers (CMake) & 6/6 & 85 & 0 & 0 & 85 (2) & 3 & 0 & 0 \\
fribidi (Meson) & 15/15 & 42 & 0 & 0 & 42 (2) & 0 & 0 & 0 \\
zlib-ng (CMake), arm64 & 10/11 & 90 & 90 & 0 & 0 & 0 & 0 & 0 \\
\quad x86\_64 & 10/10 & 110 & 110 & 0 & 0 & 0 & 0 & 0 \\
ggml (CMake), CPU & 5/42 & 28 & 0 & 0 & 28 (2) & 0 & 38\textsuperscript{a} & 0 \\
\quad arm64 & 9/12 & 60 & 0 & 0 & 60 (2) & 0 & 4\textsuperscript{a} & 1\textsuperscript{b} \\
\quad x86\_64 & 31/34 & 236 & 0 & 0 & 236 (2) & 0 & 4\textsuperscript{a} & 1\textsuperscript{b} \\
nanobind (scikit-build-core) & 2/2 & 12 & 0 & 0 & 12 (3) & 0 & 0 & 0 \\
\multicolumn{9}{@{}l}{\emph{Actions, paired by output (libHalide: release)}} \\
libHalide (CMake) & full & 4,748 & 2,814\textsuperscript{c} & 538 & 1,396 (17) & n.r. & 4,221\textsuperscript{d} & 86\textsuperscript{e} \\
\multicolumn{9}{@{}l}{\emph{Actions, paired by configuration role, type, and source or output name}} \\
flatbuffers/Bazel 8.8.1, cross & 17/49 & 49 & 0 & 0 & 49 (23) & n.r. & 13\textsuperscript{f} & 0 \\
\bottomrule
\end{tabular*}
\smallskip
\begin{minipage}{\textwidth}\footnotesize\RaggedRight
\textsuperscript{a}Objects of targets not ported.
\textsuperscript{b}An example upstream skips with run-time backends.
\textsuperscript{c}870 classified as identical.
\textsuperscript{d}4,203 \texttt{phony} edges and 18 non-phony actions; 331
install/housekeeping edges excluded.
\textsuperscript{e}75 configure-time export lists/stubs; remaining classes described below.
\textsuperscript{f}Archives: Bazel's per library, \sys's per configuration.
Full Halide includes Python bindings and apps; full-build output impact is unmeasured.
\end{minipage}
\end{table}

\paragraph{Categories of difference}
A script (\texttt{scripts/fidelity\_taxonomy.py}) reads each study's normalized records, pairs
them by unit as \cref{sec:evaluation:method} describes, and classifies every token in which a
pair differs. Spelling is what normalization removes: the compiler's path (Xcode's shim against
the resolved \texttt{clang}), \texttt{-arch} against \texttt{-{}-target} with a sysroot, paths in
the build tree, dependency-file flags, the order of flags and, outside libHalide, of link inputs,
and a tool spelled \texttt{buildlog -E} against \texttt{cmake -E}. Every other difference has a
category (\cref{tab:taxonomy}). A pair is \emph{semantic} if it has a difference of a category other than
diagnostics, include and include class; the three are inert as long as both builds read the same
headers, which the shadow check below tests. Semantic is a property of the flag's category, not a
finding that code differs: most semantic flags reach no code, as the list after the table shows.
\Cref{tab:action-fidelity} calls semantic pairs \emph{flag} pairs.
\begin{table}[ht]
  \caption{Paired units by category of difference. Spell.: identical or differing in spelling only.
  A pair can have several categories. L/I/O: link flags, inputs, other. Sem.: pairs with a semantic
  category; inert: pairs whose only differences are diagnostics and include categories.}
  \label{tab:taxonomy}
  \small\setlength{\tabcolsep}{3pt}
  \begin{tabular}{@{}l r r r r r r c r r r r r@{}}
    \toprule
    Study & Paired & Spell. & Lang. & Def. & Opt. & Codegen & L/I/O & Diag. & Incl. & Class & Sem. & Inert \\
    \midrule
    zlib & 34 & 0 & -- & 30 & -- & -- & -- & -- & -- & 4 & 30 & 4 \\
    flatbuffers & 85 & 0 & 85 & 58 & -- & -- & -- & -- & 43 & 15 & 85 & 0 \\
    fribidi & 42 & 0 & -- & -- & 42 & 18 & -- & 42 & 12 & 11 & 42 & 0 \\
    zlib-ng arm64 & 90 & 90 & -- & -- & -- & -- & -- & -- & -- & -- & 0 & 0 \\
    zlib-ng x86\_64 & 110 & 110 & -- & -- & -- & -- & -- & -- & -- & -- & 0 & 0 \\
    ggml CPU & 28 & 0 & 22 & 26 & -- & -- & -- & 6 & -- & 17 & 28 & 0 \\
    ggml arm64 & 60 & 0 & 48 & 56 & -- & -- & -- & 12 & -- & 46 & 60 & 0 \\
    ggml x86\_64 & 236 & 0 & 191 & 221 & -- & -- & -- & 45 & -- & 211 & 236 & 0 \\
    nanobind & 12 & 0 & 12 & 1 & -- & -- & --/11/-- & -- & -- & 12 & 12 & 0 \\
    Bazel slice & 49 & 0 & -- & 44 & 44 & 46 & 2/5/46 & 44 & 45 & -- & 49 & 0 \\
    \bottomrule
  \end{tabular}
\end{table}
Each semantic category, by study:
\begin{itemize}
\item \emph{zlib.} CMake defines \texttt{NO\_FSEEKO} on every compile, from the generator-expression
  bug of \cref{sec:background:departures}; no compiled file reads it.
\item \emph{flatbuffers.} The project asks for C++11. CMake keeps Apple clang's default level, C++14, because it satisfies
  the request, and spells it \texttt{-std=c++14}; \sys passes \texttt{-std=c++11}. The project
  builds and its tests pass as C++11.
  CMake also passes \texttt{FLATBUFFERS\_TEST\_PATH\_PREFIX}, which only the tests read, to 58
  compiles through a directory-wide \texttt{add\_definitions}, and an empty
  \texttt{\$BUILD/samples} include directory to 43.
\item \emph{fribidi.} \sys adds \texttt{-DNDEBUG}, which no fribidi file reads, and
  \texttt{-fPIC} to the library's objects, which is the default on Darwin. Meson adds
  \texttt{-Wall -Winvalid-pch -fdiagnostics-color} and each target's source directory.
\item \emph{ggml and nanobind.} CMake passes \texttt{-std=gnu++17} where the project asks for
  C++17, and \sys passes \texttt{-std=c++17}; CMake defines \textit{target}\texttt{\_EXPORTS}
  on shared libraries' and modules' objects, which no source reads. ggml's CMake adds
  \texttt{-Werror=implicit-function-declaration -Werror=implicit-int} to C compiles. nanobind's
  11 inputs are its library's sources, which CMake compiles in place in the interpreter's
  \texttt{site-packages}; \sys compiles only sources under the project or the build tree, so its
  rule package copies them there first.
\item \emph{Bazel slice.} Bazel's default macOS toolchain adds \texttt{-O2 -g0} (where the
  configuration asks for \texttt{-O3}), \texttt{-fstack-protector}, \texttt{\_FORTIFY\_SOURCE},
  redacted \texttt{\_\_DATE\_\_}-like macros, \texttt{-fdebug-prefix-map} and
  \texttt{-frandom-seed} for reproducibility, its own warning set, and, on the sample's target
  compiles, \texttt{DEBUG} without \texttt{-g}, where \sys passes \texttt{-g}. These change code; they are Bazel's toolchain's choices, which
  the study does not replicate.
\item \emph{libHalide.} The 438 compiles of the runtime's embedded bitcode are C++ files that
  hold one byte array each. CMake compiles them with only its configuration's flags and
  \texttt{-std=c++17}; \sys gives them libHalide's flags and C++20, as the library they belong
  to. \sys models CMake's \texttt{-O3 -DNDEBUG} on link lines and Halide's
  \texttt{wipe\_standard\_flags} (which strips them from the tests) as two properties of the
  Halide spec. The current full-build comparison below covers the test compiles without
  CMake's precompiled header; libHalide's rpath names the LLVM it was found
  with where CMake names \texttt{lld}'s.
\end{itemize}

\paragraph{What differs in the full build}
The full build's libHalide graph (release; the Python bindings, tutorials, apps and tests included) pairs
4,748 of CMake's actions with \sys's, by output and, where the directory layouts differ, by
target, action type and file name: 1,002 links, 2,236 compiles, 299 archives, 1,195 generator runs
and custom commands, 15 copies and a symbolic link. 2,814 differ at most in spelling, 538 only in
warnings or include directories, and 1,396 in a flag that can matter, 17 distinct differences. The classes: 730 test compiles
that CMake gives a precompiled header; the 438 byte-array wrappers described above; 110 compiles of
shared and module libraries that CMake gives \texttt{-D}\emph{target}\texttt{\_EXPORTS}
(34 of them \texttt{-flto}, on the Python runtime and module); 75 links of tutorials, Python modules and tests that differ in
\texttt{-headerpad\_max\_install\_names}, the rpath or \texttt{-search\_paths\_first}; 7
links where \sys also links a library's own runtime archive; libHalide's rpath; and 35 custom commands,
26 of which start a Python generator through \texttt{env}, 8 list the generator's target features
in another order, and one asks for one output fewer. 4,203 of CMake's 4,221 unpaired actions are \texttt{phony}
edges; the other 18 are tools and fixtures outside the port, tutorial generators that \sys
names differently, the precompiled header and install copies. \sys has 86 actions CMake lacks, 75
of them export lists and stubs that CMake writes at configure time.
The full-build comparison is on macOS. The scoped replay below measures wrapper and PCH
effects, not the output impact of every full-build command difference.

\paragraph{Native configuration sharing}
The native Python preset makes tools inherit \texttt{python.interpreter}, so generators
and products share one Halide instance. Previously the tool policy reset that setting to
\texttt{none}, duplicating Python-independent C++ work. The emitted graph shrinks from
6,857 to 4,874 edges: 1,983 repeated steps disappear, including 661 C++ compilations.
The single configuration also resolves path-matching ambiguities for four tutorial-generator
compiles; the paired total rises by four, all warning/include-only.
This is an explicit native policy, not automatic per-target setting projection.
The refreshed graph comparison is in \texttt{notes/data/halide-inherit}; CMake's unchanged
graph is retained from the preceding native refresh.

\paragraph{Tutorial correction and native refresh}
The pinned upstream build applies \texttt{test/}'s flag wiping and hidden visibility only
to tests, not to \texttt{tutorial/}. The port incorrectly reused \texttt{test\_defaults}
for ordinary tutorials, lesson~10's runner, tutorial generators and generated-library
runners. Removing those four dependencies restores \texttt{-O3 -DNDEBUG} in Release
compile/link commands and removes hidden-visibility flags. The rules explicitly retain
upstream's C++17 minimum, which lesson~10's runtime-only consumer requires.
Four command regressions cover these properties in the four rule families.
All 27 tutorial compiles in the current pairing
fall in the warning/include-only class. Tutorial links retain ordinary linker/rpath
differences. This refresh does not measure flag effects on performance or code size,
nor repeat the Linux execution run. The tutorial-correction evidence is in
\texttt{notes/data/author-revision}; the subsequent configuration-sharing refresh is in
\texttt{notes/data/halide-inherit}.

For the non-Halide, non-Bazel studies, baseline-flag replay rebuilds each semantic
compile pair with \sys's command and the baseline's semantic flags, at the same output
path. A repeated original command checks determinism. The changed-output counts are in
\cref{tab:action-fidelity}; they apply only to the pairs replayed.

\paragraph{Fresh current-graph replay}
We replay all 438 wrapper and 730 PCH-related
compile pairs from the completed single-composition native graph. Each current
command runs twice, then once with the semantic intervention, at the same isolated
output path. All 3,504 invocations succeed and all original repeats agree.
All 438 wrappers and 728 PCH-related compiles give identical objects; two do not.
For PCH, we rebuild CMake's exact generated header under each current command's
compiler, target and options (51 contexts), then add it to that command. This isolates
the header/PCH choice, not every upstream flag or reuse of a serialized PCH from another build.
The two differing tests, \texttt{for\_each\_element} and \texttt{halide\_buffer}, include
\texttt{HalideBuffer.h}, not \texttt{Halide.h}; adding the latter through the PCH
introduces static-initialization sections. Their text sections grow from 33,668 to
40,584 bytes and from 241,040 to 247,952 bytes, respectively; isolated reruns reproduce
all three object hashes in each case. We do not infer runtime equivalence or a
performance effect. Complete commands, hashes, compiler identities and diagnostic
objects are retained in \texttt{notes/data/halide-current-replay}.

\paragraph{Excluded Bazel actions}
In flatbuffers' Bazel slice, 254 housekeeping actions (symbolic links for include paths, runfiles
trees, build information) and 4 actions that build Bazel's toolchain wrappers have no Ninja
counterpart and are excluded from the comparison. In an optimized build \sys shares one tool
instance with the target configuration, 7 actions fewer than Bazel, which always separates the two.

\paragraph{Which comparisons keep order, and why the others may drop it}
Order matters on a command line in three places: the include search path, where the first
directory with a header wins; the link line, where a static library is searched only for
symbols already undefined; and a repeated definition, where the last \texttt{-D} wins. The
libHalide comparison keeps all three: compiles compare flags as a multiset but the include
path in order, and links, archives and generator runs are compared token by token in order;
its self-test checks that a reordered link is reported. zlib's comparison checks the include
directories in order. The other comparisons treat flags as sets and sort link inputs, for two
reasons that hold in these studies. First, the order inside one include class differs by
design, as above, so an ordered comparison would report every compile without finding anything;
what makes this safe is that both builds read the same file. The taxonomy script checks it: for
every header that two include directories of one compile command hold, it finds the file each
build reads, from the command's directories and the quoted-include rule. In every compile study
both builds read the same file, except in flatbuffers' 27 test compiles, where 8 generated
headers exist both in the build tree and checked in, \sys reads the generated copies and CMake
the checked-in ones, and each pair of copies is byte-identical. The Bazel slice and libHalide,
which compares include order, are not checked this way. \sys also enables clang's
\texttt{-Wshadow-header} where the compiler accepts it, which the Apple clang of these runs does
not. Second, every macOS link in these studies is done by Apple's linker, which keeps searching a
static library until no new symbol resolves (its manual page: ``there is no need to specify a
static library multiple times''), so the order of libraries does not decide which members are
loaded. \sys derives the order of libraries on a link line from the dependency graph
(\cref{sec:design:core}), which is what a single-pass GNU linker needs, and every study links
there on Linux; those that run tests pass them (\cref{sec:case-studies}). Set comparison would
miss a macro defined twice with different values; the studies' functional checks would catch
one that changes behavior, but a difference in performance or code size that no check observes
would go unnoticed. The taxonomy script checks this case too: in the three studies where a flag
is repeated with different values (\texttt{-O3 -O2} in zlib-ng's \sys build, \texttt{-DNDEBUG
-UNDEBUG} in ggml's two builds, \texttt{-O3 -Os} in nanobind's two), the last value is the same
in both builds.

\paragraph{An order-keeping comparison}
\label{sec:eval-supp:order}
\Cref{tab:taxonomy} folds order into spelling. A second script
(\texttt{scripts/fidelity\_order.py}) keeps order and classifies each paired command as
identical in order, different only in order (the same tokens as a multiset), different in content
only, or different in both (\cref{tab:order}). It then tests each difference in order where order
can matter: include directories that both hold a header the compile reads, a definition given
conflicting values, and static archives on a link line, or an archive's position among objects.
An include directory in a build tree is listed in that build's own tree, and two headers of one
name in different places count as the same if their bytes are equal.
Object files and dynamic libraries are left out, since Apple's linker resolves those in a
two-level namespace whatever their order.
\begin{table}[ht]
  \caption{Paired commands under an order-keeping comparison. Order: the same tokens in a
  different order. Both: order and content differ. Candidates: differences in order that could
  change the result.}
  \label{tab:order}
  \small\setlength{\tabcolsep}{3pt}
  \begin{tabular}{@{}l l r r r r r r@{}}
    \toprule
    Study & Commands & Paired & Same & Order & Content & Both & Candidates \\
    \midrule
    zlib & compile/link & 34/6 & 0/0 & 0/1 & 0/2 & 34/3 & 0 \\
    flatbuffers & compile/link & 85/6 & 0/0 & 0/1 & 0/1 & 85/4 & 0 \\
    fribidi & compile/link & 42/15 & 0/0 & 0/0 & 0/7 & 42/8 & 0 \\
    zlib-ng arm64 & compile/link & 90/10 & 0/0 & 90/1 & 0/3 & 0/6 & 0 \\
    zlib-ng x86\_64 & compile/link & 110/10 & 0/0 & 110/1 & 0/3 & 0/6 & 0 \\
    ggml CPU & compile/link & 28/5 & 0/0 & 0/0 & 0/0 & 28/5 & 0 \\
    ggml arm64 & compile/link & 60/3 & 0/0 & 0/0 & 0/0 & 60/3 & 0 \\
    ggml x86\_64 & compile/link & 236/3 & 0/0 & 0/0 & 0/0 & 236/3 & 0 \\
    nanobind & compile/link & 12/2 & 0/0 & 0/0 & 0/0 & 12/2 & 0 \\
    Bazel slice & compile & 49 & 0 & 0 & 49 & 0 & 0 \\
    \bottomrule
  \end{tabular}
\end{table}
In these comparisons no difference in order is a candidate. zlib-ng's compiles differ only in the
order of independent flag groups, and the other compile studies' differences in order interleave
independent groups beside the content differences of \cref{tab:taxonomy}. Every include order
is decided. In flatbuffers' 27 test compiles the source tree and the build tree both hold 8
generated headers, in different orders, so the two builds read different copies; the copies are
byte-identical. ggml's build directories share no header name with its source directories.
\sys orders a link line by its dependency edges, a library before what it needs, with ties broken
lexically. A test asserts
this order transitively on emitted link lines, and another links and runs a project with a
chain, a diamond and a runtime-style archive under GNU~ld on Linux. These are therefore
checks of the emitted order under a single-pass linker, separate from command-pair comparisons.
A third test builds \texttt{A} $\to$ \texttt{B} $\to$ \texttt{C} with \texttt{B} and \texttt{C} each
static, shared or header-only and private, public or interface edges (12 cases). In
all 12 the link line and whether \texttt{C}'s include directory and definition reach \texttt{A}
follow CMake's usage requirements: a static library passes on its private dependencies' archives,
a shared library hides them, and only public and interface edges pass on compile requirements.
Each case also links and runs under GNU~ld on Linux.

\paragraph{zlib-ng and zlib}
zlib-ng's pairs differ in spelling only and zlib's do not, though both are CMake builds of the
same kind of library. zlib's CMake build adds a definition through a generator-expression bug
and moves \texttt{zconf.h}'s directory between include classes; zlib-ng's adds nothing to a
compile beyond what \sys's specification states, and both builds repeat no flag differently.
The study's own comparison, which drops include flags and \texttt{-O3}, reports no difference
either; the uniform script above keeps them and agrees.

\paragraph{Halide's full build}
\sys builds libHalide and runs 941 of the 950 CTest tests that upstream's build defines at the
pinned commit, across the correctness, performance, generator, warning, fuzz and runtime tests,
the four autoschedulers' tests (Adams2019, Li2018, Mullapudi2016, Anderson2021), the C++ and
Python tutorials, the Python bindings and the apps.
With native tools inheriting the Python setting, there are 941 executions of 941 distinct
tests; the lesson-10 AOT generator no longer runs twice. The completed native build
gives 848 passes, 93 skips and no failures (arm64 macOS, Release, Homebrew's
LLVM~22, Python~3.10.19 with NumPy~2.2.6, imageio~2.38.0 and pybind11~3.1.0).
Skips cover unavailable GPU, Hexagon and WebAssembly runtimes and the Anderson2021
GPU demos. Nine distinct tests are not
modeled. Six drive a nested build inside \texttt{ctest} (lesson~15's generator build and file
check, and four Python lessons); in \sys their outputs are ordinary targets of the one graph. Two
drive a debugger, outside the port's Release execution scope.
One, \texttt{filter\_dll\_exports}, runs a CMake script that filters DLL exports,
which only MSVC builds use.
The command comparison in \cref{tab:action-fidelity} covers the whole native build,
not just the cross-build validation sample.

\paragraph{The cross build of libHalide}
We cross-built libHalide, its runtime and the 32-test sample (one test per build
feature) from arm64 macOS to Linux on AArch64, against a sysroot exported from the Linux image of
\cref{sec:eval-supp:linux}, and ran the tests in that container. CMake needed two build trees and
two configure invocations, with a partial build of the macOS tree (its exported generators,
three build-time tools and the autoschedulers) before the Linux tree could build. Beyond the
Linux build's flags, the Linux tree needed three standard settings (a toolchain file, the
sysroot, compile commands) and workarounds: re-rooting the bundled dependencies' search
path; turning off the Python bindings and the tutorials; and three hacks, a script registered as \texttt{CMAKE\_CROSSCOMPILING\_EMULATOR} that runs each
build-time program's macOS twin (prefixed to 1,303 edges), and shadow packages for flatbuffers
and for the imported generators.
\sys builds the same set in one tree from one configure: the generators and host tools for macOS
and libHalide and the tests for Linux, in one graph. The tests give 30 passes and 2 skips, as
under CMake, and the build needs no workarounds. The
sysroot is handled by rules, not by settings: the core relation \texttt{platform\_sysroot} names
it, an imported file under it keeps its root, the CMake importer maps a run path under the
sysroot to its path on the target (\texttt{/usr/lib/llvm-22/lib}, as in CMake's build), and a
\texttt{.pc} file's absolute paths under a sysroot are prefixed with it, as pkg-config's
\texttt{PKG\_CONFIG\_SYSROOT\_DIR} does. The user's preset pins the host C++ compiler, since Apple
clang and Homebrew's clang both build for macOS; without the pin the configure is an error that
lists both. \sys's x86\_64 and universal2 configurations of libHalide are only configured,
because Homebrew's LLVM is arm64-only, and CMake's builds of them fail at link.

\subsection{Halide's Cross-Built Apps}
\label{sec:eval-supp:apps}

Halide's HelloWasm and HelloAndroid apps (\cref{sec:cs:first}) are compared with CMake builds
of the same pinned sources by reducing each Ninja graph to records: compiles (target, source,
flags), generator runs (name, Halide target string, parameters), runtime runs, archives and links,
with machine-specific macOS architecture and SDK flags dropped (\cref{tab:apps}).

\begin{table}[t]
\caption{Halide's cross-built apps against CMake. ``Same'': records whose key matches; HelloAndroid's
keys collapse its two ABIs.}
\label{tab:apps}
\small
\begin{tabular}{@{}lrrrrrr@{}}
\toprule
& \multicolumn{3}{c}{HelloWasm} & \multicolumn{3}{c}{HelloAndroid} \\
\cmidrule(lr){2-4}\cmidrule(l){5-7}
Records & CMake & \sys & same & CMake & \sys & same \\
\midrule
compiles & 9 & 9 & 9 & 8 & 10 & 4 \\
generator runs & 15 & 15 & 15, identical & 4 & 4 & 4, identical \\
runtimes & 15 & 15 & 15, identical & 4 & 4 & 4, identical \\
archives & 30 & 30 & 30 & 8 & 8 & 4, identical \\
links & 6 & 6 & 6 & 2 & 4 & 1 \\
\midrule
spec lines (+ project, preset) & \multicolumn{3}{l}{138 (+13, +13)} & \multicolumn{3}{l}{61 (+13, +7)} \\
Ninja steps & \multicolumn{3}{l}{75} & \multicolumn{3}{l}{30} \\
\bottomrule
\end{tabular}
\end{table}

\paragraph{HelloWasm} Upstream's super-build builds the generator project with the host compiler
through \texttt{ExternalProject}, configures the app under Emscripten's toolchain file, and
unrolls four \texttt{wasm-32-wasmrt} variants (SIMD and threads crossed) into separate targets,
each with its own generator runs and link. In \sys the variants are configurations, where
CMake, with one toolchain per build tree, unrolls them into targets: one \texttt{index} target and its three Halide libraries, each
stated once, are built in four wasm32-emscripten configurations that differ in two settings,
\texttt{emscripten.simd} and \texttt{emscripten.threads}, beside the host configuration that
builds and runs the generator, all in one graph. The specification has 138 lines (96 non-blank,
non-comment). It states the app's own facts; the Emscripten knowledge is in the plugin: the link
gets \texttt{-O3} from the release optimization setting and \texttt{-matomics} from the threads
setting, SDL2 is a package the plugin provides (\texttt{emscripten::sdl2}, giving
\texttt{-{}-use-port=sdl2}), and the plugin states each configuration's CPU features in an open
relation of the C++ package, which the Halide rule package maps to Halide's
\texttt{wasm\_simd128} and \texttt{wasm\_threads}; it reads no Emscripten setting. Nothing is
added to the compile lines. The Halide rule package mirrors CMake's \texttt{add\_halide\_generator}
and \texttt{add\_halide\_library}, with CMake's defaults for the generator and function names
and generator parameters keyed by condition. All four modules pass \texttt{wasm-validate}, and
the generator project's three native benchmark tests pass. Compiles differ from CMake's only in
include spelling: generated and Halide headers come through \texttt{-isystem} rather than
\texttt{-I}, from one directory per library rather than per variant, since libraries are named
per configuration rather than per variant. Other recorded deviations are on host lines or outside
the graph: libpng and libjpeg come from pkg-config rather than CMake's find modules, optimization
flags are not repeated on host links, force-loaded archives and the macOS platform are spelled
differently, \texttt{RunGenMain.cpp} is a listed source because the CMake importer rejects
Halide's \texttt{INTERFACE\_SOURCES}, and the static web assets CMake copies are not modeled.
Both builds use binaryen~123, since emsdk~3.1.73's bundled version rejects what LLVM~22 emits.
The demo draws to a browser canvas. A script (\texttt{scripts/hellowasm\_check.py}) instead runs
each of the four modules under Node.js for 64 frames of the app's main loop, with a stub page whose
canvas has no WebGL, so SDL draws with its software renderer. It compares the last frame with a
native C++ program (\texttt{scripts/hellowasm\_reference.cpp}) that drives the same three
pipelines, built for the build machine by the same graph, as the main loop does. All four modules
run the 64 frames and their last frames are byte-identical; against the native frame, 236 of
1,048,576 pixels differ, each by one unit in one channel. The demo is not run in a browser, and
HelloAndroid is not run on a device.

\paragraph{HelloAndroid} Upstream's Gradle build runs its CMake build per ABI (arm64-v8a and
x86\_64, minimum API 23), which configures itself again for the host at configure time to build
the two generators. Without Gradle, the baseline configures upstream's
native \texttt{CMakeLists.txt} directly with NDK~30's toolchain file. \sys builds the library in
one configuration per ABI, and the host generators, in one graph, from a 61-line specification;
the Android system libraries it links are packages the NDK plugin provides (\texttt{ndk::log},
\texttt{ndk::android}, \texttt{ndk::atomic}). Both libraries are ELF shared objects of the right
machine, need only \texttt{libandroid}, \texttt{liblog}, \texttt{libm}, \texttt{libdl} and
\texttt{libc}, and export all eight JNI entry points. Of the extra records, two compiles and two
links are the generators, which CMake builds outside the ABI's graph; the other compiles and
links differ in the triple's spelling (\texttt{aarch64-linux-android23} against the NDK
toolchain file's \texttt{aarch64-none-linux-android23}) and in flags CMake's target model adds
(an export definition, \texttt{-fPIC}, \texttt{-pthread}). Gradle, APK packaging and running on a
device are out of scope.

\paragraph{What was added} The toolchain knowledge is in two plugins standing in for
what the NDK and Emscripten would ship, as \texttt{android.toolchain.cmake} and
\texttt{Emscripten.cmake} do for CMake. Counting non-blank lines outside comments, the NDK plugin
has 43 lines of Datalog and 104 of Python, the Emscripten plugin 51 and 73; they are not counted
as specification. The core gained three open relations (an executable suffix, a link side file
and a toolchain's link defaults), the C++ package generic cross-build hooks, among them the open
CPU-feature relation, and the Halide rule package the WebAssembly and Android targets and the
CMake-like library forms. CPU microarchitecture knowledge in general, such as the inclusion of
one level's features in another's, as Spack's archspec records it, belongs in a generic layer
between the core and the C++ package rather than in specifications; we leave it to future work.

\subsection{Further Studies in Detail}
\label{sec:eval-supp:further}

The paper (\cref{sec:cs:zlibng}) gives each of these studies in one paragraph; this is the
full account.

\paragraph{zlib-ng}
zlib-ng compiles each architecture-specific source with its own instruction-set flags and
dispatches at run time. Its CMake build examines one architecture per configure, so a universal2
build of 2.3.3 silently builds no SIMD code in either slice and compiles an ARM-only source into
the x86\_64 slice (zlib-ng\#1016). In \sys the C++ plugin knows each architecture's extensions and their spellings, the project
states which each source group needs, and an incompatible placement is an error. From one configure each universal2 slice gets the
same ISA sources and flags as a single-architecture build; the arm64 and x86\_64 builds match
CMake's translation unit for translation unit (90 and 110, flags equal as sets), and generated
headers, \texttt{.pc} files, archive members and dylib metadata are identical. GCC~13's spellings
match clang's except one AArch64 extension; MSVC is not covered.

\paragraph{ggml}
ggml builds its CPU backend once per ISA level as a module the loader scores and selects at run
time; upstream CMake copies one target per level. \sys uses one template instantiated per level
and a run-time dependency (\cref{sec:design:templates,sec:design:runtime}). The module sets (3 on arm64, 14 on
x86\_64), each level's flags, symbol counts and load commands match CMake's, and both builds load
the same module. Upstream accepts a native tuning combined with all variants on ARM, producing three identical
modules the loader cannot tell apart; \sys reports an error.

\paragraph{SPIRV-Tools}
SPIRV-Tools generates its tables from JSON grammars in the separately installed SPIRV-Headers,
which \sys imports and whose grammars it reads as package data. Five generators produce six files,
byte-identical to CMake's, and the shared library exports the same 79 symbols; CMake's optimizer
library has 2,656 extra RTTI symbols from a directory-scoped \texttt{-fno-rtti} quirk that
per-target properties do not reproduce. \texttt{why} traces a grammar read to the user's line naming
the package prefix, and pointing it at another release reruns exactly the four grammar-driven
generators.

\paragraph{nanobind}
scikit-build-core drives CMake once per interpreter; in \sys the interpreter is a configuration
setting, and the stable ABI is one derived fact deciding the \texttt{Py\_LIMITED\_API} definition,
the file name and the wheel tags together. For CPython~3.13 (abi3) and 3.11, the modules' undefined Python symbols and
exports equal the baseline's, the wheels hold the same six files with byte-identical metadata,
upstream's tests pass, and the abi3 module also loads in 3.12 and 3.14. The Python plugin builds no
source distributions or editable installs.

\paragraph{zlib's consumers}
Upstream's \texttt{find\_package} consumer test passes with either install, CMake reads \sys's
CPS export, and a consumer asking for the next major version is refused, as the compatible
versions derive from the ABI version; upstream's \texttt{AnyNewerVersion} file accepts it.

\subsection{Further Capabilities}
\label{sec:eval-supp:capabilities}
\label{sec:eval-supp:modules}

The experiments of this section exercise capabilities beyond the eleven studies of
\cref{sec:case-studies}: a toolchain built in the build, C++20 named modules and CUDA.

\paragraph{Independently packaged generators}
\label{sec:eval-supp:composition}
The runnable \texttt{examples/composition} project installs two independent libraries,
each exporting a runtime header and consumer-side generator rules, into separate prefixes.
An ordinary consumer dependency on each runtime activates both generators without naming
their rule packages; the resulting executable prints \texttt{42}.
A third valid rule package is composed but unused. Adding its search root leaves consumer
actions and needed targets unchanged, excluding Ninja regeneration inputs; its warning and
active lock entry remain absent. Explicit activation makes its warning, needed target and
lock entry appear. A conflicting second installation and invalid dormant executable code
both reject configuration. On macOS/Apple clang and Linux/GCC, each successful fresh
configuration composes once and starts
one evaluator. Single-run preparation/configure observations are retained, not presented
as repeated benchmarks. The example uses shell/sed generators, not a second industrial
DSL. Its self-request fails the full-program W5 audit described in
\cref{sec:formal-supp:consequences}; accepted composition does not imply answer-free requests.
Evidence and exact source identities are in \texttt{notes/data/review-followup}.

\paragraph{TinyCC: a toolchain built in the build}
\label{sec:eval-supp:tcc}
A toolchain built with \sys states its facts, as a library's package file does. TinyCC~0.9.28rc
bootstraps in one configure and one Ninja graph: Apple clang builds stage~1, stage~1 builds
stage~2 as compiler, linker and archiver, and stage~2 builds stage~3. Each compiling stage builds
its own runtime library. Stages~2 and~3 are byte-identical, as are their objects, and a program
and library built by stage~2 run. One fact per stage gives what the inventory probe finds for an
installed compiler, with targets instead of paths, and per target the stage that compiles it;
the user pins only Apple clang. A built toolchain is never a candidate in toolchain selection,
and its tool dependency (Invariant~\ref{prop:exec}) is derived. The study runs on arm64 macOS
only (Linux needs facts, not mechanism), and its reduced runtime lacks \texttt{tcc -run}.

\paragraph{C++20 named modules} A module's name is in its source, so which compile reads which
module interface is known only after the sources are read. \sys keeps this discovery out of
configure. Each source is scanned during the build (P1689R5); a build-time stratum of Datalog
rules in buildlog-cpp, run by the same engine per target, collates the scans with the exported
static facts into Ninja dyndep files, module maps and a provenance sidecar. The loader checks that
no configure rule reads the stratum, so information flows one way and the theorems of
\cref{sec:formal} are unchanged: configure derives a graph whose edges the build completes.
Each name has at most one visible provider; a collision, an import cycle or an import from a
target that is not a dependency is an error, and an import never adds a dependency or a link edge.
Consumers whose flags are compatible share one module interface per class; an incompatible
consumer gets its own. \texttt{import std} is supported. \texttt{why} on an edge from a module
interface to a compile joins the sidecar's derivation, whose leaf is a scan fact, with the
configure-time proof of the provider's visibility. {fmt}~11.2.0, whose \texttt{fmt.cc} is one
interface unit, builds and runs as \texttt{import fmt} with LLVM~22's clang; the specification
names no module, since the collator learns the name from the scan. Apple clang, which ships no
scanner, is an error, not a fallback.

\paragraph{CUDA} The \texttt{buildlog-cuda} plugin adds nvcc on Linux and changes
nothing in the core or buildlog-cpp. stdgpu~1.3.0, in NVIDIA's CUDA~13.4 arm64 image, was compared
with CMake's build: all 17 compiles (five library sources, three C++ and nine CUDA example
sources) are identical as sets of definitions, include directories, flags and
\texttt{-{}-generate-code} specifications. The archive has the same inputs, and every one of the
12 programs links the same objects and libraries; CMake's lines add CMake's implicit device
runtime, a second static runtime and an absolute \texttt{librt.a}, and order the system
libraries differently. The programs are not run, as no NVIDIA GPU was available. The host compiler
is the configuration's unique C++ compiler or a user pin; two candidates without a pin are an
error that lists them. \texttt{cuda.arch} names a set of architectures, and there is no
\texttt{native} detection, which would make the configured build depend on the machine's GPU.
Setting keys are ordered by declared priority, then by name, so independently written plugins
compose.

\subsection{Provenance Answers}
\label{sec:eval-supp:why}

A proof is only useful if it is short enough to read. A raw proof tree is large, so the driver
shapes it before printing. It shares repeated subproofs as a DAG, which does most of the
reduction (a 703-node tree for one flag becomes 76 nodes). It also splices out engine-internal
nodes, hides structural atoms, collapses single-premise chains of plumbing rules and, with
\texttt{-{}-leaves}, prints only the origins.
By default \texttt{why} folds the steps
that construct a configuration's name and cites a premise already shown instead of repeating it.
On the questions the studies' authors asked, the default answers are 25 lines for why the x86\_64
build needs \texttt{flatc} built for arm64 (9 with \texttt{-{}-depth 3}), 29 for where the Bazel
slice's compiler flag comes from (59 with \texttt{-{}-with-demand}, which adds the requests that
asked for it), and 16 for why an instruction-set attribute is off in a configuration (a test
bounds it at 16).
Length is not the only criterion. Why zlib-ng compiles one source with \texttt{-mavx2} takes 52
lines, because the answer shows the chain by which AVX2 requires SSE4.2, SSE4.1, SSSE3 and SSE2,
as upstream's build does, with each link's option, check, flags, header and probe answer (29
lines), beside the source set (10), the configuration (9) and the C++ plugin's ISA knowledge
(3). A test bounds the answer at 53 lines and asserts that no unrelated
premise (another configuration, a shared library, the ARMv8 condition) appears. Each answer
ends in spec, user-file, probe or rule-package lines.

Asked, in fribidi's cross build for x86\_64, why a table generator is built for arm64,
\texttt{why -{}-leaves} lists every origin of its 31-line derivation:
\begin{lstlisting}[language={},basicstyle=\ttfamily\scriptsize,breaklines=true,literate={<-}{{$\leftarrow$}}2]
$ buildlog why -C build --leaves 'needed("gen-bidi-type-tab", @arm64-macos.release.none)' project
needed("gen-bidi-type-tab", <arm64-macos.release.none>)
  config_decl("release", "arm64-macos")  <- project/spec.toml:5
  config_decl("x86_64-release", "x86_64-macos")  <- project/presets/cross-x86_64.toml:11
  config_decl_setting("release", "native.opt", "release")  <- project/spec.toml:6
  config_decl_setting("x86_64-release", "native.opt", "release")  <- project/presets/cross-x86_64.toml:12
  config_select("x86_64-release")  <- driver: config_select: --config on the configure command line
  gen_primary("table1(bidi-type, UnicodeData.txt)", "bidi-type.tab.i")  <- project/gen.toml:93; project/gen.toml:6
  gen_word("table1(bidi-type, UnicodeData.txt)", 0, "", "tool", "gen-bidi-type-tab", "")  <- project/gen.toml:87; project/gen.toml:6
  generator_label("table1(bidi-type, UnicodeData.txt)", "GENERATE")  <- project/gen.toml:86; project/gen.toml:6
  generator_owner("table1(bidi-type, UnicodeData.txt)", "tables")  <- project/gen.toml:85; project/gen.toml:6
  request_all("libfribidi")  <- project/spec.toml:7
  res_host_platform("inventory:default", "arm64-macos")  <- probe inventory request "inventory:default"
  res_platform("inventory:default", "x86_64-macos", "x86_64-apple-macos13")  <- probe inventory request "inventory:default"
  target("gen-bidi-type-tab")  <- project/gen.toml:47
  target("tables")  <- project/gen.toml:39
  target_deps("libfribidi", "public", "tables")  <- project/spec.toml:91
\end{lstlisting}
The generator template's word naming the program (\texttt{gen.toml:87}, instantiated at line~6)
makes it a tool of \texttt{tables}, which \texttt{libfribidi} needs; the inventory's answers name
the build machine. No leaf says the program is a host tool.

Presentation also has to undo an artifact of the engine. For a premise that only checks
existence (``some toolchain targets this platform''), Soufflé records one arbitrary witness, and
printing it suggests a dependency that does not exist. \texttt{why} therefore elides such
witnesses in every view, including the focused ones that list only leaves or one path; the
cross-compilation example's leaf listing is 9 lines. The lattice join binds its value, so its
candidate is a winner, not an arbitrary witness (\cref{sec:eval-supp:why-join}).

\subsection{Linux}
\label{sec:eval-supp:linux}

The results below report scoped build and execution checks. The full native Halide
command comparison is on macOS; no Linux Halide command-pair totals are reported.

\paragraph{Platform coverage}
The Linux studies use an arm64 Ubuntu 24.04 container image under GCC~13 and Clang~18,
with the macOS specifications; each study's record lists the platform's deviations.
The source studies include zlib, nlohmann/json (through a downstream build), zlib-ng, ggml,
fribidi, flatbuffers, SPIRV-Tools, nanobind, the Qt consumer and libHalide with LLVM~22.
The full Halide build and its test limitations are reported below.
All 90 of zlib-ng's translation units get upstream's flags; fribidi's 42 objects and 36 exported
symbols match Meson's; flatbuffers' and SPIRV-Tools' generated files are byte-identical; ggml's
flags match without filtering, as, under Clang, do its eight ARM variant modules (GCC 13 cannot
build two of the levels); nanobind's 12 compiles (bar one define), \texttt{Py*} symbols and wheel
file lists match scikit-build-core's (with Python 3.11.16 for 3.11.15).
The Qt consumer uses Qt 6.11.3, in debug and release, with \texttt{moc} and
\texttt{rcc} as host tools, and passes its two tests headless. In the Bazel slice (Bazel 8.8.1)
\texttt{flatc} is built only in the tool configuration, its 37 compiles match Bazel's exec
compiles and the generated header is byte-identical to Bazel's; there is no cross variant, since
it targets x86\_64 macOS and the image has no cross toolchain for x86\_64 Linux. The Halide apps'
generator runs, runtimes and archives are identical to CMake's (HelloWasm 15, 15 and 30;
HelloAndroid 4, 4 and 4), their compiles and links have the same keys or the same deviations as
on macOS, and HelloWasm's three host benchmarks pass, as on macOS. HelloWasm's host link
has CMake's \texttt{-ldl} and also \texttt{-pthread}, which CMake omits because its threads
probe finds the threads library inside glibc's C library; \sys states the operating system's
flag rather than probing. The Android NDK's host tools exist only for x86\_64,
so they run under emulation in the arm64 image, and both ABIs link. The specifications of nlohmann/json, nanobind, the Qt consumer, both Halide apps and the Bazel
slice have no Linux-specific line. The others use conditions on the platform and the compiler
family (CMake's \texttt{\$<CXX\_COMPILER\_ID>}), a check that runs a program, and an ELF version
script beside Mach-O's exported-symbols list.
Package rules hold the rest: nanobind's state ELF's link and strip flags; an imported ELF shared
library's directory becomes an rpath, as CMake's build rpath does; Qt's OpenGL package states its
link flag per platform; and a generated Halide runtime carries the operating system's
dynamic-loading library and threads flag, as \texttt{add\_halide\_library} adds them, from
per-OS facts of the C/C++ plugin that are empty on macOS, WebAssembly and Android.

\paragraph{Execution results}
The source studies were exercised in the Ubuntu image and, where
available, natively on macOS. GCC's non-integration suite gives 1,180 passes, 121 skips and one
expected failure; Clang's suite gives the same totals. Clang's C++ module tests require its matching scanner: using the installed
Clang~22 and adjacent \texttt{clang-scan-deps} gives 12 passes and three skips.
The test fixture selects this pair rather than the image's default Clang~18 without a scanner.
Native integration checks cover available-metadata provenance, build-local regeneration
certificates, fmt modules and TinyCC bootstrap.

The default GCC and Clang case harnesses build the ordinary libraries, staged consumers,
nanobind for both Python interpreters, Qt, Bazel \texttt{flatc}, HelloWasm and both Android ABIs.
The full GCC Halide build succeeds; its 877 tests report 783 passes, 92 skips and two failures.
This harness selects the native Linux configuration without the Python-binding configuration
used in the 941-distinct-test macOS fidelity study; the totals are not the same test population.
An isolated rerun passes \texttt{mullapudi2016\_mat\_mul}.
\texttt{performance\_fast\_inverse} still fails its timing inequality, also in three executions
linked against the image's independently CMake-built Halide. This is a platform performance
assertion, not evidence of equal test outcomes on every run.

In the CUDA~13.4.2 image, stdgpu builds and links; its 17 compile commands match CMake's,
with the documented runtime-library link differences. Toolkit tests give five passes with
real nvcc and 29 with a fake toolkit (two skips). No GPU runtime is available.
Commands, outcomes and limitations are recorded in
\texttt{notes/data/single-composition/validation.json}.

\subsection{Pins}
\label{sec:eval-supp:pins}

An ambiguous choice is an error that a pin resolves, so the pins a study needs depend on the
machine. We repeated the pin experiment of \cref{sec:evaluation:fidelity} in the Linux image (\cref{sec:eval-supp:linux}),
removing each of the 34 pin rows of the Linux user files of every study but Qt's in turn. With only GCC
offered, the default there, one row is needed: libHalide's LLVM linkage. With GCC and Clang both
offered, 21 are: one toolchain pin per language a study compiles plus one for the linker, 0 to 4
per study; without its pins a study reports from 0 (nlohmann/json, which compiles nothing) to 859
(libHalide) ambiguity errors. A single user-global file of three rows (C, C++ and the linker
driver, all GCC) removes every toolchain ambiguity in these studies, so those rows are
per-machine policy, not repeated project decisions. The separate LLVM-linkage pin remains.

\subsection{End-to-End Configure Comparison}
\label{sec:eval-supp:cli-cost}

This comparison asks whether the prototype's architecture permits practical configuration
costs, not whether it outperforms mature systems.
\texttt{notes/data/author-revision/configuration} retains 660 successful samples,
commands, environment, scope checks and source hashes; the exact measured harness is
\texttt{notes/data/author-revision/configure-all.measured.py}.
The Halide row is replaced by 60 new paired samples after native interpreter inheritance,
retained separately in \texttt{notes/data/halide-inherit/configuration}, using the same
harness and method. After restoring Qt 6.11.3, a further 60 paired samples in
\texttt{notes/data/qt-refresh/configuration} supply its missing row.
The published selection thus has 720 samples: ten serial trials of eleven paired
configurations and the two SDK apps on the \sys side only. These datasets are not pooled.
Runs use the machine and native tools of \cref{sec:evaluation:method}, Python~3.14.0
and the patched Souffl\'e binary. Tool order alternates by trial, study and mode.
The harness waits before each study block for load below 3.5 and no running Docker
containers; load can rise within a block (observed range 2.26--9.17).
The replacement Halide samples range from 3.11 to 6.58.
The Qt samples range from 2.95 to 4.33.
We run no concurrent test or build jobs.

Each study has private project/source copies, build directories and \sys caches.
The timer brackets the unprofiled subprocess, including startup, compiler probes and graph
generation, but excludes copying, edits, result processing, builds and tests.
Cold empties tool state, not OS filesystem caches. Warm explicitly invokes configure
unchanged. Meson uses \texttt{setup -{}-reconfigure}; Bazel uses
\texttt{build -{}-nobuild} against its live server, with a fresh server/output root cold.
Bazel's networked dependency prefetch is untimed; timed commands have networking disabled.
Edit adds one private definition to corresponding targets (an interface definition for
header-only nlohmann/json). Emitted files verify the edit; an untimed Bazel
\texttt{aquery} checks its actual compile arguments.

Each cell of \cref{tab:single-composition} gives the arithmetic mean and relative
two-sided 95\% Student-$t$ confidence half-width, with nine degrees of freedom:
$100\,t_{0.975,9}s/(\sqrt{10}\,\bar{x})$ percent.
These intervals describe repeat variability on this machine, not uncertainty across
machines or workloads. No medians, observed ranges or API-only measurements are mixed
into these intervals.

\paragraph{Scopes}
All rows use the native release configuration (Bazel's \texttt{opt}).
zlib is shared-only. zlib-ng includes both library variants and the ported C tests;
the additional upstream GTest suite is disabled. Halide includes Python bindings,
tutorials, utils, autoschedulers and tests on both sides, with the nine port exclusions
described above. Its Python binding dependencies match the refreshed native validation.
Nanobind covers Python~3.13/abi3 and Python~3.11 in one \sys invocation; the upstream
sample sums two direct CMake configures, excluding scikit-build-core and wheel packaging.
For ggml, CMake configures additional tests/examples beyond the two ported programs.
SPIRV-Tools' port uses installed GoogleTest for four tests, while the upstream source
copy lacks its GoogleTest checkout and configures without tests.
The Qt row selects native release on both sides, using the same Qt 6.11.3 SDK;
debug, x86\_64 and universal2 validation are outside its timed scope.
Thus the rows share source pins and product/toolchain modes, not identical graphs.
The command logs and compiler inventories record these boundaries.

\paragraph{SDK apps}
HelloWasm's \sys cold/warm/edit means are
\textbf{7.11$\pm$5.4\%}, \textbf{0.89$\pm$2.3\%} and \textbf{3.76$\pm$3.4\%} seconds.
HelloAndroid's are \textbf{3.11$\pm$3.8\%}, \textbf{0.13$\pm$3.1\%} and
\textbf{0.91$\pm$3.0\%}. We do not time the CMake outer shells as comparators:
HelloWasm defers its sub-configures until build time, while HelloAndroid compiles its
host generators inside configure. Neither does the same work as \sys's graph derivation.

The prototype retains fixed startup cost and workload-dependent overhead, but configures
these builds in seconds. This is feasibility evidence, not parity, superiority or graph
equivalence. The earlier five-trial, three-project CLI dataset remains separately archived
in \texttt{notes/data/configure-comparison}; it is not pooled with these runs.

\subsection{Instrumented Configure API Measurements}
\label{sec:eval-supp:single-composition}

\texttt{notes/data/single-composition} retains the raw records, source hashes and medians.
We report its 45 implementation samples: five repetitions of three projects in three modes.
Runs use Python~3.14.0, Souffl\'e \texttt{12912a9dd}, the native tools of
\cref{sec:evaluation:method} and common source dependencies.
Each repetition visits trivial, zlib and Halide; each project gets a private copy,
build directory and initially empty composition/probe cache. Cold updates only that
copy's lock; warm repeats unchanged; edit adds a real compile definition.
These Halide API runs omit the Python preset and are unaffected by its native
configuration-sharing override.
No tests or container builds run concurrently; observed one-minute load is 3.4--5.3.

The timer brackets \texttt{configure\_project}, excluding interpreter startup and result
capture. Every mode succeeds. Consequently these are not whole-process command-line times
and do not establish a timing comparison against CMake, Meson or Bazel.

\begin{table}[t]
\caption{Configure API times, seconds: median (minimum--maximum), five samples per cell.}
\small
\begin{tabular}{@{}lrrr@{}}
\toprule
Project & Cold & Warm & Edit \\
\midrule
Trivial & 1.903 (1.865--1.971) & .035 (.034--.036) & .611 (.597--.630) \\
zlib & 2.274 (2.224--2.303) & .050 (.049--.054) & .783 (.772--.809) \\
Halide & 5.987 (5.884--6.307) & .292 (.289--.307) & 3.567 (3.541--3.767) \\
\bottomrule
\end{tabular}
\end{table}

Independent wrappers count actual composition calls, evaluator starts and evaluations.
All three projects perform $1/1/4$ operations cold and $1/1/1$ after edit in every
repetition; every warm run performs $0/0/0$.
An untimed-for-comparison profile attributes .554 of .704 profiled seconds to four certificate
validations (content, directory membership, absence and ancestor symlinks). Profiling overhead
means these seconds are not another timing sample. We leave the validation checks intact.

\subsection{A Standard From the Lattice Join}
\label{sec:eval-supp:why-join}

The cross-compilation example's \texttt{app} depends privately on \texttt{util}, which states
C++20 publicly and depends publicly on \texttt{core}, which states C++17. \texttt{app} states no
standard. Here is the full answer excerpted in \cref{sec:evaluation:why}, for
\texttt{app}'s \texttt{-std=c++20} (x86\_64 release, run from the example's parent directory):
\begin{lstlisting}[language={},basicstyle=\ttfamily\scriptsize,breaklines=true,literate={<-}{{$\leftarrow$}}2]
$ buildlog why -C build --flag -std=c++20 --output obj/x86_64-macos.release.none/app/app/main.cpp.o cross
action_arg("out:obj/x86_64-macos.release.none/app/app/main.cpp.o", 10, "-std=c++", "20")  [cpp:L4_cxx.dl:33]
  cxx_object("app", <x86>, "app/main.cpp", "out:obj/x86_64-macos.release.none/app/app/main.cpp.o")  [cpp:L4_cxx.dl:23]  ^ #n51
    src_on("app", <x86>, "app/main.cpp")  [cpp:L4_cxx.dl:210]
      #n3 needed("app", <x86>)  [exists here because it is built in this configuration; --demand explains why]
      #n51 cxx_source("app", "app/main.cpp")  <- cross/spec.toml:25
      !src_grouped("app", "app/main.cpp")
  use_join("app", <x86>, "cxx.std", "20")  [L1:760]
    use_cand("app", <x86>, "cxx.std", "20", 1)  [L1:742]
      eff_dep("app", <x86>, "util", "private")  [L1:596]
        eff_ref("app", <x86>, "util", "private")  [L1:594]  ^ #n3
          #n74 dep("app", "util", "private")  <- cross/spec.toml:34
        own_ref("app", "util", "util")  [L1:706]  (via dep_resolves)  ^ #n74 (via dep_ref)
          !scoped("app")
          target("util")  <- cross/spec.toml:88  (via target_id)
          !scoped("util")
      eff_prop("util", <x86>, "cxx.std", "20", "public")  [L1:718]  (via iface)
        needed("util", <x86>)  [exists here because it is built in this configuration; --demand explains why]
        prop("util", "cxx.std", "20", "public")  <- cross/spec.toml:70
    lattice_key("cxx.std")  [L1:748]
\end{lstlisting}
Here
\texttt{<x86>} is the record \texttt{["x86\_64-macos", [["native.opt", "release"],
[["native.sanitize", "none"], [["native.linkage", "default"], nil]]]]}. The maximum itself is a
leaf: \texttt{-{}-full} adds the constraint \texttt{20 = 20} under \texttt{use\_join}, and no view
lists the candidates that lost.

The join's first premise binds the value, so the candidate printed is always one that carries the
winning value. In a variant where \texttt{app} also states C++17 privately (one added row, spec
line 71), the flag is still \texttt{-std=c++20}, and the leaf listing still cites \texttt{util}'s
line 70, not \texttt{app}'s C++17 line; a regression test checks this. The losing candidates are
one query away:
\begin{samepage}
\begin{lstlisting}[language={},basicstyle=\ttfamily\scriptsize,breaklines=true]
$ buildlog query -C build-variant 'use_cand("app", c, "cxx.std", v, r)' cross   # x86_64 rows
c      v   r
<x86>  17  0
<x86>  17  1
<x86>  20  1
\end{lstlisting}
Rank~0 is the target's own value and rank~1 an inherited one.
\end{samepage}

\subsection{An Ambiguous Package}
\label{sec:eval-supp:ambiguity}

\sys's package search finds every candidate and does not rank them. A test of the importers gives
a consumer two install prefixes that both hold CMake export files for \texttt{Greet::shout}: an
installed fixture and a copy of it. Configure stops before building anything. With the two
prefixes written as \texttt{<copy>} and \texttt{<prefix>} (sorted, as the error lists them), the
output is:
\begin{lstlisting}[language={},basicstyle=\ttfamily\scriptsize,breaklines=true]
error: ambiguous package: Greet::shout on arm64-macos: <copy>/lib/cmake/Greet/Greet-targets.cmake and <prefix>/lib/cmake/Greet/Greet-targets.cmake both provide it; add a package_pin
error: unresolved dependency: dependency Greet::shout of hello has no definition
\end{lstlisting}
The error is the same whichever order the prefixes are listed in. One line in a user file resolves it:
\begin{lstlisting}[language={},basicstyle=\ttfamily\scriptsize,breaklines=true]
package_pin = [["Greet::shout", "<copy>/lib/cmake/Greet/Greet-targets.cmake"]]
\end{lstlisting}
The build then links the copy's library. Given
the same two prefixes in \texttt{CMAKE\_PREFIX\_PATH}, CMake's \texttt{find\_package} takes the
package file under the first prefix and says nothing~\cite{cmake-find-package}. Reversing the
order of the prefixes silently changes which library is linked.

\bibliographystyle{ACM-Reference-Format}
\bibliography{refs}

\end{document}